\documentclass[12pt,english]{article}
\usepackage{geometry}
\usepackage{amsmath}
\usepackage{amssymb}
\usepackage{bm}
\usepackage{graphicx}
\usepackage{setspace}
\usepackage{xargs}[2008/03/08]
\makeatletter
\numberwithin{equation}{section}
\newcommand{\lyxaddress}[1]{
	\par {\raggedright #1
	\vspace{1.4em}
	\noindent\par}
}

\allowdisplaybreaks[1]

\usepackage{young}
\usepackage{ytableau}
\usepackage{youngtab}
\usepackage{wick}
\usepackage{here}

\usepackage{braketmod}
\usepackage[T1]{fontenc}

\usepackage[]{hyperref}
\hypersetup{
colorlinks,
linktoc=page,
citecolor=blue,
linkcolor=blue,
urlcolor=blue
}

\date{}

\makeatother

\usepackage{babel}
\usepackage[style=phys,biblabel=brackets, chaptertitle=false, pageranges=false, eprint=true]{biblatex}
\begin{document}
\begin{flushright}
{\small{}YITP-21-140}{\small\par}
\par\end{flushright}

\noindent\begin{minipage}[t]{1\columnwidth}%
\title{\textbf{3d $\mathcal{N}=3$ Generalized Giveon-Kutasov Duality}}
\author{Naotaka Kubo\,\footnotemark,~~~~Keita Nii\footnotemark}
\maketitle

\lyxaddress{\begin{center}
\vspace{-18bp}
\textit{$^*\, ^\dagger$Center for Gravitational Physics, Yukawa Institute for Theoretical Physics,}\\
\textit{Kyoto University, Sakyo-ku, Kyoto 606-8502, Japan}\vspace{-10bp}
\par\end{center}}

\begin{abstract}
We generalize the Giveon-Kutasov duality for the 3d $\mathcal{N}=3$ $U(N)_{k,k+nN}$ Chern-Simons matter gauge theory with $F$ fundamental hypermultiplets by introducing $SU(N)$ and $U(1)$ Chern-Simons levels differently. We study the supersymmetric partition functions and the superconformal indices of the duality, which supports the validity of the duality proposal.
From the duality, we can map out the low-energy phases: For example, confinement appears for $F+k-N=-n=1$ or $N=2F=k=-n=2$. For $F+k-N<0$, supersymmetry is spontaneously broken, which is in accord with the fact that the partition function vanishes. In some cases, the theory shows supersymmetry enhancement to 3d $\mathcal{N}=4$.
For $k=0$, we comment on the magnetic description dual to the so-called ``ugly'' theory, where the usual decoupled sector is still interacting with others for $n \neq 0$. We argue that the $SU(N)_0$ ``ugly-good'' duality (which corresponds to the $n \rightarrow \infty$ limit in our setup) is closely related to the S-duality of the 4d $\mathcal{N}=2$ $SU(N)$ superconformal gauge theory with $2N$ fundamental hypermultiplets. By reducing the number of flavors via real masses, we suggest possible ways to flow to the ``bad'' theories.   
\end{abstract}

%
\end{minipage}

\renewcommand{\thefootnote}{\fnsymbol{footnote}}
\footnotetext[1]{\textsf{naotaka.kubo@yukawa.kyoto-u.ac.jp}}
\footnotetext[2]{\textsf{nii.keita1209@gmail.com}}
\renewcommand{\thefootnote}{\arabic{footnote}}

\newpage{}

\newcommandx\zU[4][usedefault, addprefix=\global, 1=N, 2=F, 3=k, 4=n]{\text{\ensuremath{Z_{#2,#3,#4}^{U\left(#1\right)}}}}%

\newcommandx\zUab[2][usedefault, addprefix=\global, 1=F, 2=k]{\text{\ensuremath{Z_{#1,#2}^{U\left(1\right)}}}}%

\newcommandx\zUmag[4][usedefault, addprefix=\global, 1=N, 2=F, 3=k, 4=n]{\text{\ensuremath{Z_{#2,#3,#4}^{U\left(#1\right),{\rm mag}}}}}%

\newcommandx\zSU[3][usedefault, addprefix=\global, 1=N, 2=F, 3=k]{\text{\ensuremath{Z_{#2,#3}^{SU\left(#1\right)}}}}%

\newcommandx\zpureU[2][usedefault, addprefix=\global, 1=N, 2=k]{\text{\ensuremath{Z_{#2}^{U\left(#1\right),{\rm pure}}}}}%

\global\long\def\bra#1{\Bra{#1}}%

\global\long\def\bbra#1{\Bbra{#1}}%

\global\long\def\ket#1{\Ket{#1}}%

\global\long\def\kket#1{\Kket{#1}}%

\global\long\def\braket#1{\Braket{#1}}%

\global\long\def\bbraket#1{\Bbraket{#1}}%

\global\long\def\brakket#1{\Brakket{#1}}%

\global\long\def\bbrakket#1{\Bbrakket{#1}}%

\tableofcontents{}

\section{Introduction}
A Chern-Simons (CS) gauge field has a rich dynamics: For example, CS gauge theories describe topological invariants of knots \cite{Witten:1988hf}. Although a CS gauge field itself has no physical propagating degree of freedom, the CS gauge interaction transmutes the spin of the particle that couples with it to be fractional \cite{Polyakov:1988md, Semenoff:1988jr}. When the CS theory is defined on a manifold with boundaries, its Hilbert space becomes infinite dimensional \cite{Witten:1988hf, Elitzur:1989nr}. Yang-Mills Chern-Simons gauge theories describe a topologically massive particle \cite{Schonfeld:1980kb, Deser:1981wh, Deser:1982vy}. Recently, the non-supersymmetric CS gauge theories with bosonic/fermionic matter fields have been extensively studied since they enjoy infrared bosonization dualities \cite{Giombi:2011kc, Aharony:2011jz, Seiberg:2016gmd, Karch:2016sxi}. Also, in a supersymmetric setup, the Seiberg-like duality was proposed for the CS gauge theory \cite{Giveon:2008zn} and generalized in \cite{Niarchos:2008jb, Kapustin:2011gh, Willett:2011gp, Benini:2011mf, Aharony:2014uya, Nii:2020ikd, Amariti:2021snj}. These can be regarded as a supersymmetric generalization of the 3d bosonization duality.  

This paper focuses on the Giveon-Kutasov (GK) duality with extended supersymmetry \cite{Giveon:2008zn, Kapustin:2010mh}.
In particular, we consider the 3d $\mathcal{N}=3$ version of the generalized Giveon-Kutasov duality with a $U(N)_{k,k+nN}$ gauge group and $F$ fundamental hypermultiplets. The Chern-Simons levels are independently introduced for the $SU(N)$ and $U(1)$ subgroups, which are denoted by $k$ and $n$, respectively. We find that the dual description can be obtained by slightly modifying the 3d $\mathcal{N}=2$ generalized GK duality \cite{Nii:2020ikd, Amariti:2021snj}. By using the duality, we will show that the theory has confinement, supersymmetry-breaking and supersymmetry-enhancement phases depending on the value of $N,F,k$ and $n$.
This paper considers generalizing the Chern-Simons level of the $U(N)$ gauge group. This is not only the generalization of the CS interaction but also the unification of the unitary and special unitary gauge groups. By taking the $n \rightarrow \infty$ limit, the $U(N)_{k,k+nN}$ group is reduced to $SU(N)_k$. In this way, we can give a unified understanding of the $(S)U(N)$ Chern-Simons matter (level-rank) dualities.    

This paper also discusses the 3d $\mathcal{N}=3$ duality for the $k=0$ and $n \neq 0$ case with $F=2N-1$ flavors. This can be regarded as the generalization of the so-called ``ugly-good'' duality in the 3d $\mathcal{N}=4$ $U(N)_{0,0}$ gauge theory with $2N-1$ hypermultiplets \cite{Kapustin:2010mh, Yaakov:2013fza, Assel:2017jgo}. The theory is called ``ugly'' since the low-energy theory contains a decoupled free sector. The (dual) ``good'' theory must be attached to a corresponding free hypermultiplet. For $n \neq 0$, we argue that the (would-be) decoupled sector is still interacting with the (dual) ``good'' theory. For $n \rightarrow \infty$, we obtain the $SU(N)_0$ ``ugly-good'' duality with $2N-1$ fundamental hypermultiplets. We find that the $n \rightarrow \infty$ duality is naturally obtained from the S-duality of the 4d $\mathcal{N}=2$ $SU(N)$ gauge theory with $2N$ hypermultiplets via a circle compactification. See \cite{Benini:2017dud, Garozzo:2019xzi}, where the similar duality and the S-duality (wall) are considered in a 3d $\mathcal{N}=2$ set-up. 

The rest of this paper is organized as follows: 
In Section \ref{section:GKduality}, we propose a generalized Giveon-Kutasov duality for the 3d $\mathcal{N}=3$ $U(N)_{k,k+nN}$ Chern-Simons matter gauge theory. 
In Section \ref{subsection:ugly}, we present the generalized ``ugly-good'' duality and discuss the connection between the $SU(N)$ ``ugly-good'' duality and the 4d S-duality. We also comment on possible ways to study the ``bad'' theory.
Section \ref{sec:MatrixModel} and Section \ref{sec:Index} are devoted to some non-trivial tests of the 3d $\mathcal{N}=3$ generalized GK duality by computing the supersymmetric sphere partition functions and the superconformal indices. 
In Section \ref{sec:Summary}, we will summarize our findings and discuss future directions.
In Appendix \ref{sec:QM} and Appendix \ref{sec:MM-Duality}, we summarize our notations and formulae for the matrix models and prove the matrix integral identities which come from the Seiberg-like dualities known in the literature.

\section{The 3d $\mathcal{N}=3$ $U(N)_{k,k+n N}$ generalized GK duality}\label{section:GKduality}
In this section, we consider the 3d $\mathcal{N}=3$ supersymmetric $U(N)_{k,k+nN}$ Giveon-Kutasov duality with fundamental hypermultiplets. This duality can be obtained from the 3d $\mathcal{N}=3$ $U(N)_{k,k}$ Giveon-Kutasov duality \cite{Giveon:2008zn, Kapustin:2010mh} by slightly modifying the magnetic description \`{a} la \cite{Radicevic:2016wqn, Nii:2020ikd}. See \cite{Radicevic:2016wqn} for a non-supersymmetric version of the similar duality, which is known as the 3d non-abelian bosonization/fermionization. Therefore, the duality proposed in this paper can be regarded as a 3d $\mathcal{N}=3$ version of the bosonization duality. In order to describe the electric and magnetic theories explicitly, we use the notation of the 3d $\mathcal{N}=2$ superspace/superfield. The electric theory is given by a 3d $\mathcal{N}=2$ $U(N)_{k,k+nN}$ Chern-Simons gauge theory with $F$ (anti-)fundamental flavors (denoted by $Q$ and $\tilde{Q}$) and a single adjoint chiral multiplet (denoted by $\Phi$). The $SU(N)$ and $U(1)$ Chern-Simons levels are tuned to be $k$ and $k+nN$, respectively. For the $U(N)$ Chern-Simons levels, we use the notation adopted for example in \cite{Hsin:2016blu}. The theory includes a tree-level superpotential
\begin{align}
    W_{ele}= -\frac{k}{4\pi} \, \mathrm{tr} \left(\Phi^2 \right) - \frac{n}{4\pi} \left( \mathrm{tr}\, \Phi \right)^2  +  Q\Phi\tilde{Q},
\end{align}
which completes the electric theory such that the resulting Lagrangian possesses a 3d $\mathcal{N}=3$ supersymmetry. The bosonic global symmetry, which is explicitly realized in the 3d $\mathcal{N}=2$ notation, is $SU(F) \times U(1)_R$. The $U(1)_R$ symmetry is an R-symmetry and will be used for the localization calculus. 
As we explain below, the axial $U(1)_A$ symmetry is not a symmetry of the theory: First, notice that 3d $\mathcal{N}=4$ gauge theories without CS levels have an $SU(2)_C \times SU(2)_H$ R-symmetry and only the $U(1)_C \times U(1)_H$ subgroup is manifest in the 3d $\mathcal{N}=2$ notation. The axial $U(1)_A$ symmetry, which is defined as the difference between these two $U(1)$ R-symmetries, is explicitly broken due to the CS superpotential for $\Phi$. See Table \ref{electric} for the charge assignment of the electric elementary fields.

\begin{table}[h]\caption{The 3d $\mathcal{N}=3$ $U(N)_{k,k+nN}$ gauge theory with $F$ fundamental hypermultiplets in the 3d $\mathcal{N}=2$ notation. The theory consists of $F$ (anti-)fundamental and one adjoint chiral superfields in addition to the $U(N)_{k,k+nN}$ vector multiplet.} 
\begin{center}
\scalebox{1}{
  \begin{tabular}{|c||c|c|c| } \hline
  &$U(N)_{k,k+nN}$&$SU(F)$& $U(1)_R$  \\ \hline
$Q$&${\tiny \yng(1)}_{1}$&${\tiny \yng(1)}$&$\frac{1}{2}$ \\
$\tilde{Q}$&${\tiny \overline{\yng(1)}}_{-1}$&${\tiny \overline{\yng(1)}}$&$\frac{1}{2}$ \\
$\Phi$&$\mathbf{adj.}_{0}$&$\mathbf{1}$&$1$ \\
$W_\alpha$&$\mathbf{adj.}_{0}$&$\mathbf{1}$&$1$ \\ \hline
  \end{tabular}}
  \end{center}\label{electric}
\end{table}

Next, we move on to the magnetic dual description, which is given by a 3d $\mathcal{N}=3$ $\stackrel{+1~~~~~~~~~~~~~~~~~~~~~~~~~~~~~~~~~~~}{\protect\wick{1}{<1 U(1)_{n+1} \times >1 U ( F +k-N )_{-k,-k+(F+k-N) }}}$ gauge theory with $F$ fundamental (dual) hypermultiplets. Unlike the 3d $\mathcal{N}=2$ Seiberg-like dualities, there is no meson singlet on the magnetic side. We use lower case letters to denote the magnetic elementary fields (as shown in Table \ref{magnetic}). The $U(1)_{n+1}$ subgroup is a gauged topological $U(1)$ symmetry associated with the overall $U(1) \subset U(F+k-N)$ subgroup. The ``Wick'' contraction indicates that there is a level-$1$ mixed CS term between the abelian $U(1)_{n+1}$ and $U(1)\subset U(F+k-N)$ gauge nodes. The magnetic matter fields are neutral under the $U(1)_{n+1}$ gauge symmetry. The magnetic theory also has a tree-level superpotential
\begin{align}
    W_{mag}=- \frac{n+1}{4 \pi}\phi_{U(1)}^2 -\frac{1}{2\pi}\phi_{U(1)}  \mathrm{tr}\, \phi +  \frac{k}{4\pi} \, \mathrm{tr} \left(\phi^2 \right) - \frac{1}{4 \pi} \left( \mathrm{tr}\, \phi \right)^2 +q \phi  \tilde{q}.
\end{align}
The matter content and the quantum numbers of the magnetic fields are summarized in Table \ref{magnetic}. We here listed the Chern-Simons matter duality for the $k>0$ region. The duality with $k<0$ can be simply obtained by applying the time-reversal transformation.\footnote{For a non-zero $k$ (including the $k<0$ region), the 3d $\mathcal{N}=3$ generalized Giveon-Kutasov duality is summarized into a concise form:\begin{align}
  3d~~\mathcal{N}=3 ~~~~ U(N)_{k,k+nN} ~~ \Leftrightarrow ~~ \stackrel{+\mathbf{sgn}(k)~~~~~~~~~~~~~~~~~~~~~~~~~~~~~~~~~~~~~~~~~~~~~~~}{\protect\wick{1}{<1 U(1)_{n+\mathbf{sgn}(k)} \times >1 U ( F +|k|-N )_{-k,-k+\mathbf{sgn}(k)(F+|k|-N) }}}.
\end{align}
The $k=0$ and $n \neq 0$ duality, which will be discussed in Section \ref{subsection:ugly}, is also consistent with this form of the duality by regarding $\mathbf{sgn}(0)=0$ although the matter content of the magnetic theory is slightly modified. 
} As a quick consistency check of the duality, we can see that the duality twice takes us back to the original theory: First, by acting the duality transformation on the magnetic $U(F+k-N)$ gauge group, we obtain the $\stackrel{+1~~~~~~~~~~-1~~~~~~~~~}{\protect\wick{11}{<1 U(1)_{n+1} \times >1U(<2 1)_0 \times  >2U (N )_{k,k-N }}}$ second dual description. By integrating out the $U(1)_0$ vector multiplet, which does not couple with the magnetic matter sector and just works as a Lagrange multiplier, we reproduce the electric $U(N)_{k,k+nN}$ gauge group. This serves as a quick consistency check of our duality. 

\begin{table}[h]\caption{The magnetic $\stackrel{+1~~~~~~~~~~~~~~~~~~~~~~~~~~~~~~~~~~~}{\protect\wick{1}{<1 U(1)_{n+1} \times >1 U ( F +k-N )_{-k,-k+(F+k-N) }}}$ gauge theory dual to Table \ref{electric}. The theory consists of $F$ fundamental hyper and vector multiplets. All the fields are not charged under the $U(1)_{n+1}$ subgroup and its charge assignment is simply abbreviated in the table.} 
\begin{center}
\scalebox{1}{
  \begin{tabular}{|c||c|c|c| } \hline
  &$\stackrel{+1~~~~~~~~~~~~~~~~~~~~~~~~~~~~~~~~~~~}{\protect\wick{1}{<1 U(1)_{n+1} \times >1 U ( F +k-N )_{-k,-k+(F+k-N) }}}$&$SU(F)$& $U(1)_R$  \\ \hline
$q$&${\tiny \yng(1)}_{\, 1}$&${\tiny \yng(1)}$&$\frac{1}{2}$ \\
$\tilde{q}$&${\tiny \overline{\yng(1)}}_{-1}$&${\tiny \overline{\yng(1)}}$&$\frac{1}{2}$ \\
$\phi_{U(1)}$&$\mathbf{1}_{0}$&$\mathbf{1}$&$1$ \\
$\tilde{w}^{U(1)}_\alpha$&$\mathbf{1}_{0}$&$\mathbf{1}$&$1$ \\
$\phi$&$\mathbf{adj.}_{0}$&$\mathbf{1}$&$1$ \\
$\tilde{w}_\alpha$ & $\mathbf{adj.}_{0}$ & $\mathbf{1}$ & $1$  \\  \hline
  \end{tabular}
  }
  \end{center}\label{magnetic}
\end{table}

As another consistency check of our duality, we can derive the 3d $\mathcal{N}=2$, $\mathcal{N}=1$ and $\mathcal{N}=0$ (non-supersymmetric) level-rank dualities \cite{Naculich:1990pa} from our duality proposal with $F=0$. 
Since this can be done by integrating out non-dynamical fermions/scalars of the 3d $\mathcal{N}=3$ vector multiplet, the resulting dualities with various amounts of supersymmetry give rise to the identical description of the gapped phase. 
Firstly, we integrate out the adjoint chiral superfield $\Phi$ in the 3d $\mathcal{N}=4$ vector multiplet. The $\Phi$ field has a mass term proportional to $k$ in the superpotential. Since the superpotential mass is parity-preserving, this integration does not induce the shift of the CS level. The same argument can be applied for the magnetic side. Then, we immediately obtain the 3d $\mathcal{N}=2$ generalized level-rank (GK) duality studied in \cite{Nii:2020ikd}. Secondly, we break one more supersymmetry and flow to the 3d $\mathcal{N}=1$ theory: The $\mathcal{N}=2$ vector multiplet consists of the 3d $\mathcal{N}=1$ vector and (real) scalar multiplets. By integrating out the scalar multiplet which is massive and generates the shift of the non-abelian CS level, we obtain the 3d $\mathcal{N}=1$ generalized level-rank duality
\begin{align}
\mathcal{N}=1: ~~~ U(N)_{k+\frac{N}{2},k+nN} ~ \Leftrightarrow  ~\, \stackrel{+1~~~~~~~~~~~~~~~~}{\protect\wick{1}{<1 U(1)_{n} \times >1 U ( k )_{-N-\frac{k}{2},-N }}},   
\end{align}
where we relabeled the parameter as $k \rightarrow k+N$ and $n \rightarrow n-1$ for simplifying the duality relation. For some $n$'s, this duality has been proposed in \cite{Benini:2018umh}. Finally, we can further flow to the non-supersymmetric level-rank duality \cite{Hsin:2016blu, Radicevic:2016wqn}. By integrating out the massive gaugino of the 3d $\mathcal{N}=1$ vector multiplet, we can reproduce the (generalized) level-rank duality proposed in \cite{Radicevic:2016wqn} with $F=0$. In this way, we obtain a supersymmetric ``ladder'' of the generalized $U(N)_{k,k+nN}$ level-rank duality with various amounts of supersymmetry. The sequence of the supersymmetry reduction is schematically depicted as follows: 
\begin{align}
& \mathcal{N}=3:~~~U(N)_{k,k+nN} ~ \Leftrightarrow ~\, \stackrel{+1~~~~~~~~~~~~~~~~~~~~~~~~~}{\protect\wick{1}{<1 U(1)_{n+1} \times >1 U ( k-N )_{-k,-k+(k-N) }}} \\
& \quad  \Downarrow \nonumber \\
& \mathcal{N}=2:~~~U(N)_{k,k+nN} ~\Leftrightarrow ~ \,  \stackrel{+1~~~~~~~~~~~~~~~~~~~~~~~~~}{\protect\wick{1}{<1 U(1)_{n+1} \times >1 U ( k-N )_{-k,-k+(k-N) }}} \\
&  \quad \Downarrow \nonumber \\
& \mathcal{N}=1:~~~U(N)_{k-\frac{N}{2},k+nN} ~ \Leftrightarrow  ~\, \stackrel{+1~~~~~~~~~~~~~~~~~~~~~~~~~~~~~~}{\protect\wick{1}{<1 U(1)_{n+1} \times >1 U ( k-N )_{-\frac{k}{2}-\frac{N}{2},-k+(k-N) }}} \\
&  \quad \Downarrow \nonumber \\
&    \mathcal{N}=0:~~~U(N)_{k-N,k-N+(n+1)N} ~ \Leftrightarrow  ~\, \stackrel{+1~~~~~~~~~~~~~~~~~~~~~~~~~~}{\protect\wick{1}{<1 U(1)_{n+1} \times >1 U (k-N )_{-N,-k+(k-N) }}},
\end{align}
where we kept the shifts of the Chern-Simon levels explicitly. By inserting $k \rightarrow k+N$ and by considering some small $n$'s, one can obtain simpler and familiar forms of the level-rank dualities.

For several $n's$, the proposed duality can be recast into simpler forms: For $n=\infty$, the $U(1)$ CS gauge dynamics becomes weaker and weaker since the CS gauge interaction is characterized by the inverse square root of the CS level. By dropping off the $U(1)$ gauge dynamics on both sides, we obtain the 3d $\mathcal{N}=3$ $SU(N)_k$ duality for $n \rightarrow \infty$
\begin{align}
    \mathcal{N}=3:~~~SU(N)_{k} ~ \Leftrightarrow ~\, U(F+k-N)_{-k,-k+(F+k-N)}.\label{eq:nInf-kneq0}
\end{align}
For $n=0,-1$ and $-2$, we can integrate out the $U(1)_{n+1}$ vector multiplet by taking a linear combination of the $U(1)_{n+1}$ and $U(1) \subset U(F+k-N)$ abelian gauge symmetries. We thus obtain the simplified versions of the 3d $\mathcal{N}=3$ dualities:
\begin{align}
   & \mathcal{N}=3:~~~U(N)_{k,k} ~ \Leftrightarrow ~\, U(F+k-N)_{-k,-k}\\
    & \mathcal{N}=3:~~~U(N)_{k,k-N} ~ \Leftrightarrow ~\, SU(F+k-N)_{-k} \\
    & \mathcal{N}=3:~~~U(N)_{k,k-2N} ~ \Leftrightarrow ~\, U(F+k-N)_{-k,-k+2(F+k-N)}.
\end{align}
The first line is nothing but the 3d $\mathcal{N}=3$ version of the Giveon-Kutasov duality \cite{Giveon:2008zn, Kapustin:2010mh}. The similar simplification of the generalized GK duality was discussed for 3d $\mathcal{N}=2$ in \cite{Nii:2020ikd} and we do not repeat it here. Also for the 3d $\mathcal{N}=1$ cases, the generalized level-rank duality is simplified by integrating out the magnetic $U(1)_{n+1}$ vector multiplet. As a result, we obtain the simpler forms of the 3d $\mathcal{N}=1$ level-rank dualities
\begin{align}
    & \mathcal{N}=1:~~~SU(N)_{k+\frac{N}{2}} ~ \Leftrightarrow ~\, U(k)_{-N-\frac{k}{2},-N}\\
    & \mathcal{N}=1:~~~U(N)_{k+\frac{N}{2},k} ~ \Leftrightarrow ~\, SU(k)_{-N-\frac{k}{2}} \\
    & \mathcal{N}=1:~~~U(N)_{k+\frac{N}{2},k+N} ~ \Leftrightarrow ~\, U(k)_{-N-\frac{k}{2},-N-k} \\
     & \mathcal{N}=1:~~~U(N)_{k+\frac{N}{2},k-N} ~ \Leftrightarrow ~\, U(k)_{-N-\frac{k}{2},-N+k}. 
\end{align}
These are for example proposed and studied in \cite{Benini:2018umh}. 

Next, we comment on the duality enhancement for the abelian gauge group. Namely, we take $N=1$. As studied in \cite{Nii:2020ikd}, we can show that the 3d $\mathcal{N}=3$ $U(1)$ SQED with and without a CS level has an infinite number of magnetic dual descriptions: For $N=1$, the electric side becomes a 3d $\mathcal{N}=3$ $U(1)_{\tilde{k}=k+n}$ gauge theory with $F$ fundamental hypermultiplets. The electric theory is parametrized by a single parameter $\tilde{k}=k+n$ while the magnetic side depends on $k$ and $n$, independently. By changing $k$ and $n$ with $\tilde{k}$ fixed, we can construct an infinite number of magnetic duals. We here show some simple examples of the duality enhancement in the generalized GK duality with one flavor: 
\begin{align}
     & \mathcal{N}=3:~~~U(1)_{0} \mbox{~w/~} F=1 ~ \Leftrightarrow ~\, \stackrel{+1~~~~~~~~~~~~~~}{\protect\wick{1}{<1 U(1)_{-k+1} \times >1 U ( k)_{-k,-k+k }}}  \mbox{~w/~} F=1 \\
     & \qquad \qquad \qquad \qquad \qquad  \qquad  \Leftrightarrow ~\, \mbox{one free hypermultiplet} \nonumber \\
       & \mathcal{N}=3:~~~U(1)_{n} \mbox{~w/~} F=1 ~ \Leftrightarrow ~\, \stackrel{+1~~~~~~~~~~~~~~}{\protect\wick{1}{<1 U(1)_{n-k+1} \times >1 U ( k)_{-k,-k+k }}}  \mbox{~w/~} F=1 
\end{align}
Especially, for $\tilde{k}=k+n=0$, we can construct the CS duality for the 3d $\mathcal{N}=4$ $U(1)_0$ gauge theory. The dual description has a non-zero CS level and hence it has an explicit $\mathcal{N}=3$ supersymmetry while the electric side has an $\mathcal{N}=4$ supersymmetry for $\tilde{k}=k+n=0$. This is an example of the supersymmetry enhancement.   

Finally, we consider the confinement (or more exotic) phases of the 3d $\mathcal{N}=3$ $U(N)_{k,k+nN}$ Chern-Simons matter gauge theory with fundamental flavors. Roughly speaking, the confinement (or gapped) phases appear when the dual gauge group vanishes. When the rank of the magnetic gauge group becomes negative, we expect that supersymmetry is spontaneously broken. This is closely tied with the vanishing of the partition function as will be seen in the next section. In what follows, we study the confinement phase with massless degrees of freedom. For $n=-1$, the dual gauge group becomes $SU(F+k-N)_{-k}$ and then we expect that the electric $U(N)_{k,k-N}$ gauge theory exhibits confinement for $F+k-N=1$, where the low-energy massless degrees of freedom are described by the dual (gauge-invariant) quarks, $q$ and $\tilde{q}$. Since the magnetic theory consists only of hypermultiplets, the electric theory exhibits an enhanced $\mathcal{N}=4$ supersymmetry.
On the electric $U(N)_{k,k-N}$ side, these massless modes are defined as dressed monopole operators. Let us consider the electric Coulomb branch where the gauge group is spontaneously broken as
\begin{align}
    U(N)_{k,k-N} & \rightarrow \stackrel{-1~~~~~~~~~~~~~~~~~~~~~~}{\protect\wick{1}{<1 U(k)_{k,0} \times >1 U (N-k)_{k,k-(N-k)}}}  \\
    {\tiny \yng(1)}_{\, +1} & \rightarrow ({\tiny \yng(1)}_{\, +1},1_0)+(1_0,{\tiny \yng(1)}_{\,+1}) \\
    {\tiny \overline{ \yng(1)}}_{\, -1} & \rightarrow ({\tiny \overline{ \yng(1)}}_{\, -1},1_0)+(1_0,{\tiny \overline{ \yng(1)}}_{\,-1}) \\
\mathbf{adj.}_{0} & \rightarrow (\mathbf{adj.}_{0},1_0)+(1_0,\mathbf{adj.}_{0})+({\tiny  \yng(1)}_{\,1}, {\tiny \overline{ \yng(1)}}_{\, -1}) +({\tiny \overline{ \yng(1)}}_{\, -1},{\tiny  \yng(1)}_{\,1} ),
\end{align}
where the symbol $\mathbf{adj.}$ denotes the gaugino $W_\alpha$ superfield. Since the effective CS level for the $U(1) \subset U(k)$ subgroup is zero under this breaking, the associated Coulomb branch is exactly massless even if the original gauge theory has non-zero CS levels. As a result, there could be massless modes in the infrared physics. However, this Coulomb branch in itself is not gauge-invariant since the mixed CS level induces a non-zero $U(1) \subset U(N-k)$ charge to the bare Coulomb branch operator $V_{\pm}^{bare}$. Therefore, we need to define the dressed Coulomb branch operators 
\begin{align}
    V_{+ d} &:= V_{+}^{bare} (1_0,{\tiny \overline{ \yng(1)}}_{\,-1})^{N-k} \sim V_{+}^{bare} \tilde{Q}^{N-k} \label{V_+d} \\
    V_{- d} &:= V_{-}^{bare} (1_0,{\tiny \yng(1)}_{\,+1})^{N-k} \sim V_{-}^{bare} Q^{N-k}, \label{V_-d}
\end{align}
which describe the mixed (Coulomb-baryonic) branch of the moduli space. 
These are identified with the dual (gauge-invariant) quark hypermultiplets and the low-energy theory is in an s-confinement phase and becomes free. In the above construction of the dressed Coulomb branch, we rely on the low-energy picture of the theory and study the massless modes along the Coulomb branch. Alternatively, we can argue that the same dressed coordinates can be defined solely by using the ultra-violet degrees of freedom. For that purpose, we consider the monopole operator $V_{U(1), \pm}^{bare}$ whose insertion induces the gauge symmetry breaking
\begin{align}
    U(N)_{k,k-N} & \rightarrow \stackrel{-1~~~~~~~~~~~~~~~~~~~~~~}{\protect\wick{1}{<1 U(1)_{k-1} \times >1 U (N-1)_{k,k-(N-1)}}}  \\
    {\tiny \yng(1)}_{\, +1} & \rightarrow (+1,1_0)+(0,{\tiny \yng(1)}_{\,+1}) \\
    {\tiny \overline{ \yng(1)}}_{\, -1} & \rightarrow (-1,1_0)+(0,{\tiny \overline{ \yng(1)}}_{\,-1}) \\
\mathbf{adj.}_{0} & \rightarrow (0,1_0)+(0,\mathbf{adj.}_{0})+(+1, {\tiny \overline{ \yng(1)}}_{\, -1}) +(-1,{\tiny  \yng(1)}_{\,1} ).
\end{align}
Since this is not associated with the flat direction of the Coulomb moduli space, the $U(1)$ CS level does not vanish at all. As a result, the bare monopole operator $V_{U(1), \pm}^{bare}$ is charged under both of the $U(1)_{k-1}$ and $U(1) \subset U(N-1)$ subgroups. In order to construct gauge-invariant monopole operators, we need to dress the bare operator as
\begin{align}
    V_{+ d} &:=V_{U(1), +}^{bare} (+1, {\tiny \overline{ \yng(1)}}_{\, -1})^{k-1}(0,{\tiny \overline{ \yng(1)}}_{\,-1})^{N-k}   \sim V_{U(1), +}^{bare} W_\alpha^{k-1} \tilde{Q}^{N-k}    \\
    V_{- d} &:=V_{U(1), -}^{bare} (-1,{\tiny  \yng(1)}_{\,1} )^{k-1}  (0,{\tiny \yng(1)}_{\,+1})^{N-k} \sim V_{U(1), -}^{bare} W_\alpha^{k-1} Q^{N-k}. 
\end{align}
On the monopole background, the gaugino field behaves as a boson and hence the above composites totally become bosonic \cite{Polyakov:1988md, Dimofte:2011py, Aharony:2015pla}. These dressed operators are identified with the dual (gauge-invariant) hypermultiplets.

We can find other confinement phases by combining the proposed and known dualities: Let us consider the 3d $\mathcal{N}=3$ $U(2)_{2,2-4}$ gauge theory with one hypermultiplet, which corresponds to the $N=2F=k=-n=2$ case. The proposed duality claims that the magnetic description is given by a 3d $\mathcal{N}=4$ $U(1)_0$ SQED with one hypermultiplet, which is known to be in a confinement phase and dual to a free hypermultiplet \cite{Intriligator:1996ex, Kapustin:1999ha}. The free hypermultiplet corresponds to the bare monopole. By using this fact, we can conclude that the 3d $\mathcal{N}=3$ $U(2)_{2,2-4}$ gauge theory with one hypermultiplet exhibits confinement and supersymmetry enhancement to 3d $\mathcal{N}=4$. On the electric $U(2)_{2,2-4}$ side, this free hypermultiplet is defined as a dressed monopole operator. The bare monopole operator is associated with the gauge symmetry breaking
\begin{align}
      U(2)_{2,2-4} & \rightarrow \stackrel{-2~~~~~~}{\protect\wick{1}{<1 U(1)_{0} \times >1 U (1)_{0}}}  \\
    {\tiny \yng(1)}_{\, +1} & \rightarrow (+1,0)+(0,+1) \\
    {\tiny \overline{ \yng(1)}}_{\, -1} & \rightarrow (-1,0)+(0,-1),
\end{align}
where the tree-level CS terms are decomposed as indicated above. These CS levels are not shifted by massive components since there are equal numbers of positively and negatively charged components. Due to the non-zero mixed CS term between the two $U(1)$ subgroups, the bare monopole $V_{\pm}^{bare}$ is not gauge-invariant. As a result, a dressed monopole operator should be defined as
\begin{align}
    V_{+d} &:=V_+^{bare} (0,-1)^2 \sim V_+^{bare} \tilde{Q}^2 \\
    V_{-d} &:=V_-^{bare} (0,+1)^2 \sim V_-^{bare} Q^2.
\end{align}
These two chiral superfields consist of a free hypermultiplet and leads to an $\mathcal{N}=4$ supersymmetry in the far-infrared limit. The similar supersymmetry enhancement can be generalized for other non-confining cases: For $k=-n=2$ and $F=N-1$, the dual description becomes a 3d $\mathcal{N}=4$ $U(1)_0$ gauge theory with $N-1$ hypermultiplets. Since the magnetic side has an enhanced $\mathcal{N}=4$ supersymmetry, we claim that the low-energy dynamics of the 3d $\mathcal{N}=3$ $U(N)_{2,2-2N}$ CS gauge theory with $F=N-1$ hypermultiplets also has an enhanced $\mathcal{N}=4$ supersymmetry. This is an example of the SUSY enhancement in the infrared physics.  

\subsection{The magnetic dual for the ``ugly'' $U(N)_{0,0+nN}$ gauge theory}\label{subsection:ugly}
In this subsection, we comment on the 3d $\mathcal{N}=3$ duality for the $U(N)_{0,0+nN}$ gauge group with $F$ fundamental hypermultiplets. The 3d $\mathcal{N}=2$ version is discussed in \cite{Amariti:2021snj}. The Chern-Simons level is only introduced for the abelian subgroup and the $SU(N)$ dynamics is governed by the Yang-Mills interaction. As a result, this can be seen as a partially Seiberg-like and partially Giveon-Kutasov duality.
As studied in \cite{Kapustin:2010mh, Yaakov:2013fza, Assel:2017jgo}, the 3d $\mathcal{N}=4$ Seiberg duality with no CS level is correct only for the ``ugly'' theory (with $F=2N-1$).\footnote{The scaling dimension of the (bare) monopole operator does not become $\frac{1}{2}$, for example, in the 3d $\mathcal{N}=4$ $SU(N)$ gauge theory with $F=2N-1$ flavors and hence there is no region of ``ugly''. In this paper, we call the theory with $F=2N-1$ flavors ``ugly'' for convenience although it is a misnomer.} Even in the ugly case, there must be an additional free hypermultiplet attached to the dual (good) description. In \cite{Assel:2017jgo}, it is argued that the Seiberg duality proposed in \cite{Yaakov:2013fza} should be regarded as a local Seiberg duality for $N \le F \le 2N-2$.
Here, we propose the magnetic theory dual to the 3d $\mathcal{N}=3$ $U(N)_{0,0+nN}$ gauge theory with $F=2N-1$ flavors. The magnetic dual description for $k=0$ is given by a 3d $\mathcal{N}=3$ $U(1)_{n} \times U(F-N)_{0,0}$ gauge theory with $F=2N-1$ fundamental flavors and a $U(1)$-charged hypermultiplet denoted by $b$ and $\tilde{b}$. See Table \ref{magnetic_k=0} for the charge assignment of the elementary fields. We call the additional matter ``electron'' since it is only charged under the $U(1)_n \times U(1)_0 \subset U(F-N)$ subgroup. Notice that there is no mixed CS term for the dual gauge group \`{a} la \cite{Amariti:2021snj}.

As a simple check of the duality, the $N=1$ case just becomes a self-duality of the 3d $\mathcal{N}=3$ $U(1)_n$ with one hypermultiplet. For $n=0$, we can reproduce the 3d $\mathcal{N}=4$ $U(N)_{0,0}$ Seiberg-like duality between the ``ugly'' and ``good'' theories, where the good theory is attached to a free hypermultiplet \cite{Kapustin:2010mh, Yaakov:2013fza, Assel:2017jgo}. This can be seen as follows: Since the $U(1)_{n=0}$ gauge theory has a single flavor $(b,\tilde{b})$, the $U(1)_{n=0}$ dynamics is confined into a monopole operator \cite{Intriligator:1996ex, Kapustin:1999ha} which becomes free in the infrared limit and decouples from the other sector. As a result, we obtain the duality \cite{Kapustin:2010mh, Yaakov:2013fza, Assel:2017jgo}:
\begin{align}
    3d~~\mathcal{N}=4~~~~~U(N)_{0,0}~~~~ \Leftrightarrow ~~~~U(N-1)_{0,0}+\mbox{a free hypermultiplet}~~~~\mbox{with $2N-1$ flavors}.
\end{align}
For non-zero $n$'s, the $U(1)_n$ dynamics does not decouple from the other sector since the electron hypermultiplets are charged under the $U(1)_n \times U(1)_0 \subset U(F-N)$ gauge group. 
 
\begin{table}[h]\caption{The magnetic $U(1)_{n} \times U(F-N)_{0,0}$ gauge theory dual to Table \ref{electric} with $k=0$. The theory consists of $F$ fundamental flavors, a single $U(1)$ charged hypermultiplet and a vector multiplet. The table only shows the matter multiplets for simplicity. The duality is only valid for $F=2N-1$.} 
\begin{center}
\scalebox{1}{
  \begin{tabular}{|c||c|c|c| } \hline
  &$U(1)_{n} \times U(F-N)_{0,0}$&$SU(F)$& $U(1)_R$  \\ \hline
$q$&$(0,{\tiny \yng(1)}_{\, 1})$&${\tiny \yng(1)}$&$\frac{1}{2}$ \\
$\tilde{q}$&$(0,{\tiny \overline{\yng(1)}}_{-1})$&${\tiny \overline{\yng(1)}}$&$\frac{1}{2}$ \\
$b$&$(-1,\mathbf{1}_{-(F-N)})$&$\mathbf{1}$&$\frac{1}{2}$ \\
$\tilde{b}$&$(+1,\mathbf{1}_{+(F-N)})$&$\mathbf{1}$&$\frac{1}{2}$ \\ \hline
  \end{tabular}}
  \end{center}\label{magnetic_k=0}
\end{table}

As a non-trivial test of the duality for the ugly case $F=2N-1$, let us consider the $n \rightarrow \infty$ limit. The electric side becomes a 3d $\mathcal{N}=4$ $SU(N)_0$ gauge theory with $2N-1$ fundamental hypermultiplets. On the other hand, the magnetic side leads to a 3d $\mathcal{N}=4$ $U(N-1)_{0,0}$ gauge theory with $2N-1$ fundamental hypermultiplets and an electron hypermultiplet. For $N=2$, the duality relates the 3d $\mathcal{N}=4$ $SU(2)$ and $U(1)$ gauge theories. The electric side has an enhanced $SO(6)$ global symmetry since the $SU(2)$ doublet is pseudo-real and the fundamental hypermultiplets should be counted in terms of half-hypermultiplets. On the magnetic side for $N=2$, the fundamental and electron representations become identical since the gauge group is abelian. As a result, the magnetic flavor symmetry is also enhanced to $SU(4)$ and the non-abelian global symmetry coincides under the duality. Actually, the duality for $N=2$ was studied, for example, in \cite{Giacomelli:2019blm, Fan:2019jii, Nii:2020eui}. For a general $N \ge 3$, the 3d $\mathcal{N}=4$ duality, which relates the $SU(N)_0$ and $U(N-1)_{0,0}$ gauge theories, is presumably new to the best of our knowledge.\footnote{
\textbf{Note added.} After submitting this paper on arXiv, we learned that this duality and the corresponding equality of the partition functions \eqref{eq:SU-U-k0Dual} were studied in \cite{Dey:2021rxw}. We would like to thank Anindya Dey for pointing out his relevant work.
}
Now, the electric gauge group is special-unitary, there are baryonic operators, $B:=Q^N$ and $\tilde{B}:= \tilde{Q}^N$. These are mapped to the magnetic baryons, $\tilde{q}^{N-1} \tilde{b}$ and $q^{N-1} b$.   

Next, we argue that the $SU(N)_0$ ``ugly-good'' duality for the $n \rightarrow \infty$ case is closely related to the S-duality of the 4d $\mathcal{N}=2$ $SU(N)$ gauge theory with $2N$ fundamental hypermultiplets, which is a superconformal field theory \cite{Seiberg:1994aj, Argyres:1995wt, Argyres:1996eh, Leigh:1995ep, Hirayama:1997tw}. The 3d $\mathcal{N}=2$ version of the similar argument based on the S-duality wall can be found in \cite{Benini:2017dud, Garozzo:2019xzi}. 
The S-duality relates the strong and weak coupling regions of this theory. The electric theory consists of $F=2N$ fundamental and $F=2N$ anti-fundamental chiral multiplets. The dual side also has the same matter content. Although the dual (anti-)quarks have the same $SU(2N)$ representations as the electric (anti-)quarks, they have the opposite $U(1)_B$ charge. By dimensionally reducing the S-duality pair on a circle, we obtain the duality on $\mathbb{S}^1 \times \mathbb{R}^3$. Since the theory has eight supercharges and the (KK) monopole has too many fermion zero modes \cite{Callias:1977kg, Nye:2000eg, Poppitz:2008hr}, we expect no monopole potential to be generated under the compactification. As a result, we obtain the (infrared) 3d self-duality between the 3d $\mathcal{N}=4$ $SU(N)$ gauge theories with $2N$ fundamental hypermultiplets. Notice that this is not a trivial duality since the $U(1)_B$ flavor symmetry acts oppositely on dynamical quarks under the duality. We will test this self-duality in the next section by studying the sphere partition function. Now, the theory has an $SU(2N) \times U(1)_B$ flavor symmetry. Then, we can introduce real masses by background gauging these global symmetries. On the electric side, we introduce a real mass to the last flavor, $Q_{i=2N}$ and $\tilde{Q}^{i=2N}$. In order to keep the CS level being zero, we give the positive mass $m>0$ to $Q_{i=2N}$ and the negative mass $-m<0$ to $\tilde{Q}^{i=2N}$. At the origin of the moduli space, the electric theory flows to a 3d $\mathcal{N}=4$ $SU(N)$ gauge theory with $2N-1$ hypermultiplets. On the dual side, this deformation corresponds to giving all the magnetic quarks non-zero real masses. Thus, we find that the low-energy limit of the dual side should be taken at a non-trivial Coulomb branch 
\begin{align}
\braket{\sigma_{adj.}} = 2m \, \mathrm{diag.} 
\bigl(\, \overbrace{1,\cdots,1}^{N-1},-(N-1) \, \bigr),\label{eq:CoulombPoint}
\end{align}
which breaks the dual gauge group as $SU(N) \rightarrow SU(N-1) \times U(1) \simeq U(N-1)$. Along this Coulomb branch, the dual quarks obtain the following real masses:
\begin{gather*}
    m \left( q_{a=1,\cdots,2N-1}^{i=1,\cdots,N-1} \right) =0,~~~~~~~m \left( q_{a=2N}^{i=1,\cdots,N-1} \right) =2mN \\
    m \left( q_{a=1,\cdots,2N-1}^{i=N} \right) =-2mN,~~~~~~~m \left( q_{a=2N}^{i=N} \right) =0.
\end{gather*}
For the dual anti-quarks, the sign of the real masses are just flipped and the dual gauge group obtains no CS level shift. 
As a result, we obtain the 3d $\mathcal{N}=4$ $U(N-1)$ gauge theory with $2N-1$ fundamental hypermultiplets and an electron hypermultiplet as a low-energy description.
This is nothing but the ``ugly-good'' duality between the $SU(N)_0$ and $U(N-1)_{0,0}$ gauge groups, which we proposed above.

Finally, we comment on the ``bad'' region of the 3d $\mathcal{N}=4$ $SU(N)$ and $U(N)$ gauge theories with $N < F < 2N$ flavors. By introducing additional real masses to the remaining flavors, we can in principle construct a 3d $\mathcal{N}=4$ $SU(N)_0$/$U(N)_{0,0}$ duality for the bad theories.\footnote{Although the $F=2N-1$ case is ``ugly'' (where the monopole operator does not break the unitarity bound), the following argument is applicable for this case as well.} Since we cannot rely on exact computations of supersymmetric quantities for this region, the following discussion contains some guess works.\footnote{Especially, we cannot exactly determine the correct point of the magnetic moduli space under the mass deformation where the low-energy limit of magnetic theory should be taken such that the magnetic RG flow corresponds to the electric one. In 4d, the Coulomb branch is described by the Seiberg-Witten curves \cite{Argyres:1994xh, Klemm:1994qs, Argyres:1995wt, Hanany:1995na} for these theories while the Higgs branch is constrained by the non-renormalization theorem. As a result, we can take a correct low-energy limit of the dual theory \cite{Argyres:1996eh}.} 
We will demonstrate two plausible stories of the RG flow to a bad region of the theory. In the first case, we will see that the $SU(N)$ self-duality leads to the $U(N)$ local Seiberg duality \cite{Yaakov:2013fza, Assel:2017jgo} after reducing flavors and gauging the $U(1)_B$ symmetry. In the second case, the resulting duality might be unfortunately a ``bad-bad'' duality and still require ``redeeming''. 

Let us start with the first possibility: We consider deforming the 3d $\mathcal{N}=4$ $SU(N)$ self-duality with $2N$ flavors, which is a 3d remnant of the 4d S-duality. Now, we can introduce real masses to the $2N-F$ flavors on the electric side by background gauging the $SU(2N) \times U(1)_B$ flavor symmetry. Suppose that the $2N-F$ flavors obtain different non-zero real masses as
\begin{align}
    m_{ele} = \mathrm{diag.} \bigl(   \overbrace{0,\cdots,0}^{F},m_1,\cdots,m_{2N-F} \bigr).
\end{align}
By taking the low-energy limit of the electric theory at the origin of the moduli space, we obtain the 3d $\mathcal{N}=4$ $SU(N)$ gauge theory with $F$ flavors and no CS level shift. On the magnetic side, these real masses are delivered to all the magnetic quarks 
\begin{align}
    m_{mag} = \Biggl(\overbrace{-\frac{1}{N} \sum_{i=1}^{2N-F} m_i,\cdots,-\frac{1}{N} \sum_{i=1}^{2N-F} m_i}^{F}, m_1-\frac{1}{N} \sum_{i=1}^{2N-F} m_i,\cdots,m_{2N-F}-\frac{1}{N} \sum_{i=1}^{2N-F} m_i \Biggr).
\end{align}
The real masses for anti-quarks are given by $-m_{mag}$. Therefore, we find that the low-energy limit of the magnetic theory should be taken at a non-trivial point of the moduli space. We here assume that the correct RG flow of the magnetic side is very similar to the 4d ``baryonic-root'' story \cite{Argyres:1996eh}. Then, we take the low-energy limit of the magnetic side at a point of the Coulomb moduli space
\begin{align}
    \braket{\sigma_{mag}} = \mathrm{diag.} \Biggl( \overbrace{ \frac{1}{N} \sum_{i=1}^{2N-F} m_i, \cdots, \frac{1}{N} \sum_{i=1}^{2N-F} m_i}^{F-N},-m_1+\frac{1}{N} \sum_{i=1}^{2N-F} m_i,\cdots,-m_{2N-F}+\frac{1}{N} \sum_{i=1}^{2N-F} m_i \Biggr), \label{magCBflow}
\end{align}
which is traceless as it should be. This vacuum expectation value breaks the $SU(N)$ gauge group to $SU(F-N) \times U(1)^{2N-F}$. The low-energy massless spectrum consists of $F$ $SU(F-N)$ fundamental flavors and $2N-F$ electrons. The charge assignment of the magnetic fields is summarized in Table \ref{Mag1charge}. Notice that we used a redundant notation for the $U(1)$ charges by introducing an additional $U(1)$ gauge group. In this representation of the $U(1)$ charges, we can apply the mirror symmetry \cite{Seiberg:1996bs, Intriligator:1996ex, deBoer:1996mp, Aharony:1997bx} for the $U(1)^{2N-F}/U(1)$ sector. The resulting theory is a 3d $\mathcal{N}=4$ $SU(F-N) \times \stackrel{+1~~~~}{\protect\wick{1}{<1 U(1) \times >1 U ( 1 ) }}$ gauge theory with $F$ fundamental flavors and $2N-F$ electrons. From the charge assignment of the mirror-dual matter fields (see Table \ref{Mag2charge}), we can regard the gauge group as $\stackrel{+1~~~~~}{\protect\wick{1}{<1 U(F-N) \times >1 U ( 1 ) }}$. Notice that the mixed CS term is introduced for the resulting abelian gauge groups. This is because the $U(1)$ gauge group that is not mirror-dualized serves as a topological symmetry of the mirror $U(1)$ SQED. Since only the mixed CS level is introduced, the magnetic theory also has an $\mathcal{N}=4$ supersymmetry \cite{Brooks:1994nn, Kapustin:1999ha}. In the mirror picture, the magnetic theory has an additional $SU(2N-F)$ flavor symmetry, which is invisible in the electric description. Therefore, we are led to claim that the electric theory has decoupled sectors, which are expected to be free monopoles, and results in an emergent and enhanced $SU(2N-F)$ topological symmetry.

As a simple test of the duality, we consider the $F=2N-1$ case. On the magnetic side, the second $U(1)$ gauge sector contains a single electron. Therefore, the $U(1)$ dynamics is confined into a monopole operator. Due to the mixed CS term, the monopole hypermultiplet is charged under the overall $U(1)\subset U(N-1)$ subgroup. As a result, the magnetic side becomes a 3d $\mathcal{N}=4$ $U(N-1)_{0,0}$ gauge theory with $2N-1$ fundamental flavors and an electron hypermultiplet. This is the duality in Table \ref{magnetic_k=0} with $n \rightarrow \infty$.
Let us also check the matching of the (anti-)baryonic operators. On the magnetic side, we can consider the monopole operator $X^{U(1)}_{\pm}$ for the second $U(1)$ factor. This bare operator is not gauge-invariant since there is a mixed CS term between the two abelian subgroups. As a result, we should consider the dressed monopoles
\begin{align}
    B_{mag}:= X^{U(1)}_{-} \tilde{q}^{F-N},~~~~~\tilde{B}_{mag}:= X^{U(1)}_{+} q^{F-N}.
\end{align}
From their quantum numbers, we can naturally identify them with the electric (anti-)baryons, $B:=Q^N \sim B_{mag}$ and $\tilde{B}:=\tilde{Q}^{F-N} \sim \tilde{B}_{mag}$. By gauging the $U(1)_B$ symmetry, we can easily reproduce the $U(N)$ local Seiberg duality \cite{Yaakov:2013fza, Assel:2017jgo} with $2N-F$ free hypermultiplets. In the mirror description, the $U(1)_B$ gauging corresponds to removing the mirror-dualized $U(1)$ gauge symmetry and the $2N-F$ electrons, denoted by $b$ and $\tilde{b}$, decouple from the other sector. 
Since the magnetic RG flow considered here leads to the local Seiberg duality (not a rigorous duality) after the $U(1)_B$ gauging, this may imply that we should take a more non-trivial low-energy limit of the real-mass deformed magnetic theory. At present, we do not know how to remedy it. For example, a genuine ``bad-good'' duality must reproduce the 3d $\mathcal{N}=2$ $SU(N)$ Seiberg duality \cite{Aharony:2013dha} by breaking one half of supersymmetry \`{a} la \cite{Argyres:1996eh, Hirayama:1997tw}. At this stage, we do not know how to achieve this through the mirror dual description in Table \ref{Mag2charge}. We leave these problems to future studies.

\begin{table}[h]\caption{The charge assignment of the low-energy magnetic massless fields. In the third column, the sum of the $U(1)$ gauge fields does not couple with the matter sector and should be removed. The table only lists the chiral multiplets and the corresponding anti-chiral multiplets are implicit.} 
\begin{center}
\scalebox{1}{
  \begin{tabular}{|c||c|c|c|c|c|c| } \hline
  &$SU(F-N) \times U(1) $&$ \overbrace{U(1) \times \cdots \times U(1)}^{2N-F}/U(1) $&$SU(F)$&$U(1)_B$&$U(1)_A$&$U(1)_R$  \\ \hline 
  $q$&$ \left({\tiny \yng(1)}, \frac{1}{F-N} \right)$&$(0,\cdots,0)$& ${\tiny \yng(1)}$ &$-\frac{N}{F-N}$&$1$&$\frac{1}{2}$ \\
  $e_1$&$ \left(1, \frac{-1}{2N-F} \right)$&$(1,0,\cdots,0,-1)$&1&$0$&$1$&$\frac{1}{2}$  \\
  $e_2$&$ \left(1, \frac{-1}{2N-F} \right)$&$(-1,1,0,\cdots,0)$&1&$0$&$1$&$\frac{1}{2}$  \\
  $e_3$&$ \left(1, \frac{-1}{2N-F} \right)$&$(0,-1,1,0,\cdots,0)$&1&$0$&$1$&$\frac{1}{2}$  \\ 
  $\vdots$&$\vdots$&$\vdots$&$\vdots$ &$0$&$1$&$\frac{1}{2}$  \\
  $e_{2N-F}$&$ \left(1, \frac{-1}{2N-F} \right)$&$(0,\cdots,0,-1,1)$&1 &$0$&$1$&$\frac{1}{2}$ \\  \hline
  \end{tabular}
}
  \end{center}\label{Mag1charge}
  \end{table}

\begin{table}[h]\caption{The mirror magnetic description dual to Table \ref{Mag1charge}. If the RG flow on the magnetic side corresponds to the electric one, this is also dual to the 3d $\mathcal{N}=4$ $SU(N)$ SQCD with $N<F<2N$ flavors. In this table, we used our normalization of the $U(1) \subset U(\tilde{N})$ charge for simplicity.} 
\begin{center}
\scalebox{1}{
  \begin{tabular}{|c||c|c|c|c|c| } \hline
  &$\stackrel{+1~~~~~}{\protect\wick{1}{<1 U(F-N) \times >1 U ( 1 ) }}$&$SU(F)$& $SU(2N-F)$&$U(1)_A$&$U(1)_R$ \\ \hline 
  $q$&$({\tiny \yng(1)}_{\,1},0)$&${\tiny \yng(1)}$&1&$1$&$\frac{1}{2}$  \\ 
  $\tilde{q}$&$({\tiny \overline{\yng(1)}}_{\,-1},0)$&${\tiny \overline{\yng(1)}}$&1 & $1$ &$\frac{1}{2}$  \\ 
  $b$&$(1_0,+1)$&1&${\tiny \yng(1)}$& $-1$ & $\frac{1}{2}$  \\
  $\tilde{b}$&$(1_0,-1)$&1&${\tiny \overline{\yng(1)}}$ & $-1$ &$\frac{1}{2}$  \\ \hline
  \end{tabular}
}
  \end{center}\label{Mag2charge}
  \end{table}
We can take an alternate route of the RG flow by the real mass deformation: Let us again start with the 3d $\mathcal{N}=4$ $SU(N)$ self-duality with $2N$ flavors, which is a remnant of the 4d S-duality. Now, we can introduce equal real masses to the $2N-F$ flavors on the electric side by taking $m_1=m_2=\cdots=m_{2N-F}$. The electric theory flows to a 3d $\mathcal{N}=4$ $SU(N)$ gauge theory with $F$ flavors at the origin of the moduli space. On the magnetic side, we have to take a low-energy limit at a non-trivial point of the moduli space since all the dual quarks are again massive. 
Due to large quantum corrections to the Coulomb branch, we cannot exactly determine the fate of the low-energy magnetic theory. For $N<F<2N$, however, we propose that the low-energy limit of the magnetic theory may be taken at a point of the Coulomb moduli space, where the gauge group is spontaneously broken as
\begin{align}
    SU(N) \rightarrow S \left( U(F-N) \times U(2N-F) \right) \cong SU(F-N) \times SU(2N-F) \times U(1),
\end{align}
which is consistent with \eqref{magCBflow} for $m_1=\cdots=m_{2N-F}$. 
The matter content of the low-energy magnetic theory is summarized in Table \ref{good-magnetic-dual,2}. Notice that the magnetic theory again has an additional $SU(2N-F)$ flavor symmetry, which is invisible on the electric side. If this RG flow is the case, we may conjecture that this non-abelian flavor symmetry is an enhanced topological symmetry associated with the fact that the $2N-F$ monopole operators reach the unitarity bound. Since the gauge group contains an abelian factor, the topological symmetry would be $U(2N-F)$. Notice also that the two gauge sectors of the magnetic theory do not decouple with each other since the $U(1)$ gauge dynamics connect them. Since the $SU(2N-F)$ part only includes $2N-F$ flavors, the dual theory still looks ``bad'' although the detailed analysis will be left for a future study.
In this scenario, the $U(1)_B$ gauging is straightforward since the $U(1)_B$ global symmetry is manifest. The electric side just becomes a 3d $\mathcal{N}=4$ $U(N)$ gauge theory with $F$ flavors while the gauge group of the magnetic side becomes $U(F-N)\times U(2N-F)$. For $F=2N-1$, the duality goes back to the $SU(N)$ or $U(N)$ ``ugly-good'' duality \cite{Kapustin:2010mh, Yaakov:2013fza, Assel:2017jgo}. In this RG flow, we can easily guess how to obtain the 3d $\mathcal{N}=2$ $SU(N)$ Seiberg duality \cite{Aharony:2013dha} by giving a complex mass to the adjoint chiral multiplet \cite{Argyres:1996eh, Hirayama:1997tw}. Again, we cannot exactly determine where the low-energy limit of the magnetic theory should be taken. However, from consistency with the 3d $\mathcal{N}=2$ $SU(N)$ Seiberg duality, we can conjecture that the magnetic theory with the adjoint mass term produces a vacuum condensation   
\begin{align}
    \mathrm{rank}\,  \mathinner{\langle{\tilde{b}_i b^j}\rangle} = 2N-F-1,~~~~~(i,j=1,\cdots,2N-F).
\end{align}
The resulting low-energy theory becomes a 3d $\mathcal{N}=2$ $U(F-N)$ gauge theory with $F$ fundamental flavors $(q,\tilde{q})$ and an electron multiplet $(b, \tilde{b})$. The magnetic superpotential can be generated by the Leigh-Strassler transformation \cite{Leigh:1995ep, Strassler:1995xm, Hirayama:1997tw}. We will leave the tests of these putative $SU(N)$/$U(N)$ ``bad-good'' (or maybe ``bad-bad'') dualities to future problems since the analysis of the bad theory is outside of the scope of this paper. 

\begin{table}[H]\caption{The putative magnetic dual to the 3d $\mathcal{N}=4$ $SU(N)$ gauge theory with $F$ hypermultiplets. This table only shows the charge assignment of the fundamental hypermultiplets and omits the adjoint chiral superfields. The gauge group should be $S \left( U(F-N) \times U(2N-F) \right)$.} 
\begin{center}
\scalebox{1}{
  \begin{tabular}{|c||c|c|c|c|c|c| } \hline
  &$SU(F-N) \times SU(2N-F) \times U(1)$&$SU(F)$&$SU(2N-F)$&$U(1)_B$&$U(1)_A$& $U(1)_R$  \\ \hline 
$q$&$({\tiny \yng(1)},\cdot)_{2N-F}$&${\tiny \yng(1)}$&1&$-1$&1&$\frac{1}{2}$\\
$\tilde{q}$&$({\tiny \overline{\yng(1)}},\cdot)_{-(2N-F)}$&${\tiny \overline{\yng(1)}}$&1&$+1$&1&$\frac{1}{2}$\\
$b$&$(\cdot, {\tiny \yng(1)})_{-(F-N)}$&1&${\tiny \yng(1)}$&$-1$&1&$\frac{1}{2}$\\
$\tilde{b}$&$(\cdot, {\tiny \overline{\yng(1)}})_{F-N}$&1&${\tiny \overline{\yng(1)}}$&$+1$&1&$\frac{1}{2}$\\ \hline 
  \end{tabular}}
  \end{center}\label{good-magnetic-dual,2}
  \end{table}
\section{SUSY partition functions and matrix models\label{sec:MatrixModel}}
In this section, we study the 3d $\mathcal{N}=3$ generalized Giveon-Kutasov duality by computing the sphere partition function via supersymmetric localization \cite{Pestun:2007rz, Pestun:2016zxk, Kapustin:2009kz,Hama:2010av}. We denote the sphere partition function of the 3d $\mathcal{N}=3$ $U(N)_{k,k+nN}$ gauge theory with $F$ fundamental flavors by $\zU\left(\bm{m},\zeta\right)$.\footnote{We only consider a round sphere for simplicity.} We will see that the partition function $\zU\left(\bm{m},\zeta\right)$ is given in terms of the $n=0$ partition function $\zU[N][][k][n=0]\left(\bm{m},\eta\right)$. This observation enables us to derive the matrix integral identity of the generalized GK duality from the conventional GK duality pair. In this section, we consider the generalized GK duality with Fayet-Iliopoulos (FI) and real mass deformations, which are denoted by $\zeta$ and $m_a$, respectively. From the matching of the sphere partition functions, we find how these parameters are transformed under the duality. 

\subsection{Matrix models}
We first enumerate the matrix models of the electric and magnetic descriptions, explicitly. These matrix integrals are obtained by the supersymmetric localization technique. For a comprehensive review of the supersymmetric localization, see \cite{Pestun:2016zxk} and references therein.  
The matrix model for the 3d $\mathcal{N}=3$ $U\left(N\right)_{k,k+nN}$ gauge theory with $F$ fundamental hypermultiplets (see Table \ref{electric}) is given by 
\begin{equation}
\zU\left(\bm{m},\zeta\right)=\int\frac{d^{N}\lambda}{N!}e^{i\pi k\sum_{j}^{N}\lambda_{j}^{2}+i\pi n\left(\sum_{j}^{N}\lambda_{j}\right)^{2}+2\pi i\zeta\sum_{j}^{N}\lambda_{j}}\frac{\prod_{j<j'}^{N}\left(2\sinh\pi\left(\lambda_{j}-\lambda_{j'}\right)\right)^{2}}{\prod_{a,j}^{F,N}2\cosh\pi\left(\lambda_{j}-m_{a}\right)}, \label{eq:Zele-Def}
\end{equation}
where the parameter $\zeta$ corresponds to the FI parameter associated with the overall $U(1) \subset U\left(N\right)_{k,k+nN}$ subgroup. The parameters $m_a$ $(a=1,\cdots,F)$ are the real masses for the $F$ hypermultiplets.
These masses are associated with the background gauging of the flavor $SU(F)$ symmetry and hence these must be traceless $\sum_{i=1}^{F} m_i =0$.\footnote{
In the main text, we regard the real masses associated with the flavor $SU(F)$ symmetry as traceless. In the appendix, we will turn on the trace part of the real masses since it becomes clear how the parameters are mapped when we massage the integrands.
}
In association with this redundancy, the FI term becomes an overall phase of the supersymmetric partition function by shifting the integration variables. In the matrix model, the one-loop determinants of the $F$ fundamental hypermultiplets are represented by the $\cosh$ functions while the vector multiplets are represented by the $\sinh$ functions. The exponential factor is the Chern-Simons action on the localization locus.   

The matrix model for the 3d $\mathcal{N}=3$ magnetic $\stackrel{+1~~~~~~~~~~~~~~~~~~~~~~~~~~~~~~~~~~~}{\protect\wick{1}{<1 U(1)_{n+1} \times >1 U ( F +k-N )_{-k,-k+(F+k-N) }}}$ gauge theory with $F$ dual flavors (see Table \ref{magnetic}) is given by
\begin{align}
\zUmag\left(\bm{m},\zeta\right)= & \int d\kappa \, e^{i\pi\left(n+1\right)\kappa^{2}+2\pi i\zeta\kappa}\int\frac{d^{\tilde{N}}\lambda}{\tilde{N}!}e^{2\pi i\kappa\sum_{j}^{\tilde{N}}\lambda_{j}}\nonumber \\
 & \times e^{-i\pi k\sum_{j}^{\tilde{N}}\lambda_{j}^{2}+i\pi\left(\sum_{j}^{\tilde{N}}\lambda_{j}\right)^{2}}\frac{\prod_{j<j'}^{\tilde{N}}\left(2\sinh\pi\left(\lambda_{j}-\lambda_{j'}\right)\right)^{2}}{\prod_{a,j}^{F,\tilde{N}}2\cosh\pi\left(\lambda_{j}-m_{a}\right)}~~~\quad (k\neq 0), \label{eq:Zmag-Def}
\end{align}
where we assume $k \neq 0$ and define $\tilde{N}:=F+k-N$.
The FI parameter $\zeta$ is introduced for the $U(1)_{n+1}$ vector multiplet. In what follows, we will see that the electric FI parameter is mapped to the one for the $U(1)_{n+1}$ subgroup. The $\kappa$ integral corresponds to the path-integral of the $U(1)_{n+1}$ adjoint scalar. The integral associated with the $U(\tilde{N})$ gauge group is denoted by $d^{\tilde{N}}\lambda$. Since the matter fields are neutral under the $U(1)_{n+1}$ subgroup, the $\kappa$ integral only includes the CS action at the localization locus. Since the $\kappa$ variable just appears quadratically, we can exactly compute its integral. This results in the following matrix model
\begin{align}
&\zUmag\left(\bm{m},\zeta\right) \nonumber \\
&=  \frac{e^{\frac{\pi i}{4}{\rm sgn}(n+1)}e^{-\frac{i \pi }{n+1}\zeta^2}}{\sqrt{|n+1|}}\int\frac{d^{\tilde{N}}\lambda}{\tilde{N}!}
  e^{-i\pi k\sum_{j}^{\tilde{N}}\lambda_{j}^{2}+\frac{i\pi n}{n+1}\left(\sum_{j}^{\tilde{N}}\lambda_{j}\right)^{2}-\frac{2\pi i \zeta}{n+1}\sum _j^{\tilde{N}}\lambda_j}
  \frac{\prod_{j<j'}^{\tilde{N}}\left(2\sinh\pi\left(\lambda_{j}-\lambda_{j'}\right)\right)^{2}}{\prod_{a,j}^{F,\tilde{N}}2\cosh\pi\left(\lambda_{j}-m_{a}\right)}.
\end{align}
This is formally regarded as the sphere partition function of the 3d $\mathcal{N}=3$ $U(\tilde{N})_{-k,-k+\frac{n}{n+1}\tilde{N}}$ Chern-Simons matter gauge theory. Of course, this manipulation is field-theoretically not allowed since there exists a fractional CS level. However, at the level of matrix models, we can regard this duality as a mathematical equivalence between the CS gauge theories with quantized and fractional CS levels. 

As we studied in Section \ref{section:GKduality}, the $SU(N)$ Chern-Simons/Yang-Mills gauge theories appear for some choices of $n$.
The difference between the matrix model for the $U(N)$ gauge group \eqref{eq:Zele-Def} and the matrix model for the $SU(N)$ gauge group is whether the domain of the integration includes $\sum_{j}^{N}\lambda_{j}=0$ direction or not.
Therefore, the matrix model for the $SU(N)$ gauge group with $F$ fundamental hypermultiplets and the Chern-Simons level $k$ is given by
\begin{equation}
\zSU\left(\bm{m}\right)=\int\frac{d^{N}\lambda}{N!}e^{i\pi k\sum_{j}^{N}\lambda_{j}^{2}}\delta\left(\sum_{j}^{N}\lambda_{j}\right)\frac{\prod_{j<j'}^{N}\left(2\sinh\pi\left(\lambda_{j}-\lambda_{j'}\right)\right)^{2}}{\prod_{a,j}^{F,N}2\cosh\pi\left(\lambda_{j}-m_{a}\right)}.\label{eq:zSU-Def}
\end{equation}
Since the $U(1)_B$ symmetry is not gauged, the real masses can be introduced for the $SU(F) \times U(1)_B$ background vector multiplet. As a result, the mass parameter $m_a$ becomes tracefull.  
For example, when $n=-1$, the magnetic gauge group is reduced to $SU(F+k-N)_{-k}$. After a short calculation, we can easily find that the matrix model \eqref{eq:Zmag-Def} becomes
\begin{align}
\zUmag[][][][-1]\left(\bm{m},\zeta\right)= & e^{i\pi \left(1-\frac{k}{\tilde{N}}\right) \zeta^2}\zSU\left(\bm{m}+\frac{\zeta}{\tilde{N}}\bm{1}\right),
\end{align}
which shows that the FI term of the generalized $U(\tilde{N})$ gauge theory becomes a trace part of the real masses with a correct rescaling. 

Finally, we give a matrix model for the duality proposed in Section \ref{subsection:ugly}. For $k=0$, we proposed the 3d $\mathcal{N}=3$ $U(N)_{0,0+nN}$ duality for $F=2N-1$ flavors by generalizing the ``ugly-good'' duality \cite{Kapustin:2010mh, Yaakov:2013fza, Assel:2017jgo}. The magnetic description is given by the 3d $\mathcal{N}=3$ $U(1)_n \times U(N-1)_{0,0}$ gauge theory with $2N-1$ fundamental flavors and an electron hypermultiplet. The matrix model corresponding to this magnetic gauge theory (see Table \ref{magnetic_k=0}) is straightforwardly given by   
\begin{align}
\zUmag[][][0]\left(\bm{m},\zeta\right)=\int d\kappa \, e^{i\pi n\kappa^{2}+2\pi i\zeta\kappa}\int\frac{d^{\tilde{N}}\lambda}{\tilde{N}!}\frac{\prod_{j<j'}^{\tilde{N}}\left(2\sinh\pi\left(\lambda_{j}-\lambda_{j'}\right)\right)^{2}}{2\cosh\pi\left(\kappa+\sum_{j}^{\tilde{N}}\lambda_{j}\right)\prod_{a,j}^{F,\tilde{N}}2\cosh\pi\left(\lambda_{j}-m_{a}\right)},\label{eq:Zmagk0-Def}
\end{align}
where there is no mixed CS term  between the two $U(1)$ gauge groups. The $\cosh \pi(\kappa + \sum \lambda_j)$ factor is a one-loop determinant of the electron hypermultiplet which is charged under the abelian $U(1)_n$ and $ U(1) \subset U(N-1)_{0,0}$ subgroups. Since the real masses are associated with the background $SU(F)$ vector multiplet, the electron hypermultiplet is massless.  

\subsection{Derivation of the generalized GK duality}\label{Derivation}
In this subsection, we will prove the integral identity
\begin{align}
\zU\left(\bm{m},\zeta\right)
=i^{-\frac{1}{2}}e^{i\Theta_{F,k,N}+i\pi k\sum_{a}^{F}m_{a}^{2}}
\zUmag\left(\bm{m},\zeta\right),\label{eq:GKdual-MM}
\end{align}
which claims the equivalence of the sphere partition functions between the 3d $\mathcal{N}=3$ $U(N)_{k,k+nN}$  and $\stackrel{+1~~~~~~~~~~~~~~~~~~~~~~~~~~~~~~~~~~~}{\protect\wick{1}{<1 U(1)_{n+1} \times >1 U ( F +k-N )_{-k,-k+(F+k-N) }}}$ Chern-Simons matter gauge theories with $F$ fundamental hypermultiplets up to a phase factor.
We divide the phase factor into two parts:
The first phase is defined as
\begin{align}
\Theta_{F,k,N} =\begin{cases}
i^{\frac{1}{2}N^{2}-\frac{1}{2}F^{2}+\frac{1}{2}\tilde{N}^{2}}e^{i\theta_{\tilde{N}}+i\theta_{k-N}+i\theta_{N-F}+i\theta_{N}} & \left(F\leq N\right)\\
i^{\frac{1}{2}\tilde{N}^{2}-\frac{1}{2}\left(F-N\right)^{2}}e^{i\theta_{2N-F}+i\theta_{\tilde{N}-N}+i\theta_{k}} & \left(F>N\right)
\end{cases}, \quad ~~
\theta_{N}=\frac{\pi}{12k}\left(N^{3}-N\right), \label{eq:Phase-Def}
\end{align}
and depends on $N,F$ and $k$.
This phase is associated with the fact that the CS partition function computed by the supersymmetric localization uses an unusual framing \cite{Kapustin:2009kz, Kapustin:2010mh}. 
The second phase depends on the real mass parameter in addition to the non-abelian Chern-Simons level $k$.

In deriving the above identity, it is important to first rewrite the matrix model for the $U\left(N\right)_{k,k+nN}$ gauge group by disentangling the two contributions of the (non-)abelian Chern-Simons levels proportional to $k$ and $n$. By inserting $1=\int d\phi\, \delta\left(\sum_{j}^{N}\lambda_{j}-\phi\right)$ into the matrix model \eqref{eq:Zele-Def}, we obtain
\begin{align}
\zU\left(\bm{m},\zeta\right) &=\int\frac{d^{N}\lambda}{N!}e^{i\pi k\sum_{j}^{N}\lambda_{j}^{2}+i\pi n\left(\sum_{j}^{N}\lambda_{j}\right)^{2}+2\pi i\zeta\sum_{j}^{N}\lambda_{j}}\frac{\prod_{j<j'}^{N}\left(2\sinh\pi\left(\lambda_{j}-\lambda_{j'}\right)\right)^{2}}{\prod_{a,j}^{F,N}2\cosh\pi\left(\lambda_{j}-m_{a}\right)} \nonumber \\
&=\int d\phi \, e^{i\pi n\phi^{2}+2\pi i\zeta\phi}  \int\frac{d^{N}\lambda}{N!}\delta\left(\sum_{j}^{N}\lambda_{j}-\phi\right)e^{i\pi k\sum_{j}^{N}\lambda_{j}^{2}}\frac{\prod_{j<j'}^{N}\left(2\sinh\pi\left(\lambda_{j}-\lambda_{j'}\right)\right)^{2}}{\prod_{a,j}^{F,N}2\cosh\pi\left(\lambda_{j}-m_{a}\right)}.
\end{align}
Next, by using the Fourier transform of the delta function, $\delta\left(x\right)=\int d\eta \, e^{2\pi i\eta x}$, we can further reshape the integrand
\begin{align}
\zU\left(\bm{m},\zeta\right)= & \int d\eta\int d\phi \, e^{i\pi n\phi^{2}+2\pi i\left(\zeta-\eta\right)\phi}\nonumber \\
 & \times\int\frac{d^{N}\lambda}{N!}e^{i\pi k\sum_{j}^{N}\lambda_{j}^{2}+2\pi i\eta\sum_{j}^{N}\lambda_{j}}\frac{\prod_{j<j'}^{N}\left(2\sinh\pi\left(\lambda_{j}-\lambda_{j'}\right)\right)^{2}}{\prod_{a,j}^{F,N}2\cosh\pi\left(\lambda_{j}-m_{a}\right)}.
\end{align}
As a result, the partition function $\zU\left(\bm{m},\zeta\right)$ with a non-zero $n$ is related with the electric partition function $\zU[N][][k][0]\left(\bm{m},\eta\right)$ of the conventional GK duality with $n=0$:
\begin{align}
\zU\left(\bm{m},\zeta\right)=\int d\eta\int d\phi \, e^{i\pi n\phi^{2}+2\pi i\left(\zeta-\eta\right)\phi}\zU[][][][0]\left(\bm{m},\eta\right).\label{eq:Diag-woDiag3}
\end{align}
This is more than just the reformulation of the integrand for the electric matrix model. Actually, by introducing two additional vector multiplets, we can regard the $U(N)_{k,k+nN}$ gauge group as
\begin{align}
    \stackrel{-1~~~~\,~~~~+1~~~~~~~}{\protect\wick{11}{<1 U(1)_{n} \times >1U(<2 1)_0 \times  >2U (N)_{k,k}}}, \label{expanded_form}
\end{align}
where the matter fields are only charged under the $U(N)_{k,k}$ subgroup. Since the second $U(1)$ factor has zero CS level and appears linearly in the action, the associated $U(1)_0$ vector multiplet simply acts as a Lagrange multiplier. By integrating out the $U(1)_0$ vector multiplet, we obtain the delta functional forcing the remaining $U(1)$ gauge fields to be equivalent and reproduce the original $U(N)_{k,k+nN}$ gauge group with the matter multiplets unchanged. The above reformulation of the matrix model corresponds to this equivalence of the two differently-looking gauge groups.

Now, the electric theory is written in terms of the conventional Giveon-Kutasov duality, namely $\zU[N][][k][0]\left(\bm{m},\eta\right)$.
By using the integral identity of the conventional GK duality \eqref{eq:LRwoDiag}, which we will review and prove in the appendix, we obtain
\begin{align}
 \zU\left(\bm{m},\zeta\right)
=e^{i\Theta_{F,k,N}+i\pi k\sum_{a}^{F}m_{a}^{2}}\int d\eta\int d\kappa \, e^{i\pi n\kappa^{2}+2\pi i\left(\zeta-\eta\right)\kappa}e^{-i\pi\eta^{2}}\zU[\tilde{N}][][-k][0]\left(\bm{m},-\eta\right). \label{ele:final}
\end{align}
Notice that the phase factor $e^{-i \pi \eta^2}$ appears under the duality transformation. In the conventional GK duality, this is just an FI parameter and serves as a constant phase. In the current example, however, this is associated with the CS action of the dynamical $U(1)_0$ multiplet and must be integrated. In  what follows, we will show that the expression \eqref{ele:final} is equal to $\zUmag\left(\bm{m},\zeta\right)$ up to constant phases.

Next, we massage the matrix model \eqref{eq:Zmag-Def} of the magnetic dual gauge theory. Since the magnetic matter fields are neutral under the $U(1)_{n+1}$ subgroup, we can write the magnetic partition function as 
\begin{align}
\zUmag\left(\bm{m},\zeta\right)=\int d\kappa\, e^{i\pi\left(n+1\right)\kappa^{2}+2\pi i\zeta\kappa}\zU[\tilde{N}][][-k][1]\left(\bm{m},\kappa\right),
\end{align}
As in the electric side, we can disentangle the $U(1)$ Chern-Simons level into two parts by adding two $U(1)$ vector multiplets. By using \eqref{eq:Diag-woDiag3}, we find 
\begin{align}
\zUmag\left(\bm{m},\zeta\right)= & \int d\kappa\, e^{i\pi\left(n+1\right)\kappa^{2}+2\pi i\zeta\kappa}\int d\eta\int d\phi\, e^{i\pi\phi^{2}+2\pi i\left(\kappa-\eta\right)\phi}\zU[\tilde{N}][][-k][0]\left(\bm{m},\eta\right)\nonumber \\
= &i^{\frac{1}{2}} \int d\kappa\int d\eta\, e^{i\pi\left(n+1\right)\kappa^{2}+2\pi i\zeta\kappa}e^{-i\pi\left(\kappa-\eta\right)^{2}}\zU[\tilde{N}][][-k][0]\left(\bm{m},\eta\right)\nonumber \\
= & i^{\frac{1}{2}}\int d\kappa\int d\eta\, e^{i\pi n\kappa^{2}+2\pi i\left(\zeta-\eta\right)\kappa}e^{-i\pi\eta^{2}}\zU[\tilde{N}][][-k][0]\left(\bm{m},-\eta\right). \label{mag:final}
\end{align}
The final expression is associated with the fact that the magnetic gauge group can be recast in to the following form:
\begin{align}
    \stackrel{-1~~~~\,~~~~~-1~~~~~~~~~~~~~~~~\,~~~~~~~}{\protect\wick{11}{<1 U(1)_{n} \times >1U(<2 1)_{-1} \times  >2U (F+k-N)_{-k,-k}}}.
\end{align}
The last lines in \eqref{ele:final} and \eqref{mag:final} are identical up to a phase factor. Thus, we finally arrive at the integral identity \eqref{eq:GKdual-MM}, which claims that the sphere partition functions of the electric and magnetic descriptions coincide in the presence of the FI term $\zeta$ and the real mass parameters $m_a$. In addition, this analysis shows that the FI parameter is mapped as $\zeta \rightarrow \zeta$ without flipping the sign. Note that when $F+k-N=0$, the electric theory is dual to a topological  ${\rm U}\left(1\right)_{n+1}$ gauge theory without matter since the second $U(\tilde{N})$ group vanishes. This can be seen by using \eqref{eq:eleZ-triv} as
\begin{align}
 \zU\left(\bm{m},\zeta\right)& =e^{i\Theta_{F,k,N}+i\pi k\sum_{a}^{F}m_{a}^{2}}\int d\eta\int d\kappa\, e^{i\pi n\kappa^{2}+2\pi i\left(\zeta-\eta\right)\kappa}e^{-i\pi\eta^{2}}\nonumber \\
 & =e^{i\Theta_{F,k,N}+i\pi k\sum_{a}^{F}m_{a}^{2}}\int d\kappa\, e^{i\pi\left(n+1\right)\kappa^{2}+2\pi i\zeta\kappa}.
\end{align}

\subsection{The generalized GK duality for the $U(N)_{0,0+nN}$ ``ugly'' theory}
In this subsection, we will show the matrix integral identity of the ``ugly-good'' duality for the 3d $\mathcal{N}=3$ $U(N)_{0,0+nN}$ Giveon-Kutasov duality with $2N-1$ fundamental flavors described in Section \ref{subsection:ugly}. The dual description is given by the 3d $\mathcal{N}=3$ $U(1)_n \times U(N-1)_{0,0}$ gauge theory with $2N-1$ dual flavors and an electron hypermultiplet. We will show that the duality for this case is easily obtained from the $n=0$ duality known in the literature \cite{Kapustin:2010mh, Yaakov:2013fza, Assel:2017jgo} and the expanded form of the electric gauge group \eqref{expanded_form}.
The duality implies the following integral identity: 
\begin{equation}
\zU[][2N-1][0]\left(\bm{m},\zeta\right)=\zUmag[][2N-1][0]\left(\bm{m},\zeta\right). \label{good-ugly_n}
\end{equation}
In this identity, there is no phase factor such as $\theta_{N}$, which is proportional to the inverse of $k$. This would be related to the fact that the duality equates the Chern-Simons gauge theory with $k=0$. 

It is straightforward to prove the identity \eqref{good-ugly_n} since we can use the same argument in Section \ref{Derivation}.   
By using the expanded form of the electric partition function \eqref{eq:Diag-woDiag3}, the left-hand side is explicitly given by
\begin{align}
\zU[][2N-1][0]\left(\bm{m},\zeta\right)=\int d\eta\int d\phi\, e^{i\pi n\phi^{2}+2\pi i\left(\zeta-\eta\right)\phi}\zU[][2N-1][0][0]\left(\bm{m},\eta\right).\label{eq:k0Aux}
\end{align}
By using the matrix identity \eqref{eq:GUwoDiag}  of the conventional ``ugly-good'' duality with $n=0$ (which will be proven in the appendix) and imposing the traceless condition of the real masses, we can transform \eqref{eq:k0Aux} as follows:
\begin{align}
&\zU[][2N-1][0]\left(\bm{m},\zeta\right) \nonumber \\
& =\int d\eta\int d\phi\, e^{i\pi n\phi^{2}+2\pi i\left(\zeta-\eta\right)\phi}\frac{1}{2\cosh\pi\eta}\zU[N-1][2N-1][0][0]\left(\bm{m},-\eta\right)\nonumber \\
 & =\int d\eta\int d\kappa\int\frac{d^{N-1}\lambda}{\left(N-1\right)!}e^{i\pi n\kappa^{2}+2\pi i\zeta\kappa}\frac{e^{-2\pi i\eta\left(\kappa+\sum_{j}^{N-1}\lambda_{j}\right)}}{2\cosh\pi\eta}\frac{\prod_{j<j'}^{N-1}\left(2\sinh\pi\left(\lambda_{j}-\lambda_{j'}\right)\right)^{2}}{\prod_{a,j}^{2N-1,N-1}2\cosh\pi\left(\lambda_{j}-m_{a}\right)}\nonumber \\
 & =\int d\kappa\int\frac{d^{N-1}\lambda}{\left(N-1\right)!}e^{i\pi n\kappa^{2}+2\pi i\zeta\kappa}\frac{1}{2\cosh\pi\left(\kappa+\sum_{j}^{N-1}\lambda_{j}\right)}\frac{\prod_{j<j'}^{N-1}\left(2\sinh\pi\left(\lambda_{j}-\lambda_{j'}\right)\right)^{2}}{\prod_{a,j}^{2N-1,N-1}2\cosh\pi\left(\lambda_{j}-m_{a}\right)}. \label{Final_form}
\end{align}
On the fourth line above, we performed the $\eta$ integral, which is just a Fourier transform of the $\cosh$ function. 
Notice that the FI parameter $\eta$ is just a parameter in the conventional ``ugly-good'' duality \cite{Kapustin:2010mh, Yaakov:2013fza, Assel:2017jgo} with $n=0$ while it now becomes an integration variable. This explains why the $n \neq 0$ ``ugly-good'' duality includes an additional $U(1)$ charged field on the magnetic side.  
The last expression \eqref{Final_form} is nothing but a matrix model of the magnetic description \eqref{eq:Zmagk0-Def} and proves the integral identity \eqref{good-ugly_n}. Note that the FI parameter is simply mapped to the magnetic FI parameter for the $U(1)_n$ gauge group without flipping the sign.

\subsection{The generalized GK duality in the $n \rightarrow \infty$ limit}

\subsubsection{The $k \neq 0$ case}
In this subsection, we investigate the $n\rightarrow\infty$ limit of the generalized GK duality with $k \neq 0$, which we discussed around \eqref{eq:nInf-kneq0}, by using sphere partition functions. We will see that the $n\rightarrow\infty$ limits of both sides of the duality have an identical decoupling $U(1)_n$ CS sector in the matrix model \eqref{eq:GKdual-MM}.
For the electric side with $n\rightarrow\infty$, we can decouple the $U(1) \subset U(N)_{k,k+nN}$ dynamics since the abelian CS interaction becomes weaker and weaker. At the level of partition functions, we need to take this limit carefully since the expression includes a vanishing factor.
The expression \eqref{eq:Diag-woDiag3} is useful for taking this limit.
By evaluating the $\phi$ integral, we obtain
\begin{equation}
\zU\left(\bm{m},\zeta\right)=\frac{i^{\frac{1}{2}}}{\sqrt{n}}\int d\eta \,  e^{\frac{i\pi}{n}\left(\zeta-\eta\right)^{2}}\zU[][][][0]\left(\bm{m},\eta\right),
\end{equation}
where the prefactor corresponds to the partition function of the 3d $\mathcal{N}=3$ $U(1)_n$ pure CS gauge theory. 
Since the $\phi$ variable comes from the $U(1)_n$ vector multiplet, this manipulation corresponds to formally regarding the electric gauge group as
\begin{align}
     \stackrel{-1~~~~\,~~~~+1~~~~~~~}{\protect\wick{11}{<1 U(1)_{n} \times >1U(<2 1)_0 \times  >2U (N)_{k,k}}}  ~~~~ \Longrightarrow ~~~~ \stackrel{+1~~~~~~~~\,~}{\protect\wick{1}{<1 U(1)_{-\frac{1}{n}} \times >1 U (N)_{k,k}}}.
\end{align}
Although this transformation is field-theoretically ill-defined due to the appearance of the fractional CS level, the matrix model perspective becomes clear and it becomes easy to take the $n \rightarrow \infty$ limit.  
Since the partition function includes the $\sqrt{n}$ factor in the denominator, it is natural to consider the normalized partition function for the $n\rightarrow\infty$ limit:
\begin{align}
\lim_{n\rightarrow\infty}\sqrt{n}\, \zU\left(\bm{m},\zeta\right) =i^{\frac{1}{2}}\int d\eta \, \zU[][][][0]\left(\bm{m},\eta\right)
  =i^{\frac{1}{2}}  \zSU\left(\bm{m}\right). \label{ntoinf_ele}
\end{align}
This normalization is equivalent to dividing the partition function by the $U(1)_n$ CS partition function. 
On the second equality, the $\eta$ integral is performed and it becomes a delta function $\delta (\sum \lambda_j)$. The $SU(N)$ partition function $\zSU\left(\bm{m}\right)$ is defined in \eqref{eq:zSU-Def}.
The resulting expression \eqref{ntoinf_ele} is nothing but a sphere partition function of the 3d $\mathcal{N}=3$ $SU(N)_k$ Chern-Simons matter gauge theory with $F$ fundamental flavors. Notice that the dependence of the FI parameter $\zeta$ correctly vanishes in the $n \rightarrow \infty$ limit as expected.

On the magnetic side, the original expression of the partition function \eqref{eq:Zmag-Def} is useful to take the $n\rightarrow\infty$ limit. 
We have to rescale the integration variable $\kappa \rightarrow \frac{1}{\sqrt{n}} \kappa$ to take a sensible $n\rightarrow\infty$ limit. The resulting integral becomes
\begin{align}
\zUmag\left(\bm{m},\zeta\right)= & \frac{1}{\sqrt{n}}\int d\kappa \, e^{i\pi\left(1+\frac{1}{n}\right)\kappa^{2}+\frac{2\pi i}{\sqrt{n}}\zeta\kappa}\int\frac{d^{\tilde{N}}\lambda}{\tilde{N}!}e^{\frac{2\pi i}{\sqrt{n}}\kappa\sum_{j}^{\tilde{N}}\lambda_{j}}\nonumber \\
 & \times e^{-i\pi k\sum_{j}^{\tilde{N}}\lambda_{j}^{2}+i\pi\left(\sum_{j}^{\tilde{N}}\lambda_{j}\right)^{2}}\frac{\prod_{j<j'}^{\tilde{N}}\left(2\sinh\pi\left(\lambda_{j}-\lambda_{j'}\right)\right)^{2}}{\prod_{a,j}^{F,\tilde{N}}2\cosh\pi\left(\lambda_{j}-m_{a}\right)}.
\end{align}
As in the electric case, it is natural to consider the rescaled partition function $\sqrt{n}\zUmag$. This confirms that the $U(1)_n$ CS dynamics correctly decouples from both the electric and magnetic theories in the $n \rightarrow \infty$ limit.  
Now, we can safely take the decoupling limit of the magnetic partition function 
\begin{align}
\lim_{n\rightarrow\infty}\sqrt{n} \, \zUmag\left(\bm{m},\zeta\right) & =\int d\kappa \, e^{i\pi\kappa^{2}}\int\frac{d^{\tilde{N}}\lambda}{\tilde{N}!}e^{-i\pi k\sum_{j}^{\tilde{N}}\lambda_{j}^{2}+i\pi\left(\sum_{j}^{\tilde{N}}\lambda_{j}\right)^{2}}\frac{\prod_{j<j'}^{\tilde{N}}\left(2\sinh\pi\left(\lambda_{j}-\lambda_{j'}\right)\right)^{2}}{\prod_{a,j}^{F,\tilde{N}}2\cosh\pi\left(\lambda_{j}-m_{a}\right)}\nonumber \\
 & =i^{\frac{1}{2}}\zU[\tilde{N}][][-k][1]\left(\bm{m},0\right).
\end{align}
Notice that the FI parameter $\zeta$ vanishes in this limit as it should be.
As a result, we obtain the 3d $\mathcal{N}=3$ $SU(N)_k$ Giveon-Kutasov duality \eqref{eq:nInf-kneq0}, which relates the 3d $\mathcal{N}=3$ $SU(N)_k$ and $U(F+k-N)_{-k,-k+(F+k-N)}$ gauge theories, at the level of sphere partition functions:
\begin{align}
\zSU\left(\bm{m}\right)=i^{-\frac{1}{2}}e^{i\Theta_{F,k,N}+i\pi k\sum_{a}^{F}m_{a}^{2}}\zU[\tilde{N}][][-k][1]\left(\bm{m},0\right).\label{eq:SU-U-Dual}
\end{align}

Notice that both the electric and magnetic theories have an additional deformation parameter: On the electric side, the $U(1)_B$ baryon symmetry appears in the $n \rightarrow \infty$ limit and one can introduce the corresponding real mass. As a result, the mass parameter $\bm{m}$ can be promoted to a tracefull one. On the magnetic side, we have a $U(1)_T$ topological symmetry and one can introduce the corresponding parameter, which is nothing but an FI parameter. These two parameters are identified with each other.
To obtain a refined integral identity with these parameters, we should start with
\begin{align}
 & \zU\left(\bm{m}+\tilde{m}\bm{1},\zeta\right)\nonumber \\
 & =i^{-\frac{1}{2}}e^{i\Theta_{F,k,N}+i\pi k\sum_{a}^{F}m_{a}^{2}+i\pi F^{2}\tilde{m}^{2}}\int d\kappa e^{i\pi\left(n+1\right)\kappa^{2}+2\pi i\left(\zeta-F\tilde{m}\right)\kappa}\zU[\tilde{N}][][-k][1]\left(\bm{m}+\tilde{m}\bm{1},\kappa-F\tilde{m}\right),
\end{align}
instead of \eqref{eq:GKdual-MM}, where $\bm{m}$ is traceless, $\sum_{i=1}^{F} m_i =0$.
This relation can be obtained in a way similar to \eqref{eq:GKdual-MM} by taking care of the phase depending on the trace of real masses in \eqref{eq:LRwoDiag}.
After taking the $n\rightarrow \infty$ limit in a similar way, we obtain
\begin{equation}
\zSU\left(\bm{m}+\tilde{m}\bm{1}\right)
=i^{-\frac{1}{2}}e^{i\Theta_{F,k,N}+i\pi k\sum_{a}^{F}m_{a}^{2}+i\pi F^{2}\tilde{m}^{2}}\zU[\tilde{N}][][-k][1]\left(\bm{m},-F\tilde{m}\right),
\end{equation}
which shows that the electric real mass associated the $U(1)_B$ symmetry is mapped to the magnetic FI parameter as $\zeta=-F\tilde{m}$.

\subsubsection{The $k=0$ case}
Next, we consider the $n \rightarrow \infty$ limit of the ``ugly-good'' duality for the $k=0$ case with $F=2N-1$. 
As discussed in Section \ref{subsection:ugly}, the electric side becomes a 3d $\mathcal{N}=4$ $SU(N)_0$ gauge theory with $2N-1$ fundamental hypermultiplets, while the magnetic side leads to a 3d $\mathcal{N}=4$ $U\left(N-1\right)_{0,0}$ gauge theory with $2N-1$ fundamental hypermultiplets and an electron hypermultiplet.
Without loss of generality, we assume $n>0$ and then take the $n\rightarrow\infty$ limit in the matrix model representation of the ``ugly-good'' duality \eqref{good-ugly_n}.
In order to take this limit on the electric side, we use the expanded form of the partition function \eqref{eq:k0Aux}. After evaluating the integral over $\phi$, we obtain
\begin{equation}
\zU[][2N-1][0]\left(\bm{m},\zeta\right)=\frac{i^{\frac{1}{2}}}{\sqrt{n}}\int d\eta \, e^{-\frac{i\pi}{n}\left(\zeta-\eta\right)^{2}}\zU[][2N-1][0][0]\left(\bm{m},\eta\right).
\end{equation}
As a result, the rescaled partition function is computed as follows: 
\begin{align}
\lim_{n\rightarrow\infty}\sqrt{n}\, \zU[][2N-1][0]\left(\bm{m},\zeta\right)  =i^{\frac{1}{2}}\int d\eta \, \zU[][2N-1][0][0]\left(\bm{m},\eta\right)
  =i^{\frac{1}{2}}\zSU[][2N-1][0]\left(\bm{m}\right),\label{eq:nInf-ele}
\end{align}
which has no $\zeta$ dependence as it should be. This is nothing but a sphere partition function of the 3d $\mathcal{N}=4$ $SU(N)_0$ gauge theory with $2N-1$ fundamental hypermultiplets.

On the other hand, the sphere partition function of the magnetic side \eqref{eq:Zmagk0-Def} is modified into
\begin{align}
&\zUmag[][2N-1][0]\left(\bm{m},\zeta\right) \nonumber \\
& =\int d\kappa \, e^{i\pi n\kappa^{2}+2\pi i\zeta\kappa}\int\frac{d^{N-1}\lambda}{\left(N-1\right)!}\frac{\prod_{j<j'}^{N-1}\left(2\sinh\pi\left(\lambda_{j}-\lambda_{j'}\right)\right)^{2}}{2\cosh\pi\left(\kappa+\sum_{j}^{N-1}\lambda_{j}\right)\prod_{a,j}^{2N-1,N-1}2\cosh\pi\left(\lambda_{j}-m_{a}\right)}\nonumber \\
 & =\frac{1}{\sqrt{n}}\int d\kappa \, e^{i\pi\kappa^{2}+\frac{2\pi i}{\sqrt{n}}\zeta\kappa}\int\frac{d^{N-1}\lambda}{\left(N-1\right)!}\frac{\prod_{j<j'}^{N-1}\left(2\sinh\pi\left(\lambda_{j}-\lambda_{j'}\right)\right)^{2}}{2\cosh\pi\left(\frac{\kappa}{\sqrt{n}}+\sum_{j}^{N-1}\lambda_{j}\right)\prod_{a,j}^{2N-1,N-1}2\cosh\pi\left(\lambda_{j}-m_{a}\right)},
\end{align}
where we shifted the integration variable as $\kappa \rightarrow \frac{1}{\sqrt{n}}\kappa$ in order to factor out the partition function of the 3d $\mathcal{N}=3$ $U(1)$ pure CS gauge theory. By rescaling the partition function in the same way as the electric side, we arrive at 
\begin{align}
\lim_{n\rightarrow\infty}\sqrt{n}\zUmag[][2N-1][0]\left(\bm{m},\zeta\right) 
  =i^{\frac{1}{2}} Z^{U(N-1)_{0,0}}_{2N-1,\rm electron}\left(\bm{m}\right),\label{eq:nInf-mag}
\end{align}
where the right-hand side is explicitly given by
\begin{align}
&Z^{U(N-1)_{0,0}}_{2N-1,\rm electron}\left(\bm{m}+\tilde{m}\bm{1}\right)\nonumber \\
&=\int\frac{d^{N-1}\lambda}{\left(N-1\right)!}\frac{\prod_{j<j'}^{N-1}\left(2\sinh\pi\left(\lambda_{j}-\lambda_{j'}\right)\right)^{2}}{2\cosh\pi\left(\sum_{j}^{N-1}\lambda_{j}-\tilde{m}\right)\prod_{a,j}^{2N-1,N-1}2\cosh\pi\left(\lambda_{j}-m_{a}+\tilde{m}\right)}.
\end{align}
This is nothing but a sphere partition function of the 3d $\mathcal{N}=3$ $U(N-1)_{0,0}$ gauge theory with $F=2N-1$ flavors and an electron hypermultiplet.
Notice that the magnetic $U(N-1)_{0,0}$ gauge theory can have a trace part of the real masses. This is because the magnetic theory has two types $U(1)$ symmetries which rotate the two sets of hypermultiplets, independently. One of them is gauged and the other serves as the $U(1)_B$ symmetry as shown in Table \ref{good-magnetic-dual,2} with $F=2N-1$. The electric theory also has the $U(1)_B$ symmetry, which is emergent in the $n \rightarrow \infty$ limit. In the next subsection, we will see that the duality between the $SU(N)_0$ and $U(N)_{0,0}$ gauge theories can be obtained from the 3d $\mathcal{N}=4$ $SU(N)_0$ self-duality at the level of partition functions, where we will derive a refined integral identity
\begin{equation}
\zSU[][2N-1][0]\left( \bm{m}+\tilde{m}\bm{1} \right)
=Z^{U(N-1)_{0,0}}_{2N-1,\rm electron}\left(\bm{m}+\tilde{m}\bm{1} \right).\label{eq:SU-U-k0Dual}
\end{equation}
This relation with $\tilde{m}=0$ can be obtained from \eqref{good-ugly_n}, \eqref{eq:nInf-ele} and \eqref{eq:nInf-mag}.

\subsection{3d $\mathcal{N}=4$ $SU(N)_0$ self-duality}
In this subsection, we investigate the self-duality of the 3d $\mathcal{N}=4$ $SU\left(N\right)_0$ gauge theory with $2N$ hypermultiplets, which is a remnant of the 4d S-duality as discussed in Section \ref{subsection:ugly}.
The duality claims that the theory is self-dual up to the sign of the trace part for the $SU(2N) \times U(1)_B$ real masses. Under the duality, the real mass associated with the $SU(2N)$ flavor symmetry does not change while the sign of the $U(1)_B$ real mass is flipped. At the level of the sphere partition functions, we show the following integral identity
\begin{equation}
\zSU[][2N][0]\left(\bm{m}+\tilde{m}\bm{1}\right)=\zSU[][2N][0]\left(\bm{m}-\tilde{m}\bm{1}\right),\label{eq:SelfDual-MM}
\end{equation}
where $\bm{m}=(m_1,\cdots,m_{2N})$ are the $SU(2N)$ real masses and satisfy the traceless condition $\sum_{a=1}^{2N} m_a =0$. The $U(1)_B$ real mass is denoted by $\tilde{m}$. Although this duality symmetry is far from trivial, the proof of this integral identity is very straightforward from our massaged form of the matrix models. First, we should notice that the $SU\left(N\right)$ matrix model can be written in terms of the $U\left(N\right)$ matrix model as
\begin{align}
\zSU[][2N][0]\left(\bm{m}+\tilde{m}\bm{1}\right) & =\int d\zeta\int\frac{d^{N}\lambda}{N!}e^{2\pi i\zeta\sum_{j}^{N}\lambda_{j}}\frac{\prod_{j<j'}^{N}\left(2\sinh\pi\left(\lambda_{j}-\lambda_{j'}\right)\right)^{2}}{\prod_{a,j}^{F,N}2\cosh\pi\left(\lambda_{j}-m_{a}-\tilde{m}\right)}\nonumber \\
 & =\int d\zeta \, e^{2\pi iN\tilde{m}\zeta}\zU[][2N][][0]\left(\bm{m},\zeta\right),
\end{align}
where we shifted the integration variables as $\lambda_{j}\rightarrow\lambda_{j}+\tilde{m}$. The $U\left(N\right)$ matrix model with $2N$ hypermultiplets is invariant when flipping the sign of the FI parameter
\begin{equation}
\zU[][2N][0][0]\left(\bm{m},\zeta\right)=\zU[][2N][0][0]\left(\bm{m},-\zeta\right).
\end{equation}
This relation can be easily seen by using \eqref{Z_final_form_ungly_bad} since the determinant satisfies
\begin{align}
\det\left(\begin{array}{c}
\left[\bbraket{t_{N,\ell}-2\pi\zeta|m_{a}}\right]_{\ell,a}^{N\times2N}\\
\left[\bbraket{t_{N,\ell}|m_{a}}\right]_{\ell,a}^{N\times2N}
\end{array}\right) 
&=\det\left(\begin{array}{c}
\left[\bbraket{t_{N,\ell}|m_{a}}\right]_{\ell,a}^{N\times2N}\\
\left[\bbraket{t_{N,\ell}+2\pi\zeta|m_{a}}\right]_{\ell,a}^{N\times2N}
\end{array}\right) \nonumber \\
&=\left(-1\right)^{N}\det\left(\begin{array}{c}
\left[\bbraket{t_{N,\ell}+2\pi\zeta|m_{a}}\right]_{\ell,a}^{N\times2N}\\
\left[\bbraket{t_{N,\ell}|m_{a}}\right]_{\ell,a}^{N\times2N}
\end{array}\right),
\end{align}
where we used the fact that the real masses $\bm{m}$ are traceless.
Notice that this equality between the two determinants is possible only for $F=2N$ flavors since the above matrices are decomposed into $N$ and $F-N$ row vectors and only the $F=2N$ case has the same number of upper and lower row vectors. 
The phase factor $\left(-1\right)^{N}$ vanishes because the same phase factor arise from the cosine hyperbolic factor due to the fact that $F-1$ is odd.
As a result, for $F=2N$ flavors, the integral identity \eqref{eq:SelfDual-MM} holds.

\subsubsection{An RG flow to the $SU(N)_0$ ``ugly-good'' duality}
In this subsection, we show that the ``ugly-good'' duality \eqref{eq:SU-U-k0Dual} can be obtained from the self-duality \eqref{eq:SelfDual-MM} by turning on large real masses and by correctly flowing to the Coulomb branch on the magnetic side as explained in Section \ref{subsection:ugly} . In order to have the same argument at the level of partition functions, we first parametrize the real masses as
\begin{equation}
m_{a}=M_{a}+\Lambda\quad\left(1\leq a\leq2N-1\right),\quad m_{2N}=-\sum_{a}^{2N-1}M_{a}-\left(2N-1\right)\Lambda,\quad\tilde{m}=-\Lambda.
\end{equation}
Note that $\bm{M}=(M_1,\cdots,M_{2N-1})$ are not traceless. We here suppose that $M_{a}$ are small parameters and $\Lambda$ is large compared to $M_a$. This corresponds to giving a large real mass to the last flavor as discussed in Section \ref{subsection:ugly}.
For this choice of the real masses, the matrix integral of the electric theory is given by
\begin{align}
\zSU[][2N][0]\left(\bm{m}+\tilde{m}\bm{1}\right)= & \int\frac{d^{N}\lambda}{N!}\delta\left(\sum_{j}^{N}\lambda_{j}\right)\frac{\prod_{j<j'}^{N}\left(2\sinh\pi\left(\lambda_{j}-\lambda_{j'}\right)\right)^{2}}{\prod_{a,j}^{2N-1,N}2\cosh\pi\left(\lambda_{j}-m_{a}\right)}\nonumber \\
 & \times\frac{1}{\prod_{j}^{N}2\cosh\pi\left(\lambda_{j}+\sum_{a}^{2N-1}M_{a}+2N\Lambda\right)}.
\end{align}
The dominant contribution of this matrix integral comes from near the origin of the $\lambda_{j}$ variables. Therefore, we can safely take the large $\Lambda$ limit in this expression. This is consistent with the fact that the electric theory flows to the 3d $\mathcal{N}=4$ $SU\left(N\right)$ gauge theory with $2N-1$ hypermultiplets at the origin of the moduli space. After taking the large $\Lambda$ limit, the integrand becomes
\begin{equation}
\lim_{\Lambda\rightarrow\infty}\frac{1}{\prod_{j}^{N}2\cosh\pi\left(\lambda_{j}+\sum_{a}^{2N-1}M_{a}+2N\Lambda\right)}=\frac{e^{-\pi N\sum_{a}^{2N-1}M_{a}}e^{-\pi\sum_{j}^{2N}\lambda_{j}}}{e^{2\pi N^{2}\Lambda}}.
\end{equation}
As a result, the electric partition function approaches asymptotically to 
\begin{equation}
\lim_{\Lambda\rightarrow\infty}\zSU[][2N][0]\left(\bm{m}+\tilde{m}\bm{1}\right)=\frac{e^{-N\sum_{a}^{2N-1}M_{a}}}{e^{2N^{2}\Lambda}}\zSU[][2N][0]\left(\bm{M}\right).\label{eq:SelfDualLimEle}
\end{equation}

On the magnetic side, we have to be careful of the region where the dominant contributions of the matrix integral reside. 
The magnetic matrix integral is written as
\begin{align}
\zSU[][2N][0]\left(\bm{m}-\tilde{m}\bm{1}\right)= & \int\frac{d^{N}\lambda}{N!}\delta\left(\sum_{j}^{N}\lambda_{j}\right)\frac{\prod_{j<j'}^{N}\left(2\sinh\pi\left(\lambda_{j}-\lambda_{j'}\right)\right)^{2}}{\prod_{a,j}^{2N-1,N}2\cosh\pi\left(\lambda_{j}-M_{a}-2\Lambda\right)}\nonumber \\
 & \times\frac{1}{\prod_{j}^{N}2\cosh\pi\left(\lambda_{j}+\sum_{a}^{2N-1}M_{a}+\left(2N-2\right)\Lambda\right)}.
\end{align}
According to the argument around \eqref{eq:CoulombPoint}, the magnetic theory has to flow to a non-trivial point of the Coulomb moduli space. Then, it is natural to shift the integration variables as
\begin{equation}
\lambda_{j}\rightarrow\lambda_{j}+2\Lambda\quad\left(1\leq j\leq N-1\right),\quad\lambda_{N}\rightarrow\lambda_{N}-\left(2N-2\right)\Lambda.
\end{equation}
In this redefinition, the matrix integral can be written as
\begin{align}
&\zSU[][2N][0]\left(\bm{m}-\tilde{m}\bm{1}\right)\nonumber \\
&=  \int\frac{d^{N}\lambda}{N!}\delta\left(\sum_{j}^{N}\lambda_{j}\right)\frac{\prod_{j<j'}^{N-1}\left(2\sinh\pi\left(\lambda_{j}-\lambda_{j'}\right)\right)^{2}}{\prod_{a,j}^{2N-1,N-1}2\cosh\pi\left(\lambda_{j}-M_{a}\right)}\frac{1}{\prod_{j}^{N}2\cosh\pi\left(\lambda_{N}+\sum_{a}^{2N-1}M_{a}\right)}\nonumber \\
 &\quad \times\frac{\prod_{j}^{N-1}\left(2\sinh\pi\left(\lambda_{j}-\lambda_{N}+2N\Lambda\right)\right)^{2}}{\prod_{a}^{2N-1}2\cosh\pi\left(\lambda_{N}-M_{a}-2N\Lambda\right)}\frac{1}{\prod_{j}^{N-1}2\cosh\pi\left(\lambda_{j}+\sum_{a}^{2N-1}M_{a}+2N\Lambda\right)}.
\end{align}
At the large $\Lambda$ limit, the third line asymptotically approaches to
\begin{equation}
\frac{e^{-\pi N\sum_{a}^{2N-1}M_{a}}e^{\pi\sum_{j}^{2N}\lambda_{j}}}{e^{2\pi N^{2}\Lambda}}.
\end{equation}
Notice that although we chose the $\lambda_{N}$ variable to be a special direction, we can also shift the other $\lambda_{j}$ as $\lambda_{j}\rightarrow\lambda_{j}-\left(2N-2\right)\Lambda$. This means that there are $N$ dominant contributions from the above integral. Therefore, we have to take this multiplicity into account by collecting all the dominant contributions. As a result, we find
\begin{equation}
\lim_{\Lambda\rightarrow\infty}\zSU[][2N][0]\left(\bm{m}-\tilde{m}\bm{1}\right)=\frac{e^{-N\sum_{a}^{2N-1}M_{a}}}{e^{2N^{2}\Lambda}}Z_{2N-1,{\rm electron}}^{U\left(N-1\right)_{0,0}}\left(\bm{M}\right),\label{eq:SelfDualLimMag}
\end{equation}
where we shifted the integration variables as $\lambda_{j}\rightarrow\lambda_{j}+\frac{2}{2N-1}\sum_a^{2N-1}M_a$.
By using \eqref{eq:SelfDual-MM}, \eqref{eq:SelfDualLimEle} and \eqref{eq:SelfDualLimMag}, we arrive at \eqref{eq:SU-U-k0Dual}.


\section{A test of the duality via superconformal indices\label{sec:Index}}
In this section, we investigate the superconformal indices \cite{Bhattacharya:2008bja, Kim:2009wb, Imamura:2011su, Kapustin:2011jm} of the 3d $\mathcal{N}=3$ $U(N)_{k,k+nN}$ generalized Giveon-Kutasov duality. Firstly, we analyze confinement phases of the $U(N)_{k,k+nN}$ gauge theory, where the low-energy dynamics is described only by gauge-invariant chiral multiplets. This can be regarded as an example of s-confinement \cite{Csaki:1996sm, Csaki:1996zb}. Unlike the theories with four supercharges, non-perturbative superpotential, which describes the confinement phase, is now highly restricted.  
We expect that the electric theory exhibits confinement (or gapped) phases when the dual gauge group vanishes. For the $n=-1$ case, where the dual gauge group is simplified into $SU(F+k-N)$, the electric theory is then expected to confine for $F+k-N=1$. For the 3d $\mathcal{N}=4$ $U(1)_0$ SQED case, which corresponds to $N=1$ and $n=-k$ in our duality, the electric theory is known to be confining when $F=1$ \cite{Intriligator:1996ex, Kapustin:1999ha} and is dual to a free hypermultiplet, which corresponds to a massless monopole operator. When the dual gauge group becomes negative, the electric theory spontaneously breaks supersymmetry, which corresponds to the divergence of the supersymmetric quantities.

We start with the confinement phase of the 3d $\mathcal{N}=4$ $U(1)_0$ SQED with one hypermultiplet \cite{Intriligator:1996ex, Kapustin:1999ha}. The moduli space is described only by a monopole operator since there is no Higgs branch for one flavor. The scaling dimension of the monopole operator $V_{\pm}$ is $\frac{1}{2}$. Therefore, the theory is expected to be dual to a free hypermultiplet representing this monopole. The superconformal indices are expanded as    
\begin{align}
    I_{F=1}^{U(1)_0}&=1+2 x^{1/2}+3 x+2 x^{3/2}+x^2+2 x^{5/2}+4 x^3+4 x^{7/2}-2 x^{9/2}+3 x^5+8 x^{11/2} \nonumber \\
    &\quad +5 x^6-4 x^{13/2}-4 x^7+8 x^{15/2}+11 x^8-2 x^{17/2}-12 x^9+2 x^{19/2}+22 x^{10}+\cdots \nonumber \\
    &= \left( \frac{(x^{3/2};x^2)_\infty}{(x^{1/2};x^2)_\infty} \right)^2.
\end{align}
The second term $2 x^{1/2}$ corresponds to the monopole operator which consists of two gauge-singlet chiral superfields $V_{\pm}$ and forms a single hypermultiplet. On the last line, we rewrite the index expansion as the two contributions of the chiral superfields $V_{\pm}$. This confirms that the 3d $\mathcal{N}=4$ SQED with one hypermultiplet is in the confinement phase and dual to a free hypermultiplet. Since the electric gauge group is abelian, we can construct an infinite number of magnetic dual descriptions whose gauge group is given by $\stackrel{+1~~~~~~~~~~~~~~}{\protect\wick{1}{<1 U(1)_{-k+1} \times >1 U ( k)_{-k,-k+k }}} $. For $k=1$, we observed that the magnetic index gives the same expansion as above.  

We can check other confinement phases by studying the superconformal indices. For $n=-1$, the dual gauge group is simplified into $SU(F+k-N)_{-k}$ and hence the electric theory exhibits confinement when $F+k-N=1$ is satisfied. In this case, the dual side consists of the gauge-singlet hypermultiplets, $q$ and $\tilde{q}$. Since the dual side solely consists of the gauge-singlet hypermultiplets, the supersymmetry is enhanced to $\mathcal{N}=4$. As a test of the duality proposal, we consider the $(N,F,k,n)=(2,1,2,-1)$ and $(2,2,1,-1)$ cases. For $(N,F,k,n)=(2,1,2,-1)$, the electric superconformal index is expanded as 
\begin{align}
    I_{F=1}^{U(2)_{2,2-2}}&=1+2 x^{1/2}+3 x+2 x^{3/2}+x^2+2 x^{5/2}+4 x^3+4 x^{7/2}-2 x^{9/2}+3 x^5+8 x^{11/2}+5 x^6 \nonumber \\
    &\quad -4 x^{13/2}-4 x^7+8 x^{15/2}+11 x^8-2 x^{17/2}-12 x^9+2 x^{19/2}+22 x^{10}+\cdots \\
    &=  \left( \frac{(x^{3/2};x^2)_\infty}{(x^{1/2};x^2)_\infty}  \right)^2.
\end{align}
The second term $2x^{1/2}$ corresponds to the dressed Coulomb branch operators, \eqref{V_+d} and \eqref{V_-d}. From the magnetic viewpoint, this is just a gauge-invariant dual quark. On the last line, the index is summed into a single hypermultiplet $\left\{q, \tilde{q}  \right\}$. Similarly, the superconformal index for $(N,F,k,n)=(2,2,1,-1)$ is expanded as
\begin{align}
    I_{F=2}^{U(2)_{1,1-2}}&=1+4 x^{1/2}+10 x+16 x^{3/2}+19 x^2+20 x^{5/2}+26 x^3+40 x^{7/2}+49 x^4+40 x^{9/2}+26 x^5\nonumber \\
    &\quad +40 x^{11/2}+84 x^6+100 x^{13/2}+52 x^7+8 x^{15/2}+64 x^8+172 x^{17/2}\nonumber \\
    &\quad +150 x^9-16 x^{19/2}-61 x^{10}+\cdots \\
    &= \left( \frac{(x^{3/2};x^2)_\infty}{(x^{1/2};x^2)_\infty}  \right)^4. 
\end{align}
In this case, the dressed monopoles become the (anti-)fundamental representations under the $SU(2)$ flavor symmetry and contribute to the index as $4 x^{1/2}$. On the last line, the above expansion is combined into the index of the four chiral superfields, $V_{\pm d}$. 

Next, we move on to the non-confining examples and pick out the 3d $\mathcal{N}=3$ $U(1)_n$ SQED with one hypermultiplet. This corresponds to taking $(N,F,k,n)=(1,1,1,n)$ in our duality, where we shifted $n \rightarrow n-1$ for simplifying the duality expression. The dual description is given by a 3d $\mathcal{N}=3$ $\stackrel{+1~\,~~~~}{\protect\wick{1}{<1 U(1)_{n} \times >1 U ( 1 )_{0}}}$ gauge theory with one flavor. For $n=1$, the electric and magnetic theories have the following expansion of the index: 
\begin{align}
    I_{F=1}^{U(1)_1}&=1+x-3 x^2+4 x^3-4 x^4+x^5+5 x^6-8 x^7+11 x^8-20 x^9 \nonumber \\
    &\quad +30 x^{10}-31 x^{11} + 27 x^{12}-29 x^{13}+25 x^{14}-6 x^{15}+\cdots.
\end{align}
The first term $x$ corresponds to the adjoint chiral superfield $\Phi$, which is gauge-singlet for the $U(1)$ gauge group. At the index level, the meson singlet $Q\tilde{Q}$ is canceled with the fermion component $\Psi_\Phi$ of the adjoint scalar $\Phi$. This is consistent with the analysis along the Higgs branch, where there is no Higgs moduli space for one flavor. The third term $-3 x^2$ consists of $\Phi^2+Q\psi_Q+\tilde{Q}\psi_{\tilde{Q}}+V_+\tilde{Q} +V_- Q=(1-1-1-1-1)x^2$, where we omitted the vanishing contributions that are canceled between $Q\tilde{Q}$ and $\psi_{\Phi}$. Notice that the dressed monopoles behave as fermionic states since the bosonic fields $Q$ and $\tilde{Q}$ are attached to the monopole flux and their spins are transmuted \cite{Polyakov:1988md, Dimofte:2011py, Aharony:2015pla}. On the magnetic side, there are two types of the monopole operators for the $U(1)_1\times U(1)_0$ gauge group. The gauge-invariant states are defined by $V^{U(1)_1}_{+} V^{U(1)_0}_- q + V^{U(1)_1}_{-} V^{U(1)_0}_+ \tilde{q}$, which are identified with $V_+\tilde{Q} +V_- Q$.    

For $n=2$, the duality relates the 3d $\mathcal{N}=3$ $U(1)_2 \Leftrightarrow \stackrel{+1~\,~~~~}{\protect\wick{1}{<1 U(1)_{2} \times >1 U ( 1 )_{0}}}$ gauge theories with one hypermultiplet. The electric and magnetic superconformal indices are expanded as
\begin{align}
    I_{F=1}^{U(1)_2}&=1+x-x^2+2 x^3+2 x^{7/2}-2 x^4-2 x^{9/2}+x^5+2 x^{11/2}+x^6-2 x^{13/2}-2 x^7 \nonumber \\
    &\quad+2 x^{15/2}+3 x^8-6 x^9-2 x^{19/2}+10 x^{10}+4 x^{21/2}-9 x^{11}-8 x^{23/2}+9 x^{12}+14 x^{25/2}\nonumber \\
    &\qquad -11 x^{13}-18 x^{27/2}+9 x^{14}+22 x^{29/2}-2 x^{15}+\cdots.
\end{align}
In this case, the second term $-x^2$ can be regarded as $\Phi^2+Q\psi_Q+\tilde{Q}\psi_{\tilde{Q}}=(+1-1-1)x^2$ and there is no dressed monopole at this order. The leading monopole states $V_+ \tilde{Q}^2+V_- Q^2$ appear as $2 x^{7/2}$ in the index. On the magnetic side, the gauge-invariant monopole states are defined by $V^{U(1)_2}_{+} \left( V^{U(1)_0}_- \right)^2 q + V^{U(1)_2}_{-} \left( V^{U(1)_0}_+\right)^2 \tilde{q}$, which contributes to the index as $2 x^{7/2}$. Note that the dual matter fields behave as bosonic states under the monopole background $V^{U(1)_2}_{\pm} \left( V^{U(1)_0}_\mp \right)^2$. 

For $n=3$, the duality claims the equivalence between the 3d $\mathcal{N}=3$ $U(1)_3$ and $\stackrel{+1~\,~~~~}{\protect\wick{1}{<1 U(1)_{3} \times >1 U ( 1 )_{0}}}$ gauge theories with one flavor. The electric and magnetic indices are expanded as
\begin{align}
    I_{F=1}^{U(1)_3}&=1+x-x^2+2 x^3-2 x^4-x^5+3 x^6-4 x^7+5 x^8-8 x^9+12 x^{10} \nonumber \\
    &\quad-13 x^{11}+11 x^{12}-11 x^{13}+9 x^{14}+\cdots.
\end{align}
The sixth term $-x^5$ consists of the products of the adjoint scalar $\Phi$ and leading monopole operators $\Phi^5+V_+ \tilde{Q}^3 +V_- Q^3=(+1-1-1)x^5$. Notice that the matter fields behave as fermionic states under the monopole background $V_\pm$. On the magnetic side, the dressed monopoles are represented by $V^{U(1)_3}_{+} \left( V^{U(1)_0}_- \right)^3 q + V^{U(1)_3}_{-} \left( V^{U(1)_0}_+\right)^3 \tilde{q}$. 
As a $U(2)$ example, we study the 3d $\mathcal{N}=3$ $U(2)_{1,1+2n}$ gauge theory with two flavors, which is dual to a 3d $\mathcal{N}=3$ $\stackrel{+1~~~~~~}{\protect\wick{1}{<1 U(1)_{n+1} \times >1 U ( 1 )_{0}}} $ gauge theory with two dual flavors. As studied above, the $n=-1$ case shows the confinement phase, where the four monopoles \eqref{V_+d} and \eqref{V_-d} become free. The superconformal indices for various $n$'s are summarized in Table \ref{SCI} and show the validity of the duality. The second term $4x$ is identified with $\mathrm{tr}\,\Phi +Q\tilde{Q} \sim \phi_{U(1)_0} +q \tilde{q} \sim x+3x$. The trace part of $Q\tilde{Q}$ is eliminated by the superpotential $W \sim \Phi Q \tilde{Q}$. In the index level, the trace part is canceled by the fermion component $\psi_{\mathrm{tr} \, \Phi}$ from the adjoint chiral multiplet. The similar cancellation occurs on the magnetic side. The terms with fractional powers of $x$ are the contributions from the states with non-zero GNO charges. We here consider the (dressed) monopoles for the $n=0$ case. The insertion of the bare monopoles $V_{\pm}$ induces the gauge symmetry breaking 
\begin{align}
   U(2)_{1,1+2n} & \rightarrow \stackrel{+n~~~~~~~~}{\protect\wick{1}{<1 U(1)_{1+n} \times >1 U ( 1 )_{1+n}}} \\
   {\tiny \yng(1)}_{+1}  & \rightarrow (1,0) +(0,1)  \\
    {\tiny \overline{\yng(1)}}_{-1}  & \rightarrow (-1,0) +(0,-1)  \\
    \mathbf{adj.} & \rightarrow (0,0)+(0,0) +(+1,-1)+(-1,+1).
\end{align}
Since there are non-zero (mixed) CS levels, the bare monopoles $V_{\pm}$ are charged under the $U(1) \times U(1)$ gauge group. For $n=0$, naive leading monopoles, which are dressed by matter fields and made gauge-invariant, are defined as follows:
\begin{align}
    V_+ (+1,0) +V_+ (+1,-1)(0,+1) \sim V_+ Q +V_+ W_\alpha Q =  (-2+2)x^{-\frac{3}{2}}  
\end{align}
Notice that the component $(+1,0) \sim Q$ is transmuted to a fermionic state on the monopole background \cite{Polyakov:1988md, Dimofte:2011py, Aharony:2015pla} and these two states are canceled with each other.
The first non-vanishing contribution, which is $-4x^{5/2}$ for $n=0$, comes from
\begin{align}
      V_+ (1,0)(0,1)(0,-1)  +V_+(1,-1)(0,1)^2 (0,-1) \sim V_+ Q^2 \tilde{Q}+V_+ W_\alpha Q^2\tilde{Q} \sim (-8+6) x^{\frac{5}{2}}\\
       V_- (-1,0)(0,-1)(0,1)  +V_-(-1,1)(0,-1)^2 (0,1) \sim V_+ \tilde{Q}^2 Q+V_- W_\alpha \tilde{Q}^2Q \sim (-8+6) x^{\frac{5}{2}}.
\end{align}

\begin{table}[h]\caption{The superconformal indices for $(N,F,k)=(2,2,1)$ with various $n$'s} 
\begin{center}
\scalebox{0.8}{
  \begin{tabular}{|c|c||c| } \hline
 electric & magnetic & superconformal indices  \\ \hline
 $U(2)_{1,7}$&$\stackrel{+1~~~~~~}{\protect\wick{1}{<1 U(1)_{4} \times >1 U ( 1 )_{0}}}$&\shortstack{$1+4 x+x^2+4 x^3+7 x^4-12 x^5+26 x^6-16 x^7+6 x^8$\\ $\qquad +4 x^{17/2}+20 x^9-4 x^{19/2}-47 x^{10}+\cdots$} \\ \hline
$U(2)_{1,5}$ &$\stackrel{+1~~~~~~}{\protect\wick{1}{<1 U(1)_{3} \times >1 U ( 1 )_{0}}}$& \shortstack{$1+4 x+x^2+4 x^3+7 x^4-12 x^5+26 x^6-4 x^{13/2}-16 x^7$\\ $\qquad +4 x^{15/2}+6 x^8-4 x^{17/2}+20 x^9+4 x^{19/2}-47 x^{10}+\cdots$} \\ \hline
 $U(2)_{1,3}$&$\stackrel{+1~~~~~~}{\protect\wick{1}{<1 U(1)_{2} \times >1 U ( 1 )_{0}}}$& \shortstack{ $1+4 x+x^2+4 x^3+7 x^4+4 x^{9/2}-12 x^5-4 x^{11/2}+26 x^6+4 x^{13/2} $ \\ $\qquad-16 x^7-4 x^{15/2}+6 x^8+4 x^{17/2}+20 x^9+8 x^{19/2}-47 x^{10}+\cdots$ } \\ \hline
  $U(2)_{1,1}$&$\stackrel{+1~~~~~~}{\protect\wick{1}{<1 U(1)_{1} \times >1 U ( 1 )_{0}}}$& \shortstack{ $1+4 x+x^2-4 x^{5/2}+4 x^3+4 x^{7/2}+7 x^4-4 x^{9/2}-12 x^5-8 x^{11/2}+26 x^6 $ \\ $\qquad +24 x^{13/2}-10 x^7-44 x^{15/2}-2 x^8+52 x^{17/2}+30 x^9-52 x^{19/2}-63 x^{10}+\cdots$}\\ \hline
$U(2)_{1,-1}$&$\stackrel{+1~~~~~~}{\protect\wick{1}{<1 U(1)_{0} \times >1 U ( 1 )_{0}}}$& $\left( \frac{(x^{3/2};x^2)_\infty}{(x^{1/2};x^2)_\infty}  \right)^4=1+4 \sqrt{x}+10 x+16 x^{3/2}+19 x^2+20 x^{5/2}+26 x^3+40 x^{7/2}+49 x^4+\cdots $  \\[8pt] \hline
$U(2)_{1,-3}$&$\stackrel{+1~~~~~~}{\protect\wick{1}{<1 U(1)_{-1} \times >1 U ( 1 )_{0}}}$& \shortstack{$1+4 x+x^2-4 x^{5/2}+4 x^3+4 x^{7/2}+7 x^4-4 x^{9/2}-12 x^5-8 x^{11/2}+26 x^6 $\\ $\qquad+24 x^{13/2}-10 x^7-44 x^{15/2}-2 x^8+52 x^{17/2}+30 x^9-52 x^{19/2}-63 x^{10}+\cdots$} \\ \hline
$U(2)_{1,-5}$&$\stackrel{+1~~~~~~}{\protect\wick{1}{<1 U(1)_{-2} \times >1 U ( 1 )_{0}}}$& \shortstack{ $1+4 x+x^2+4 x^3+7 x^4+4 x^{9/2}-12 x^5-4 x^{11/2}+26 x^6+4x^{13/2}$\\ $ \qquad -16 x^7-4 x^{15/2}+6 x^8+4 x^{17/2}+20 x^9+8 x^{19/2}-47 x^{10}+\cdots$} \\ \hline
  \end{tabular}}
  \end{center}\label{SCI}
\end{table}
Let us finally consider the $U(N)_{0,0+nN}$ ``ugly-good'' duality proposed in Section \ref{subsection:ugly}. In this duality, the $SU(N)$ dynamics is governed by the Yang-Mills interaction and there is an associated Coulomb branch. For simplicity, we only consider the $N=2$ case, which claims an infrared equivalence:
\begin{align}
    3d~\mathcal{N}=3~~~U(2)_{0,0+2n}~~~~ \Leftrightarrow ~~~~ U(1)_n \times U(1)_0 ~~~~\mbox{with $3$ flavors}.
\end{align}
Notice that there is an additional electron hypermultiplet and no mixed CS term on the magnetic side. For $n=0$, the duality goes back to the 3d $\mathcal{N}=4$ Seiberg duality for the ``ugly'' case \cite{Kapustin:2010mh, Yaakov:2013fza, Assel:2017jgo}. We here focus on the non-zero $n$ cases. For $n=1$, the superconformal indices of the electric and magnetic descriptions are expanded as
\begin{align}
    I^{U(2)_{0,2}}_{F=3} &= 1+10 x+28 x^2-6 x^{5/2}+30 x^3-30 x^{7/2}+73 x^4-6 x^{9/2}+92 x^5-60 x^{11/2}-18 x^6 \nonumber \\
    &-126 x^{13/2}+346 x^7+186 x^{15/2}-11 x^8-570 x^{17/2}-18 x^9+360 x^{19/2}+1168 x^{10}+\cdots.
\end{align}
The second term $10 x$ consists of three gauge-invariant operators: The first one is a meson composite $M:=Q\tilde{Q}$ which contributes to the index as $8x$. Notice that the trace part of $M$ is eliminated by the F-term condition. The second one is a trace part of the $U(2)$ adjoint scalar $\mathrm{Tr}\,\Phi$, which is available since the gauge group is unitary. The third one is a monopole operator $Y_{SU(2)}$ which is associated with the $SU(2)_0$ Coulomb branch and its insertion leads to the gauge symmetry breaking $U(2) \simeq SU(2) \times U(1)  \rightarrow U(1) \times U(1)$. Since the $SU(2)$ CS level is zero, the bare monopole $Y_{SU(2)}$ is gauge-invariant and becomes a part of the moduli space. On the magnetic side, the second term $10 x$ consists of $\phi_{U(1)_n}+\phi_{U(1)_0}+q\tilde{q}=(1+1+8)x$. The trace part of $q \tilde{q}$ is canceled by the fermion state $\psi_{\phi_0}$ while the contribution of $b\tilde{b}$ is canceled by $\psi_{\phi_n}$. The other monopoles appear as $-6 x^{5/2}$ in the index. These are constructed from the monopole operators $V_{\pm}$ which correspond to $\mathrm{diag.} (+1,0)$ and $\mathrm{diag.} (0,-1)$ generators of the $U(2)$ gauge group. On the background with the insertion of the bare monopole $V_{\pm}$, the gauge group is spontaneously broken as 
\begin{align}
    U(2)_{0,2} & \rightarrow \stackrel{+1~~~~~\,}{\protect\wick{1}{<1 U(1)_{1} \times >1 U ( 1 )_{1}}} \\
   {\tiny \yng(1)}_{+1}  & \rightarrow (1,0) +(0,1)  \\
    {\tiny \overline{\yng(1)}}_{-1}  & \rightarrow (-1,0) +(0,-1)  \\
    \mathbf{adj.} & \rightarrow (0,0)+(0,0) +(+1,-1)+(-1,+1).
\end{align}
Since there are (mixed) CS terms in the presence of the monopole $V_{\pm}$, the bare monopoles are not gauge-invariant. Therefore, we need to define the following gauge-invariant dressed states and they contribute to the index as the fourth term $-6 x^{5/2}$:
\begin{align}
  V_+(+1,0)(0,+1)+V_+(+1,-1)(0,+1)^2 \sim  V_+ Q^2 +V_+ W_\alpha Q^2 =(-9+6)x^{-\frac{5}{2}} \\
   V_-(-1,0)(0,-1)+V_-(-1,+1)(0,-1)^2 \sim  V_- \tilde{Q}^2 +V_- W_\alpha \tilde{Q}^2 =(-9+6)x^{-\frac{5}{2}}.
\end{align}
Notice that the spins of the components $(\pm1,0)$ are transmuted to be fermionic under the monopole $V_{\pm}$ background \cite{Polyakov:1988md, Dimofte:2011py, Aharony:2015pla}. On the magnetic side, the fourth term $-6 x^{5/2}$ is constructed from the $U(1)_n$ monopoles as the dressed states $V_+^{U(1)_n} \tilde{b} \tilde{q}+V_-^{U(1)_n} b q\sim (-3-3)x^{5/2}$.
Similarly, we observed a nice agreement of the electric and magnetic indices for $n=2$ and $n=3$:
\begin{align}
    I^{U(2)_{0,4}}_{F=3} &=I^{U(1)_2 \times U(1)_0}_{F=3+electron} \nonumber \\
    &=1+10 x+28 x^2+30 x^3+73 x^4+12 x^{9/2}+92 x^5+42 x^{11/2}-18 x^6 \nonumber \\
    &\qquad -12 x^{13/2}+334 x^7+102 x^{15/2}-53 x^8+48 x^{17/2}-6 x^9-120 x^{19/2}+1066 x^{10}+\cdots\\
    I^{U(2)_{0,6}}_{F=3} &=I^{U(1)_3 \times U(1)_0}_{F=3+electron} \nonumber\\
    &= 1+10 x+28 x^2+30 x^3+73 x^4+92 x^5-18 x^6-20 x^{13/2} \nonumber \\
    & \qquad +334 x^7-54 x^{15/2}-53 x^8+34 x^{17/2}-6 x^9-160 x^{19/2}+1066 x^{10}+\cdots.
\end{align}
In these examples, the leading monopole operator is again $Y_{SU(2)}$ and represented by $x$ in the index. The dressed monopoles constructed from $V_{\pm}$ are represented by $12 x^{9/2}$ and $-20 x^{13/2}$, respectively. These analyses of the superconformal indices shows the validity of our duality proposal. 

\section{Summary and discussion\label{sec:Summary}}

In this paper, we proposed the 3d $\mathcal{N}=3$ version of the $U(N)_{k,k+nN}$ Giveon-Kutasov duality with $F$ fundamental hypermultiplets. This duality can be regarded as a natural generalization along various directions: Firstly, the proposed duality generalizes the Chern-Simons level by introducing the unbalanced CS couplings between the abelian and non-abelian subgroups. Secondly, the proposed duality generalizes the less-supersymmetric $U(N)_{k,k+nN}$ Seiberg-like/level-rank/bosonization duality to the 3d $\mathcal{N}=3$ system. Thirdly, our duality includes the $SU(N)_k$ Giveon-Kutasov duality in the $n \rightarrow \infty$ limit and thus it offers a unified description of the $SU(N)$ and $U(N)$ CS dualities. For the abelian case ($N=1$), we argued that the duality is enhanced and there is an infinite number of magnetic descriptions. By using the proposed duality, we mapped out the phase diagram of the electric theory, which exhibits confinement and supersymmetry breaking/enhancement.


We also proposed the duality for the 3d $\mathcal{N}=3$ $U(N)_{0,0+nN}$ gauge theory with $2N-1$ fundamental flavors, which is also the generalization of the so-called $U(N)_{0,0}$ ``ugly-good'' duality \cite{Kapustin:2010mh, Yaakov:2013fza, Assel:2017jgo}. In the conventional ``ugly-good'' duality, the monopole hypermultiplet becomes a free field and decouples from the other sector. As a result, the good side of the duality includes a free decoupled hypermultiplet. However, in the $U(N)_{0,0+nN}$ ``ugly-good'' duality, the would-be decoupled sector is still interacting with the other elementary fields and does not decouple. In the $n \rightarrow \infty$ limit, we constructed the $SU(N)_0$ ``ugly-good'' duality, which is surprisingly related with the S-duality of the 4d $\mathcal{N}=2$ $SU(N)$ gauge theory with $2N$ hypermultiplets \cite{Seiberg:1994aj, Argyres:1995wt, Argyres:1996eh, Leigh:1995ep, Hirayama:1997tw}. By reducing the number of flavors, we proposed possible ways to redeem or study the ``bad'' theories.  
 
We tested the proposed duality by computing the sphere partition functions via supersymmetric localization. Since we constructed the duality in a 3d $\mathcal{N}=3$ setup, we can use the power of various matrix-model techniques. By using those technologies, we exactly proved the matrix integral identities including the $U(1)$ phases, which relates the sphere partition functions of the electric and magnetic theories. 
We also studied the superconformal indices of the 3d $\mathcal{N}=3$ generalized $U(N)_{k,k+nN}$ CS gauge theory and found a nice agreement under the duality. In this paper, we explicitly computed the expansion of the indices for the $N=1$ and $N=2$ cases. For higher ranks, it becomes difficult to compute the series of the superconformal index at a non-trivial order to test the duality since it requires very huge computer power. It is important to study the superconformal indices in more elegant ways, for example, by using the factorization property of the index \cite{Hwang:2012jh, Hwang:2015wna}. 

It is important to give other tests of the proposed Giveon-Kutasov duality. For example, it is possible to study the supersymmetric Wilson loops, which can be also exactly computed by using the localization technique \cite{Kapustin:2009kz}. It is unclear how the electric Wilson loops in various representations are mapped to the magnetic side. This analysis enables us to understand the Wilson loop algebra \cite{Witten:1993xi, Kapustin:2013hpk}.
It is also important to consider the quiver-type generalization where each quiver gauge group has a generalized Chern-Simons level. It is interesting to study when those quiver-type generalized CS gauge theories show supersymmetry enhancement \`{a} la ABJ(M) theory \cite{Aharony:2008ug, Aharony:2008gk}.  
As another future direction, it is possible to introduce boundaries in the proposed duality. In case of the duality with flavors, we expect that the duality is modified such that the theory includes the 2d and 3d dynamics. In the $F=0$ case with boundaries, it is expected that the 3d $\mathcal{N}=3$ generalized dualities lead to the 2d supersymmetric WZW dualities \cite{Elitzur:1989nr, Naculich:1990pa, Nakanishi:1990hj}. We will come back to these problems in the near future.   


\section*{Acknowledgement}
We would like to thank Tomoki Nosaka, Yuya Tanizaki and Seiji Terashima for valuable discussions. We are especially grateful to Shigeki Sugimoto for various discussions and for teaching us physics.  
The work of NK is supported by Grant-in-Aid for JSPS Fellows No.20J12263.

\appendix

\section{Notations and formulae for quantum mechanics\label{sec:QM}}
In this appendix, we provide notations and formulae of quantum mechanics, which will be extensively used in the next section. The commutation relation of the position operator $\hat{q}$ and the associated momentum operator $\hat{p}$ is given by $\left[\hat{q},\hat{p}\right]=i\hbar$. The other commutation relations are zero as usual. In the main text, we relate the reduced Planck constant with the Chern-Simons level as $\hbar=2\pi k$. In the appendix, $\hbar$ is used in many places. The usual ket symbol $\ket{\cdot}$ denotes the position eigenvector. We also introduce a double-ket symbol $\kket{\cdot}$ to denote the momentum eigenvector. The inner products of these eigenvectors are defined as follows:
\begin{align}
\begin{gathered}
  \braket{q_{1}|q_{2}}=\delta\left(q_{1}-q_{2}\right),\quad\bbrakket{p_{1}|p_{2}}=\delta\left(p_{1}-p_{2}\right), \\
  \brakket{q|p}=\frac{1}{\sqrt{2\pi\hbar}}e^{\frac{i}{\hbar}pq},\quad\bbraket{p|q}=\frac{1}{\sqrt{2\pi\hbar}}e^{-\frac{i}{\hbar}pq}.
  \end{gathered}
  \label{eq:Normalization}
\end{align}
We also enumerate some useful formulae which are frequently used:
\begin{align}
\begin{gathered}
  e^{-\frac{ia}{\hbar}\hat{q}}f\left(\hat{p}\right)e^{\frac{ia}{\hbar}\hat{q}}=f\left(\hat{p}+a\right),\quad   e^{-\frac{i}{2\hbar}\hat{p}^{2}}e^{-\frac{i}{2\hbar}\hat{q}^{2}}f\left(\hat{p}\right)e^{\frac{i}{2\hbar}\hat{q}^{2}}e^{\frac{i}{2\hbar}\hat{p}^{2}}=f\left(\hat{q}\right).
\end{gathered}
 \label{eq:OpSim}
\end{align}
For the canonical coordinates eigenvectors, we have the following formulae:
\begin{align}
\begin{gathered}
  e^{\frac{ia}{\hbar}\hat{q}}\kket p=\kket{p+a},\quad\bbra pe^{-\frac{ia}{\hbar}\hat{q}}=\bbra{p+a}, \\
  e^{-\frac{i}{2\hbar}\hat{p}^{2}}e^{-\frac{i}{2\hbar}\hat{q}^{2}}\kket p=\frac{1}{\sqrt{i}}e^{\frac{i}{2\hbar}p^{2}}\ket p,\quad\bbra pe^{\frac{i}{2\hbar}\hat{q}^{2}}e^{\frac{i}{2\hbar}\hat{p}^{2}}=\sqrt{i}e^{-\frac{i}{2\hbar}p^{2}}\bra p.
 \end{gathered}
 \label{eq:VecSim}
\end{align}

\section{Integral identities from matrix models\label{sec:MM-Duality}}
This section is devoted to the proof of the matrix integral identities that are obtained from the Seiberg-like dualities known in the literature. 
Although the results presented in this section have been already proven, we here reproduce them in the Fermi gas formalism \cite{Marino:2011eh}. For deriving various integral identities, we extensively use power of the operator formalism developed in \cite{Moriyama:2015asx,Moriyama:2016kqi,Moriyama:2017gye,Kubo:2020qed}.
The operator formalism simplifies the computation of matrix models a lot.
After applying the Fermi gas formalism, the connection between various dualities becomes obvious.
The guideline of computing matrix integrals comes from the brane configuration. The operator formalism clarifies the relation between the brane configuration and the matrix integral.
In this section, we make the real masses tracefull for keeping track of the transformation of those parameters.
In Section \ref{Z_CGK}, we will show that the sphere partition functions of the 3d $\mathcal{N}=3$ $U(N)_{k}$ Chern-Simons matter gauge theories coincide under the conventional Giveon-Kutasov duality \cite{Giveon:2008zn, Kapustin:2010mh, Assel:2014awa}. In Section \ref{Z_CUG}, we derive an integral identity obtained from the 3d $\mathcal{N}=4$ Seiberg-like duality for the so-called ``ugly'' theory with $F=2N-1$ flavors \cite{Gaiotto:2008ak, Kapustin:2010mh, Yaakov:2013fza, Assel:2017jgo}.


\subsection{The conventional GK duality with $n=0$}\label{Z_CGK}
As a warm-up of deriving various integral identities from the 3d $\mathcal{N}=3$ Seiberg-like dualities, we first study the 3d $\mathcal{N}=3$ version of the conventional Giveon-Kutasov duality \cite{Giveon:2008zn, Kapustin:2010mh, Assel:2014awa}, which claims the equivalence between the 3d $\mathcal{N}=3$ $U\left(N\right)_{k,k}$ and $U\left(F+k-N\right)_{-k,-k}$ gauge theories with $F$ fundamental hypermultiplets. In our notation for the sphere partition functions, we will show an integral identity
\begin{align}
\zU[][][][0]\left(\bm{m},\zeta\right)=e^{i\Theta_{F,k,N}}e^{-i\pi\zeta^{2}+i\pi k\sum_{a}^{F}m_{a}^{2}+2\pi i\zeta\sum_{a}^{F}m_{a}}\zU[\tilde{N}][][-k][0]\left(\bm{m},-\zeta\right),\label{eq:LRwoDiag}
\end{align}
where the phase factor is divided into two parts: 
The first phase $\Theta_{F,k,N}$, which is defined in \eqref{eq:Phase-Def}, depends only on $N,F$ and $k$ while the second one depends on the FI and real mass parameters, $\zeta$ and $\bm{m}$. The second phase is important for us to prove the integral identity of the generalized Giveon-Kutasov duality in the main text. In the equality \eqref{eq:LRwoDiag}, we keep the real masses trace-full although the trace part is redundant. In this subsection, we always assume $k>0$ for simplicity and use the following notation for the rank of the dual theory:
\begin{align}
\tilde{N}&=F+k-N.
\end{align}
To show the integral identity \eqref{eq:LRwoDiag}, we study the $F\leq N$ and $F>N$ cases, separately. 
The duality for the $k=0$ case will be discussed in the next subsection.

\subsubsection{The $F\protect\leq N$ case}
\begin{figure}
\begin{centering}
\includegraphics[scale=0.75]{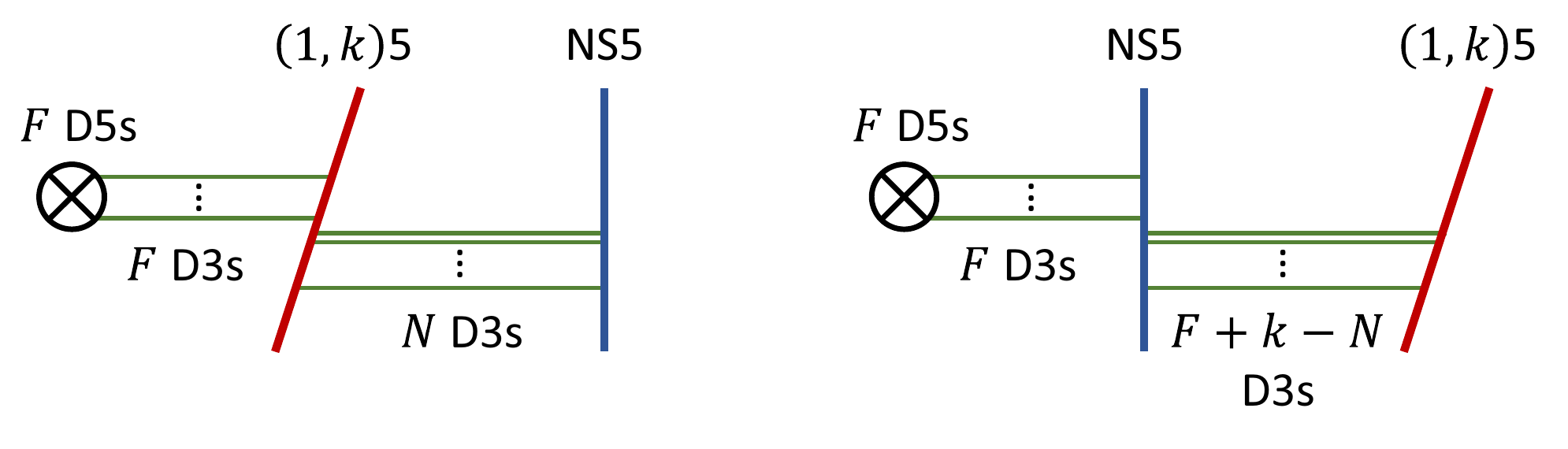}
\par\end{centering}
\caption{The brane configurations which correspond to \eqref{eq:MM-FGF1} (left) and \eqref{eq:MM-FGF2} (right).
Both the brane configurations are related with each other by the Hanany-Witten move.
The NS5-brane and the D5-brane span the $(0,1,2,3,4,5)$ directions and $(0,1,2,7,8,9)$ directions, respectively.
The $(1,k)$5-brane spans the $(0,1,2,3/7,4/8,5/9)$ directions with appropriate angles.
The D3-branes span the $(0,1,2,3)$ directions.\label{fig:FlessN}}
\end{figure}

We first consider the $F \le N$ case and start with the left-hand side $\zU[][][][0]\left(\bm{m},\zeta\right)$ of the integral identity \eqref{eq:LRwoDiag}. By changing the integration variables as $\lambda_j \rightarrow \frac{\lambda_j}{\hbar}$, where we take $\hbar=2\pi k$, the partition function is written as
\begin{align}
\zU[][][][0]\left(\bm{m},\zeta\right)= & \int\frac{d^{N}\lambda}{\hbar^{N}N!}e^{\frac{i}{2\hbar}\sum_{j}^{N}\lambda_{j}^{2}+\frac{2\pi i\zeta}{\hbar}\sum_{j}^{N}\lambda_{j}}\frac{\prod_{j<j'}^{N}\left(2\sinh\frac{\lambda_{j}-\lambda_{j'}}{2k}\right)^{2}}{\prod_{a,j}^{F,N}2\cosh\frac{\lambda_{j}-\hbar m_{a}}{2k}}.
\end{align}
First of all, we massage the integrand by introducing fictitious $\kappa_a$ integrals such that we can easily rewrite it in terms of the quantum mechanical terminology. By inserting the identity $1=\int d^{F}\kappa\prod_{a}\delta\left(\kappa_a-\hbar m_{a}\right)$ in the above matrix integral, we obtain
\begin{align}
\zU[][][][0]\left(\bm{m},\zeta\right)= & \frac{1}{\prod_{a<a'}^{F}2\sinh\pi\left(m_{a}-m_{a'}\right)}\int d^{F}\kappa\int\frac{d^{N}\lambda}{\hbar^{N}N!}e^{\frac{i}{2\hbar}\sum_{j}^{N}\lambda_{j}^{2}+\frac{2\pi i\zeta}{\hbar}\sum_{j}^{N}\lambda_{j}}\nonumber \\
 & \times\prod_{a}^{F}\delta\left(\kappa_{a}-\hbar m_{a}\right)\frac{\prod_{a<a'}^{F}2\sinh\frac{\kappa_{a}-\kappa_{a'}}{2k}\prod_{j<j'}^{N}2\sinh\frac{\lambda_{j}-\lambda_{j'}}{2k}}{\prod_{a,j}^{F,N}2\cosh\frac{\kappa_{a}-\lambda_{j}}{2k}}\prod_{j<j'}^{N}2\sinh\frac{\lambda_{j}-\lambda_{j'}}{2k}.\label{eq:MM-FGF1}
\end{align}
Note that this expression has a nice brane interpretation by using the Hanany-Witten setup in the type IIB string theory \cite{Hanany:1996ie, Giveon:1998sr}.
The Chern-Simons gauge theory is realized as the worldvolume theory of D3-branes between an NS5-brane and a $(1,k)$5-brane.
The $F$ fundamental hypermultiplets can be added by putting $F$ D5-branes between the NS5-brane and the $(1,k)$5-brane.
We can move these D5-branes to the left of the $(1,k)$5-brane.\footnote{
In the brane picture, it is necessary to remove the D5-branes from the segment between the NS5-brane and the $(1,k)$5-brane since the conventional Giveon-Kutasov duality is realized by crossing the NS5-brane and the $(1,k)$5-brane as in figure \ref{fig:FlessN}.
}
It is known that when an $(1,k)$5-brane (or an NS5-brane) and a D5-brane cross each other, one D3-brane is created.
This phenomenon is called Hanany-Witten effect \cite{Hanany:1996ie}.
As a result, we obtain the brane configuration in Figure \ref{fig:FlessN}.
The massaged matrix model \eqref{eq:MM-FGF1} can be regarded as representing this brane configuration.
In the localization formulae, the integration variables are the diagonal elements of the adjoint scalar in the vector multiplet.
Therefore, the $\kappa_a$ integral can be regarded as a $U(F)$ vector multiplet living on the $F$ D3-branes at the left of the $(1,k)$5-brane, and the $\lambda_j$ integral can be regarded as a $U(N)$ vector multiplet living on the $N$ D3-branes at the segment between the $(1,k)$5-brane and the NS5-brane.
Furthermore, in the localization formulae, the hyperbolic sine functions in the numerator come from the 1-loop determinants of the vector multiplets while the hyperbolic cosine functions in the denominator come from the bi-fundamental hypermultiplets.
Therefore, the hyperbolic sine functions including the $\kappa_a$ variables can be interpreted as the ``half'' of the vector multiplet living on the $F$ D3-branes at the left, while the hyperbolic cosine functions that include the $\kappa_a$ and $\lambda_j$ variables can be interpreted as the bi-fundamental matter multiplet describing the dynamics of the fundamental strings across the $(1,k)$5-brane and ending on the D3-branes.

Now, we can rewrite the integrand of \eqref{eq:MM-FGF1} by using the determinant formula
\begin{align}
 & \frac{\prod_{a<a'}^{N}2\sinh\frac{\kappa_{a}-\kappa_{a'}}{2k}\prod_{b<b'}^{N+L}2\sinh\frac{\lambda_{b}-\lambda_{b'}}{2k}}{\prod_{a}^{N}\prod_{b}^{N+L}2\cosh\frac{\kappa_{a}-\lambda_{b}}{2k}}=\det\left(\begin{array}{c}
\left[\hbar\braket{\kappa_{a}|\frac{1}{2\cosh\frac{\hat{p}-i\pi L}{2}}|\lambda_{b}}\right]_{a,b}^{N\times\left(N+L\right)}\\
\left[\frac{\hbar}{\sqrt{k}}\bbraket{t_{L,\ell}|\lambda_{b}}\right]_{\ell,b}^{L\times\left(N+L\right)}
\end{array}\right),\nonumber \\
 & \frac{\prod_{a<a'}^{N+L}2\sinh\frac{\kappa_{a}-\kappa_{a'}}{2k}\prod_{b<b'}^{N}2\sinh\frac{\lambda_{b}-\lambda_{b'}}{2k}}{\prod_{a}^{N+L}\prod_{b}^{N}2\cosh\frac{\kappa_{a}-\lambda_{b}}{2k}}\nonumber \\
 & =\det\left(\begin{array}{cc}
\left[\hbar\braket{\kappa_{a}|\frac{1}{2\cosh\frac{\hat{p}+i\pi L}{2}}|\lambda_{b}}\right]_{a,b}^{\left(N+L\right)\times N} & \left[\frac{\hbar}{\sqrt{k}}\brakket{\kappa_{a}|-t_{L,\ell}}\right]_{a,\ell}^{\left(N+L\right)\times L}\end{array}\right),\label{eq:CauchyDet}
\end{align}
where we defined a pure imaginary parameter
\begin{align}
t_{N,\ell}=2\pi i\left(\frac{N+1}{2}-\ell\right).
\end{align}
$\left[\mathbf{a}_{ab}\right]_{a,b}^{N\times L}$ denotes an $N\times L$ matrix whose $\left(a,b\right)$ element is $\mathbf{a}_{ab}$.
Note that each determinant corresponds to one 5-brane.
This is because the left-hand side of this formula is the combination of the 1-loop determinants of the hypermultiplet describing the fundamental strings across the 5-brane and ``half'' of the vector multiplets describing the fundamental strings living on the D3-branes on the left and right side of the 5-brane. 
The proof of this formula is in \cite{Kubo:2020qed}. By using the above determinant formula, the one-loop determinant of the matrix model can be written as
\begin{align}
\zU[][][][0]\left(\bm{m},\zeta\right)= & \frac{1}{\prod_{a<a'}^{F}2\sinh\pi\left(m_{a}-m_{a'}\right)}\int d^{F}\kappa\int\frac{d^{N}\lambda}{\hbar^{N}N!}e^{\frac{i}{2\hbar}\sum_{j}^{N}\lambda_{j}^{2}+\frac{2\pi i\zeta}{\hbar}\sum_{j}^{N}\lambda_{j}}\prod_{a}^{F}\braket{\hbar m_{a}|\kappa_{a}}\nonumber \\
 & \times\det\left(\begin{array}{c}
\left[\hbar\braket{\kappa_{a}|\frac{1}{2\cosh\frac{\hat{p}-i\pi\left(N-F\right)}{2}}|\lambda_{j}}\right]_{a,j}^{F\times N}\\
\left[\frac{\hbar}{\sqrt{k}}\bbraket{t_{N-F,\ell}|\lambda_{j}}\right]_{\ell,j}^{\left(N-F\right)\times N}
\end{array}\right)\det\left(\left[\frac{\hbar}{\sqrt{k}}\brakket{\lambda_{j}|-t_{N,\ell}}\right]_{j,\ell}^{N\times N}\right).
\end{align}
By putting the exponential factor $e^{\frac{i}{2\hbar}\sum_{j}^{N}\lambda_{j}^{2}+\frac{2\pi i\zeta}{\hbar}\sum_{j}^{N}\lambda_{j}}$, which are a CS action and a FI term at the localization locus, inside the first determinant, we obtain
\begin{align}
&\zU[][][][0]\left(\bm{m},\zeta\right)\nonumber \\
&=  \frac{1}{\prod_{a<a'}^{F}2\sinh\pi\left(m_{a}-m_{a'}\right)}\int d^{F}\kappa\int\frac{d^{N}\lambda}{\hbar^{N}N!}\prod_{a}^{F}\braket{\hbar m_{a}|\kappa_{a}}\nonumber \\
 &\quad \times\det\left(\begin{array}{c}
\left[\hbar\braket{\kappa_{a}|\frac{1}{2\cosh\frac{\hat{p}-i\pi\left(N-F\right)}{2}}e^{\frac{2\pi i\zeta}{\hbar}\hat{q}}e^{\frac{i}{2\hbar}\hat{q}^{2}}|\lambda_{j}}\right]_{a,j}^{F\times N}\\
\left[\frac{\hbar}{\sqrt{k}}\bbraket{t_{N-F,\ell}|e^{\frac{2\pi i\zeta}{\hbar}\hat{q}}e^{\frac{i}{2\hbar}\hat{q}^{2}}|\lambda_{j}}\right]_{\ell,j}^{\left(N-F\right)\times N}
\end{array}\right)\det\left(\left[\frac{\hbar}{\sqrt{k}}\brakket{\lambda_{j}|-t_{N,\ell}}\right]_{j,\ell}^{N\times N}\right).
\end{align}
By performing the similarity transformation 
\begin{align}
\int d\kappa\ket{\kappa}\bra{\kappa} & =\int d\kappa\, e^{\frac{2\pi i\zeta}{\hbar}\hat{q}}e^{\frac{i}{2\hbar}\hat{q}^{2}}e^{\frac{i}{2\hbar}\hat{p}^{2}}\ket{\kappa}\bra{\kappa}e^{-\frac{i}{2\hbar}\hat{p}^{2}}e^{-\frac{i}{2\hbar}\hat{q}^{2}}e^{-\frac{2\pi i\zeta}{\hbar}\hat{q}},\nonumber \\
\int d\lambda\ket{\lambda}\bra{\lambda} & =\int d\lambda\, e^{\frac{i}{2\hbar}\hat{p}^{2}}\ket{\lambda}\bra{\lambda}e^{-\frac{i}{2\hbar}\hat{p}^{2}},
\end{align}
and using the formulae \eqref{eq:OpSim} and \eqref{eq:VecSim}, we obtain
\begin{align}
\zU[][][][0]\left(\bm{m},\zeta\right)= & \frac{i^{\frac{1}{2}\left(N-F\right)}e^{i\theta_{N}+i\theta_{N-F}-\frac{i\pi}{k}\left(N-F\right)\zeta^{2}}e^{i\pi k\sum_{a}^{F}m_{a}^{2}+2\pi i\zeta\sum_{a}^{F}m_{a}}}{\prod_{a<a'}^{F}2\sinh\pi\left(m_{a}-m_{a'}\right)}\nonumber \\
 & \times\int d^{F}\kappa\int\frac{d^{N}\lambda}{\hbar^{N}N!}\prod_{a}^{F}\braket{\hbar m_{a}|e^{\frac{i}{2\hbar}\hat{p}^{2}}|\kappa_{a}}\nonumber \\
 & \times\det\left(\begin{array}{c}
\left[\hbar\braket{\kappa_{a}|\frac{1}{2\cosh\frac{\hat{q}+2\pi\zeta-i\pi\left(N-F\right)}{2}}|\lambda_{j}}\right]_{a,j}^{F\times N}\\
\left[\frac{\hbar}{\sqrt{k}}\braket{t_{N-F,\ell}-2\pi\zeta|\lambda_{j}}\right]_{\ell,j}^{\left(N-F\right)\times N}
\end{array}\right)\det\left(\left[\frac{\hbar}{\sqrt{k}}\brakket{\lambda_{j}|-t_{N,\ell}}\right]_{j,\ell}^{N\times N}\right),
\end{align}
where $\theta_N$ is defined in \eqref{eq:Phase-Def}.
To further simplify the integrand, let us consider diagonalizing the matrix in the first determinant. For any anti-symmetric functions $f(\bm{\lambda})=f(\lambda_1,\cdots,\lambda_N)$, the following relation holds: 
\begin{align}
\int \frac{d^{N}\lambda}{N!}\det\left(\left[g_{a}\left(\lambda_{j}\right)\right]_{a,j}^{N\times N}\right)f\left(\bm{\lambda}\right)=\int d^{N}\lambda\prod_{j}^{N}g_{j}\left(\lambda_{j}\right)f\left(\bm{\lambda}\right).\label{eq:DiagForm}
\end{align}
This formula is available since the other part of the integrand is an anti-symmetric function in $\lambda_j$. 
After using this formula, we can perform the $d^N \lambda$ integration by using $N$ dimensional delta functions appear from the first determinant.
Note that, for $k\neq 0$, the delta functions always appear in the matrix corresponding to the $\left(1,k\right)$5-brane.
Thus, we obtain
\begin{align}
\zU[][][][0]\left(\bm{m},\zeta\right)= & \frac{1}{k^{\frac{N-F}{2}}}\frac{i^{\frac{1}{2}\left(N-F\right)}e^{i\theta_{N}+i\theta_{N-F}-\frac{i\pi}{k}\left(N-F\right)\zeta^{2}}e^{i\pi k\sum_{a}^{F}m_{a}^{2}+2\pi i\zeta\sum_{a}^{F}m_{a}}}{\prod_{a<a'}^{F}2\sinh\pi\left(m_{a}-m_{a'}\right)}\nonumber \\
 & \times\int d^{F}\kappa\prod_{a}^{F}\braket{\hbar m_{a}|e^{\frac{i}{2\hbar}\hat{p}^{2}}|\kappa_{a}}\prod_{a}^{F}\frac{1}{2\cosh\frac{\kappa_{a}+2\pi\zeta-i\pi\left(N-F\right)}{2}}\prod_{j<j'}^{N}2\sinh\frac{\lambda_{j}-\lambda_{j'}}{2k},
\end{align}
where $\lambda_j$ are given in terms of the other integration variables and parameters 
\begin{equation}
\lambda_{j}=\begin{cases}
\kappa_{j} & \left(1\leq j\leq F\right)\\
t_{N-F,j-F}-2\pi\zeta. & \left(F+1\leq j\leq N\right)
\end{cases}.
\end{equation}
By inserting these expressions into the integrand, we arrive at
\begin{align}
\zU[][][][0]\left(\bm{m},\zeta\right)=&  \zpureU[N-F]\frac{i^{\frac{1}{2}N^{2}-\frac{1}{2}F^{2}}e^{i\theta_{N}+i\theta_{N-F}-\frac{i\pi}{k}\left(N-F\right)\zeta^{2}}e^{i\pi k\sum_{a}^{F}m_{a}^{2}+2\pi i\zeta\sum_{a}^{F}m_{a}}}{\prod_{a<a'}^{F}2\sinh\pi\left(m_{a}-m_{a'}\right)}\int d^{F}\kappa \nonumber \\
 & \times\prod_{a}^{F}\braket{\hbar m_{a}|e^{\frac{i}{2\hbar}\hat{p}^{2}}|\kappa_{a}}\prod_{a}^{F}i^{F-N}\frac{\prod_{j}^{N-F}2\sinh\frac{\kappa_{a}+2\pi\zeta-t_{N-F,j}}{2k}}{2\cosh\frac{\kappa_{a}+2\pi\zeta-i\pi\left(N-F\right)}{2}}\prod_{a<a'}^{F}2\sinh\frac{\kappa_{a}-\kappa_{a'}}{2k},\label{eq:eleFGF}
\end{align}
where we factored out
\begin{align}
\zpureU & =\frac{1}{\left|k\right|^{\frac{N}{2}}}\prod_{j<j'}^{N}2\sin\frac{\pi}{\left|k\right|}\left(j'-j\right).
\end{align}
This factor can be regarded as the sphere partition function of the 3d $\mathcal{N}=3$ $U(N)_k$ pure CS gauge theory in the standard framing \cite{Witten:1988hf,Kapustin:2009kz, Kapustin:2010mh,Marino:2011nm}. From this expression, we can see that the pure CS partition function vanishes
\begin{equation}
\zpureU=0~~~~~\mbox{for}~~~~N>k.
\end{equation}
Therefore, we can conclude that  
\begin{equation}
\zU[][][][0]\left(\bm{m},\zeta\right)=0 ~~~~~\mbox{for}~~~~ F+k-N<0.
\end{equation}
From the observation of the sphere partition function, we can see that the 3d $\mathcal{N}=3$ $U(N)_{k,k+nN}$ Chern-Simons gauge theory with $F$ fundamental hypermultiplets spontaneously breaks supersymmetry when the inequality $F+k-N<0$ is satisfied.

Next, we reformulate the magnetic matrix model in the right-hand side of \eqref{eq:LRwoDiag} by using the quantum mechanical description. As in the electric side, by using the determinant formula \eqref{eq:CauchyDet}, the magnetic partition function $\zU[\tilde{N}][][-k][0]\left(\bm{m},-\zeta\right)$ is massaged into
\begin{align}
\zU[\tilde{N}][][-k][0]\left(\bm{m},-\zeta\right)= & \frac{1}{\prod_{a<a'}^{F}2\sinh\pi\left(m_{a}-m_{a'}\right)}\int d^{F}\kappa\int\frac{d^{\tilde{N}}\lambda}{\hbar^{\tilde{N}}\tilde{N}!}\prod_{a}^{F}\braket{\hbar m_{a}|\kappa_{a}}\nonumber \\
 &\times \Phi\left(\kappa,\lambda\right)\det\left(\left[\frac{\hbar}{\sqrt{k}}\brakket{\lambda_{j}|e^{-\frac{i}{2\hbar}\hat{q}^{2}}e^{-\frac{2\pi i\zeta}{\hbar}\hat{q}}|-t_{\tilde{N},\ell}}\right]_{j,\ell}^{\tilde{N}\times\tilde{N}}\right),\label{eq:MM-FGF2}
\end{align}
where the first determinant factor is defined by
\begin{align}
&\Phi\left(\kappa,\lambda\right) \nonumber \\
& =\frac{\prod_{a<a'}^{F}2\sinh\frac{\kappa_{a}-\kappa_{a'}}{2k}\prod_{j<j'}^{\tilde{N}}2\sinh\frac{\lambda_{j}-\lambda_{j'}}{2k}}{\prod_{a,j}^{F,\tilde{N}}2\cosh\frac{\kappa_{a}-\lambda_{j}}{2k}}\nonumber \\
 & =\begin{cases}
\det\left(\begin{array}{c}
\left[\hbar\braket{\kappa_{a}|\frac{1}{2\cosh\frac{\hat{p}-i\pi\left(k-N\right)}{2}}|\lambda_{j}}\right]_{a,j}^{F\times\tilde{N}}\\
\left[\frac{\hbar}{\sqrt{k}}\bbraket{t_{k-N,\ell}|\lambda_{j}}\right]_{\ell,j}^{\left(k-N\right)\times\tilde{N}}
\end{array}\right) & \left(k\geq N\right)\\
\det\left(\begin{array}{cc}
\left[\hbar\braket{\kappa_{a}|\frac{1}{2\cosh\frac{\hat{p}+i\pi\left(N-k\right)}{2}}|\lambda_{j}}\right]_{a,j}^{F\times\tilde{N}} & \left[\frac{\hbar}{\sqrt{k}}\brakket{\kappa_{a}|-t_{N-k,\ell}}\right]_{a,\ell}^{F\times\left(N-k\right)}\end{array}\right) & \left(k<N\right)
\end{cases}.
\end{align}
Note that this expression also has a brane interpretation as shown in Figure \ref{fig:FlessN}.
By performing the similarity transformation 
\begin{align}
\int d\kappa\ket{\kappa}\bra{\kappa}  =\int d\kappa\, e^{\frac{i}{2\hbar}\hat{p}^{2}}\ket{\kappa}\bra{\kappa}e^{-\frac{i}{2\hbar}\hat{p}^{2}}, \quad
\int d\lambda\ket{\lambda}\bra{\lambda}  =\int d\lambda\, e^{\frac{i}{2\hbar}\hat{p}^{2}}\ket{\lambda}\bra{\lambda}e^{-\frac{i}{2\hbar}\hat{p}^{2}},
\end{align}
and using \eqref{eq:VecSim},
we obtain
\begin{align}
\zU[\tilde{N}][][-k][0]\left(\bm{m},-\zeta\right)= & \frac{i^{-\frac{1}{2}\tilde{N}}e^{-i\theta_{k-N}-i\theta_{\tilde{N}}+\frac{i\pi}{k}\tilde{N}\zeta^{2}}}{\prod_{a<a'}^{F}2\sinh\pi\left(m_{a}-m_{a'}\right)}\int d^{F}\kappa\int\frac{d^{\tilde{N}}\lambda}{\hbar^{\tilde{N}}\tilde{N}!}\prod_{a}^{F}\braket{\hbar m_{a}|e^{\frac{i}{2\hbar}\hat{p}^{2}}|\kappa_{a}}\nonumber \\
 & \times\frac{\prod_{a<a'}^{F}2\sinh\frac{\kappa_{a}-\kappa_{a'}}{2k}\prod_{j<j'}^{\tilde{N}}2\sinh\frac{\lambda_{j}-\lambda_{j'}}{2k}}{\prod_{a,j}^{F,\tilde{N}}2\cosh\frac{\kappa_{a}-\lambda_{j}}{2k}}\det\left(\left[\frac{\hbar}{\sqrt{k}}\braket{\lambda_{j}|-t_{\tilde{N},\ell}-2\pi\zeta}\right]_{j,\ell}^{\tilde{N}\times\tilde{N}}\right).
\end{align}
After diagonalizing the matrix in the determinant and performing the $d^{\tilde{N}}\lambda$ integral through the delta functions, we obtain
\begin{align}
\zU[\tilde{N}][][-k][0]\left(\bm{m},-\zeta\right)= & \frac{1}{k^{\frac{\tilde{N}}{2}}}\frac{i^{-\frac{1}{2}\tilde{N}}e^{-i\theta_{k-N}-i\theta_{\tilde{N}}+\frac{i\pi}{k}\tilde{N}\zeta^{2}}}{\prod_{a<a'}^{F}2\sinh\pi\left(m_{a}-m_{a'}\right)}\int d^{F}\kappa\prod_{a}^{F}\braket{\hbar m_{a}|e^{\frac{i}{2\hbar}\hat{p}^{2}}|\kappa_{a}}\nonumber \\
 & \times\frac{\prod_{a<a'}^{F}2\sinh\frac{\kappa_{a}-\kappa_{a'}}{2k}\prod_{j<j'}^{\tilde{N}}2\sinh\frac{\lambda_{j}-\lambda_{j'}}{2k}}{\prod_{a,j}^{F,\tilde{N}}2\cosh\frac{\kappa_{a}-\lambda_{j}}{2k}}.
\end{align}
where $\lambda_j$ are given in terms of the other parameters 
\begin{equation}
\lambda_{j}=-t_{\tilde{N},j}-2\pi\zeta.
\end{equation}
After substituting these values for the integrand, we arrive at
\begin{align}
\zU[\tilde{N}][][-k][0]\left(\bm{m},-\zeta\right)= & \zpureU[\tilde{N}][-k]\frac{i^{-\frac{1}{2}\tilde{N}^{2}}e^{-i\theta_{k-N}-i\theta_{\tilde{N}}+\frac{i\pi}{k}\tilde{N}\zeta^{2}}}{\prod_{a<a'}^{F}2\sinh\pi\left(m_{a}-m_{a'}\right)}\int d^{F}\kappa\prod_{a}^{F}\braket{\hbar m_{a}|e^{\frac{i}{2\hbar}\hat{p}^{2}}|\kappa_{a}}\nonumber \\
 & \times\prod_{a}^{F}\frac{1}{\prod_{j}^{\tilde{N}}2\cosh\frac{\kappa_{a}+2\pi\zeta-t_{\tilde{N},j}}{2k}}\prod_{a<a'}^{F}2\sinh\frac{\kappa_{a}-\kappa_{a'}}{2k}.
\end{align}

Now, we can obtain the equality between the electric and magnetic partition functions of the conventional Giveon-Kutasov duality, which corresponds to the $n=0$ case in our duality proposal. First, it is known that the supersymmetric version of the level-rank duality claims the equivalence between the 3d $\mathcal{N}=3$ $U(N)_{k,k}$ and $U(k-N)_{-k,-k}$ pure Chern-Simons gauge theories. The equivalence of the partition functions implies
\begin{equation}
\zpureU[N]=\zpureU[k-N][-k].\label{eq:pureCS-Dual}
\end{equation}
The direct proof of this identity is in \cite{Kapustin:2010mh}.\footnote{
The partition function of the 3d $\mathcal{N}=3$ $U(N)_k$ pure CS gauge theory is normalized as
$\zpureU[k]=1$.
}
By using the identity of the products of hyperbolic functions\footnote{
This identity is equivalent to
\begin{equation}
i^{L}2\cosh\frac{q-i\pi L}{2}=\left(\prod_{j}^{L}2\sinh\frac{q-t_{L,j}}{2k}\right)\left(\prod_{j}^{k-L}2\cosh\frac{q-t_{k-L,j}}{2k}\right),
\end{equation}
which can be easily proved using $z^k-1=\prod_j^k\left(z-e^{\frac{2\pi i}{k}j}\right)$.
}
\begin{equation}
\frac{1}{\prod_{j}^{N}2\cosh\frac{\kappa+2\pi\zeta-t_{N,j}}{2k}}
=i^{N-k}\frac{\prod_{j}^{k-N}2\sinh\frac{\kappa+2\pi\zeta-t_{k-N,j}}{2k}}{2\cosh\frac{\kappa+2\pi\zeta-i\pi (k-N )}{2}},\label{eq:HW-1k}
\end{equation}
we can rewrite the magnetic partition function as
\begin{align}
\zU[\tilde{N}][][-k][0]\left(\bm{m},-\zeta\right)= & \zpureU[N-F]\frac{i^{-\frac{1}{2}\tilde{N}^{2}}e^{-i\theta_{k-N}-i\theta_{\tilde{N}}+\frac{i\pi}{k}\tilde{N}\zeta^{2}}}{\prod_{a<a'}^{F}2\sinh\pi\left(m_{a}-m_{a'}\right)}\int d^{F}\kappa\prod_{a}^{F}\braket{\hbar m_{a}|e^{\frac{i}{2\hbar}\hat{p}^{2}}|\kappa_{a}}\nonumber \\
 & \times\prod_{a}^{F}i^{F-N}\frac{\prod_{j}^{N-F}2\sinh\frac{\kappa_{a}+2\pi\zeta-t_{N-F,j-F}}{2k}}{2\cosh\frac{\kappa_{a}+2\pi\zeta-i\pi\left(N-F\right)}{2}}\prod_{a<a'}^{F}2\sinh\frac{\kappa_{a}-\kappa_{a'}}{2k}\nonumber \\
 = & i^{-\frac{1}{2}\tilde{N}^{2}-\frac{1}{2}N^{2}+\frac{1}{2}F^{2}}e^{-i\theta_{k-N}-i\theta_{\tilde{N}}-i\theta_{N}-i\theta_{N-F}}e^{i\pi\zeta^{2}-i\pi k\sum_{a}^{F}m_{a}^{2}-2\pi i\zeta\sum_{a}^{F}m_{a}}\zU[][][][0]\left(\bm{m},\zeta\right).
\end{align}
As a result, we arrive at the integral identity \eqref{eq:LRwoDiag} for the $F \le N$ case, which exhibits the validity of the conventional Giveon-Kutasov duality.

Note that the electric partition function $\zU[][][][0]\left(\bm{m},\zeta\right)$ becomes trivial up to some phases for $F+k-N=0$.
By using \eqref{eq:pureCS-Dual} and \eqref{eq:HW-1k} in the expression \eqref{eq:eleFGF}, we obtain
\begin{align}
\zU[][][][0]\left(\bm{m},\zeta\right)= & \frac{i^{\frac{1}{2}N^{2}-\frac{1}{2}F^{2}}e^{i\theta_{N}+i\theta_{N-F}-i\pi\zeta^{2}}e^{i\pi k\sum_{a}^{F}m_{a}^{2}+2\pi i\zeta\sum_{a}^{F}m_{a}}}{\prod_{a<a'}^{F}2\sinh\pi\left(m_{a}-m_{a'}\right)}\nonumber \\
 & \times\int d^{F}\kappa\prod_{a}^{F}\braket{\hbar m_{a}|\kappa_{a}}\det\left(\left[\frac{\hbar}{\sqrt{k}}\brakket{\kappa_{a}|e^{\frac{i}{2\hbar}\hat{p}^{2}}|-t_{F,\ell}}\right]_{a,\ell}^{F\times F}\right),
\end{align}
where we used the determinant formula \eqref{eq:CauchyDet} and put the exponential factor $e^{\frac{i}{2\hbar}\hat{p}^{2}}$ inside the determinant. The exponential factor $e^{\frac{i}{2\hbar}\hat{p}^{2}}$ becomes a phase $e^{-i\theta_{F}}$. In this expression, we can perform all the integrals. The resulting determinant is completely the same as the term in the denominator after using the determinant formula \eqref{eq:CauchyDet}. As a result, we arrive at
\begin{equation}
\zU[][][][0]\left(\bm{m},\zeta\right)=i^{\frac{1}{2}N^{2}-\frac{1}{2}F^{2}}e^{i\theta_{N}+i\theta_{N-F}-i\theta_{F}-i\pi\zeta^{2}}e^{i\pi k\sum_{a}^{F}m_{a}^{2}+2\pi i\zeta\sum_{a}^{F}m_{a}}.\label{eq:eleZ-triv}
\end{equation}
This is consistent with the duality prediction since the magnetic theory for this case is also vanishing.

\subsubsection{The $F>N$ case}
Next, we consider the $F>N$ case, where supersymmetry is not broken and then the partition function is always well-defined since the inequality $F+k-N\geq 0$ is automatically satisfied for any integer $k$. The strategy for proving the integral identity \eqref{eq:LRwoDiag} is almost the same as the previous case. 
We start with the calculation of the electric partition function. After the shift of the integration variables $\lambda_j \rightarrow\frac{\lambda_j}{\hbar}$, where we take $\hbar=2\pi k$, the electric side of \eqref{eq:LRwoDiag} becomes
\begin{align}
\zU[][][][0]\left(\bm{m},\zeta\right)= & \int\frac{d^{N}\lambda}{\hbar^{N}N!}e^{\frac{i}{2\hbar}\sum_{j}^{N}\lambda_{j}^{2}+\frac{2\pi i\zeta}{\hbar}\sum_{j}^{N}\lambda_{j}}\frac{\prod_{j<j'}^{N}\left(2\sinh\frac{\lambda_{j}-\lambda_{j'}}{2k}\right)^{2}}{\prod_{a,j}^{F,N}2\cosh\frac{\lambda_{j}-\hbar m_{a}}{2k}}.
\end{align}
As in the previous case, we massage the integrand by inserting delta function identities. In this case, we need to insert two sets of delta functions, $1=\int d^{N}\kappa\prod_{a}^{N}\delta\left(\kappa_{a}-\hbar m_{a}\right)$ and $1=\int d^{F-N}\rho\prod_{a}^{F-N}\delta\left(\rho_{a}-\hbar m_{N+a}\right)$. The integrand is now changed as follows:
\begin{align}
\zU[][][][0]\left(\bm{m},\zeta\right)= & \frac{1}{\prod_{a<a'}^{N}2\sinh\pi\left(m_{a}-m_{a'}\right)\prod_{a<a'}^{F-N}2\sinh\pi\left(m_{N+a}-m_{N+a'}\right)}\nonumber \\
 & \times\int d^{N}\kappa\int\frac{d^{N}\lambda}{\hbar^{N}N!}\int d^{F-N}\rho\, e^{\frac{i}{2\hbar}\sum_{j}^{N}\lambda_{j}^{2}+\frac{2\pi i\zeta}{\hbar}\sum_{j}^{N}\lambda_{j}}\nonumber \\
 & \times\prod_{a}^{N}\delta\left(\kappa_{a}-\hbar m_{a}\right)\frac{\prod_{a<a'}^{N}2\sinh\frac{\kappa_{a}-\kappa_{a'}}{2k}\prod_{j<j'}^{N}2\sinh\frac{\lambda_{j}-\lambda_{j'}}{2k}}{\prod_{a,j}^{N,N}2\cosh\frac{\kappa_{a}-\lambda_{j}}{2k}}\nonumber \\
 & \times\frac{\prod_{j<j'}^{N}2\sinh\frac{\lambda_{j}-\lambda_{j'}}{2k}\prod_{a<a'}^{F-N}2\sinh\frac{\rho_{a}-\rho_{a'}}{2k}}{\prod_{j,a}^{N,F-N}2\cosh\frac{\lambda_{j}-\rho_{a}}{2k}}\prod_{a}^{F-N}\delta\left(\rho_{a}-\hbar m_{N+a}\right).\label{eq:MM-FGF3}
\end{align}
Note that this expression also has a nice brane interpretation as in the previous case.
\begin{figure}
\begin{centering}
\includegraphics[scale=0.65]{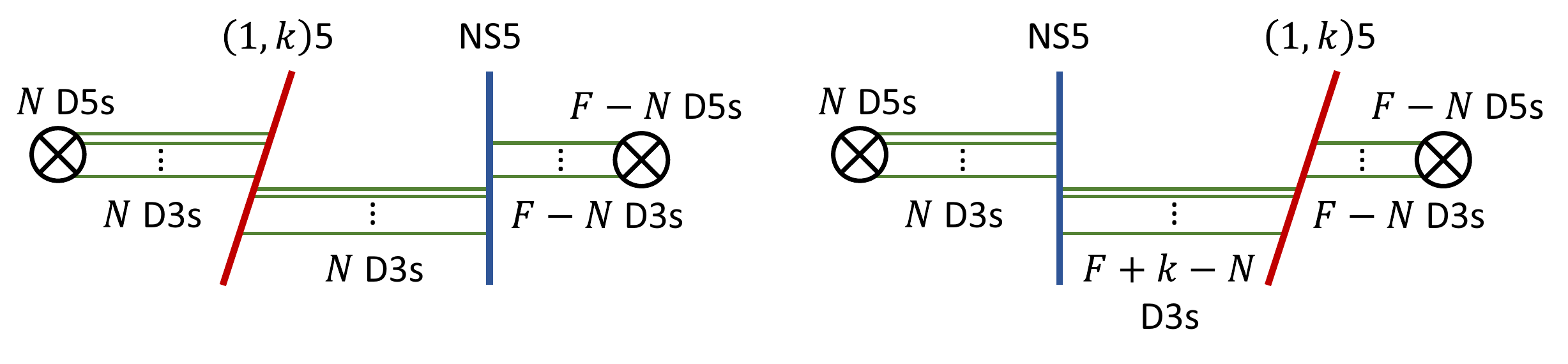}
\par\end{centering}
\caption{The brane configurations which correspond to \eqref{eq:MM-FGF3} (left) and \eqref{eq:MM-FGF4} (right). These two brane configurations are related  with each other via the Hanany-Witten move.
\label{fig:FgreatN}}
\end{figure}
The difference is that, in the present case, we move $N$ D5-branes to the left of the $(1,k)5$-brane and $F-N$ D5-branes to the right of the NS5-brane as drawn in Figure \ref{fig:FgreatN}.
The vector multiplets living on the D3-branes on the left and right sides correspond to the $\kappa$ and $\rho$ integrals, respectively.

By using the determinant formula \eqref{eq:CauchyDet} in the operator formalism, the above expression is transformed into an integrand with two determinant factors 
\begin{align}
\zU[][][][0]\left(\bm{m},\zeta\right)= & \frac{1}{\prod_{a<a'}^{N}2\sinh\pi\left(m_{a}-m_{a'}\right)\prod_{a<a'}^{F-N}2\sinh\pi\left(m_{N+a}-m_{N+a'}\right)}\nonumber \\
 & \times\int d^{N}\kappa\int\frac{d^{N}\lambda}{\hbar^{N}N!}\int d^{F-N}\rho\, e^{\frac{i}{2\hbar}\sum_{j}^{N}\lambda_{j}^{2}+\frac{2\pi i\zeta}{\hbar}\sum_{j}^{N}\lambda_{j}}\prod_{a}^{N}\braket{\hbar m_{a}|\kappa_{a}}\nonumber \\
 & \times\det\left(\left[\hbar\braket{\kappa_{a}|\frac{1}{2\cosh\frac{\hat{p}}{2}}|\lambda_{j}}\right]_{a,j}^{N\times N}\right)\Psi\left(\lambda,\rho\right)\prod_{a}^{F-N}\braket{\rho_a|\hbar m_{N+a}},
\end{align}
where the second determinant factor $\Psi (\lambda,\rho)$ is defined by 
\begin{align}
&\Psi\left(\lambda,\rho\right) \nonumber \\
& =\frac{\prod_{j<j'}^{N}2\sinh\frac{\lambda_{j}-\lambda_{j'}}{2k}\prod_{a<a'}^{F-N}2\sinh\frac{\rho_{a}-\rho_{a'}}{2k}}{\prod_{j,a}^{N,F-N}2\cosh\frac{\lambda_{j}-\rho_{a}}{2k}}\nonumber \\
 & =\begin{cases}
\det\left(\begin{array}{cc}
\left[\hbar\braket{\lambda_{j}|\frac{1}{2\cosh\frac{\hat{p}-i\pi\left(2N-F\right)}{2}}|\rho_{a}}\right]_{j,a}^{N\times\left(F-N\right)} & \left[\frac{\hbar}{\sqrt{k}}\brakket{\lambda_{j}|-t_{2N-F,\ell}}\right]_{j,\ell}^{N\times\left(2N-F\right)}\end{array}\right) & \left(F\leq2N\right)\\
\det\left(\begin{array}{c}
\left[\hbar\braket{\lambda_{j}|\frac{1}{2\cosh\frac{\hat{p}+i\pi\left(F-2N\right)}{2}}|\rho_{a}}\right]_{j,a}^{N\times\left(F-N\right)}\\
\left[\frac{\hbar}{\sqrt{k}}\bbraket{t_{F-2N,\ell}|\rho_{a}}\right]_{\ell,a}^{\left(F-2N\right)\times\left(F-N\right)}
\end{array}\right) & \left(2N<F\right)
\end{cases}.
\end{align}
After bringing the exponential factor $e^{\frac{i}{2\hbar}\sum_{j}^{N}\lambda_{j}^{2}+\frac{2\pi i\zeta}{\hbar}\sum_{j}^{N}\lambda_{j}}$ to the inside of the first determinant and rewriting it into the operator form, we obtain
\begin{align}
\zU[][][][0]\left(\bm{m},\zeta\right)= & \frac{1}{\prod_{a<a'}^{N}2\sinh\pi\left(m_{a}-m_{a'}\right)\prod_{a<a'}^{F-N}2\sinh\pi\left(m_{N+a}-m_{N+a'}\right)}\nonumber \\
 & \times\int d^{N}\kappa\int\frac{d^{N}\lambda}{\hbar^{N}N!}\int d^{F-N}\rho\prod_{a}^{N}\braket{\hbar m_{a}|\kappa_{a}}\nonumber \\
 & \times\det\left(\left[\hbar\braket{\kappa_{a}|\frac{1}{2\cosh\frac{\hat{p}}{2}}e^{\frac{2\pi i\zeta}{\hbar}\hat{q}}e^{\frac{i}{2\hbar}\hat{q}^{2}}|\lambda_{j}}\right]_{a,j}^{N\times N}\right)\Psi\left(\lambda,\rho\right)\prod_{a}^{F-N}\braket{\rho_{a}|\hbar m_{N+a}},
\end{align}
By performing the similarity transformations
\begin{align}
\int d\kappa\ket{\kappa}\bra{\kappa} & =\int d\kappa\, e^{\frac{2\pi i\zeta}{\hbar}\hat{q}}e^{\frac{i}{2\hbar}\hat{q}^{2}}e^{\frac{i}{2\hbar}\hat{p}^{2}}\ket{\kappa}\bra{\kappa}e^{-\frac{i}{2\hbar}\hat{p}^{2}}e^{-\frac{i}{2\hbar}\hat{q}^{2}}e^{-\frac{2\pi i\zeta}{\hbar}\hat{q}},\nonumber \\
\int d\lambda\ket{\lambda}\bra{\lambda} & =\int d\lambda\, e^{\frac{i}{2\hbar}\hat{p}^{2}}\ket{\lambda}\bra{\lambda}e^{-\frac{i}{2\hbar}\hat{p}^{2}},\quad
\int d\lambda\ket{\rho}\bra{\rho}  =\int d\rho\, e^{\frac{i}{2\hbar}\hat{p}^{2}}\ket{\rho}\bra{\rho}e^{-\frac{i}{2\hbar}\hat{p}^{2}},
\end{align}
and by using \eqref{eq:OpSim} and \eqref{eq:VecSim}, we obtain
\begin{align}
\zU[][][][0]\left(\bm{m},\zeta\right)= & \frac{e^{i\theta_{2N-F}}e^{i\pi k\sum_{a}^{N}m_{a}^{2}+2\pi i\zeta\sum_{a}^{N}m_{a}}}{\prod_{a<a'}^{N}2\sinh\pi\left(m_{a}-m_{a'}\right)\prod_{a<a'}^{F-N}2\sinh\pi\left(m_{N+a}-m_{N+a'}\right)}\nonumber \\
 & \times\int d^{F}\kappa\int\frac{d^{N}\lambda}{\hbar^{N}N!}\int d^{F-N}\rho\prod_{a}^{N}\braket{\hbar m_{a}|e^{\frac{i}{2\hbar}\hat{p}^{2}}|\kappa_{a}}\nonumber \\
 & \times\det\left(\left[\hbar\braket{\kappa_{a}|\frac{1}{2\cosh\frac{\hat{q}+2\pi\zeta}{2}}|\lambda_{j}}\right]_{a,j}^{N\times N}\right)\Psi\left(\lambda,\rho\right)\prod_{a}^{F-N}\braket{\rho_{a}|e^{-\frac{i}{2\hbar}\hat{p}^{2}}|\hbar m_{N+a}}.
\end{align}
After diagonalizing the $N \times N$ matrix in the first determinant by using \eqref{eq:DiagForm}, we can perform the $d^N \lambda$ integral since it just becomes an $N$-dimensional delta function. As a result, we arrive at
\begin{align}
\zU[][][][0]\left(\bm{m},\zeta\right)= & \frac{e^{i\theta_{2N-F}}e^{i\pi k\sum_{a}^{N}m_{a}^{2}+2\pi i\zeta\sum_{a}^{N}m_{a}}}{\prod_{a<a'}^{N}2\sinh\pi\left(m_{a}-m_{a'}\right)\prod_{a<a'}^{F-N}2\sinh\pi\left(m_{N+a}-m_{N+a'}\right)}\int d^{F}\kappa\int d^{F-N}\rho\nonumber \\
 & \times\prod_{a}^{N}\braket{\hbar m_{a}|e^{\frac{i}{2\hbar}\hat{p}^{2}}|\kappa_{a}}\prod_{a}^{N}\frac{1}{2\cosh\frac{\kappa_{a}+2\pi\zeta}{2}}\Psi\left(\kappa,\rho\right)\prod_{a}^{F-N}\braket{\rho_{a}|e^{-\frac{i}{2\hbar}\hat{p}^{2}}|\hbar m_{N+a}}. \label{Final_Z_ele_NFk0}
\end{align}
In what follows, we will see that this expression is directly compared with the magnetic partition function.

Let us move on to the calculation of the magnetic partition function $\zU[\tilde{N}][][-k][0]\left(\bm{m},-\zeta\right)$ for the $F>N$ case. 
As in the electric side, by using the determinant formula \eqref{eq:CauchyDet}, the magnetic side of the integral identity \eqref{eq:LRwoDiag} can be reformulated into
\begin{align}
&\zU[\tilde{N}][][-k][0]\left(\bm{m},-\zeta\right)\nonumber \\
&=\frac{1}{k^{\frac{k}{2}}} \frac{1}{\prod_{a<a'}^{N}2\sinh\pi\left(m_{a}-m_{a'}\right)\prod_{a<a'}^{F-N}2\sinh\pi\left(m_{N+a}-m_{N+a'}\right)}\nonumber \\
 &\quad \times\int d^{N}\kappa\int\frac{d^{\tilde{N}}\lambda}{\tilde{N}!}\int d^{F-N}\rho \prod_{a}^{N}\braket{\hbar m_{a}|\kappa_{a}}\Psi'\left(\kappa,\lambda\right)\nonumber \\
 &\quad \times\det\left(\left[\braket{\lambda_{j}|e^{-\frac{i}{2\hbar}\hat{q}^{2}}e^{-\frac{2\pi i\zeta}{\hbar}\hat{q}}\frac{1}{2\cosh\frac{\hat{p}+i\pi k}{2}}|\rho_{a}}\right]_{j,a}^{\tilde{N}\times\left(\tilde{N}-k\right)}\left[\brakket{\lambda_{j}|e^{-\frac{i}{2\hbar}\hat{q}^{2}}e^{-\frac{2\pi i\zeta}{\hbar}\hat{q}}|-t_{k,\ell}}\right]_{j,\ell}^{\tilde{N}\times k}\right)\nonumber \\
 &\quad \times\prod_{a}^{F-N}\braket{\rho_a|\hbar m_{N+a}}, \label{eq:MM-FGF4}
\end{align}
where the first determinant factor is defined as
\begin{align}
&\Psi'\left(\kappa,\lambda\right) \nonumber \\
& =\frac{\prod_{a<a'}^{N}2\sinh\frac{\kappa_{a}-\kappa_{a'}}{2k}\prod_{j<j'}^{\tilde{N}}2\sinh\frac{\lambda_{j}-\lambda_{j'}}{2k}}{\prod_{a,j}^{N,\tilde{N}}2\cosh\frac{\kappa_{a}-\lambda_{j}}{2k}}\nonumber \\
 & =\begin{cases}
\det\left(\begin{array}{c}
\left[\hbar\braket{\kappa_{a}|\frac{1}{2\cosh\frac{\hat{p}-i\pi\left(\tilde{N}-N\right)}{2}}|\lambda_{j}}\right]_{a,j}^{N\times\tilde{N}}\\
\left[\frac{\hbar}{\sqrt{k}}\bbraket{t_{\tilde{N}-N,\ell}|\lambda_{j}}\right]_{\ell,j}^{\left(\tilde{N}-N\right)\times\tilde{N}}
\end{array}\right) & \left(\tilde{N}\geq N\right)\\
\det\left(\begin{array}{cc}
\left[\hbar\braket{\kappa_{a}|\frac{1}{2\cosh\frac{\hat{p}+i\pi\left(N-\tilde{N}\right)}{2}}|\lambda_{j}}\right]_{a,j}^{N\times\tilde{N}} & \left[\frac{\hbar}{\sqrt{k}}\brakket{\kappa_{a}|-t_{N-\tilde{N},\ell}}\right]_{a,\ell}^{N\times\left(N-\tilde{N}\right)}\end{array}\right) & \left(\tilde{N}<N\right)
\end{cases}.
\end{align}
Note that this expression also has a brane interpretation as shown in Figure \ref{fig:FgreatN}.
By performing the similarity transformation 
\begin{align}
\int d\kappa\ket{\kappa}\bra{\kappa} & =\int d\kappa\, e^{\frac{i}{2\hbar}\hat{p}^{2}}\ket{\kappa}\bra{\kappa}e^{-\frac{i}{2\hbar}\hat{p}^{2}},\quad
\int d\lambda\ket{\lambda}\bra{\lambda}  =\int d\lambda\, e^{\frac{i}{2\hbar}\hat{p}^{2}}\ket{\lambda}\bra{\lambda}e^{-\frac{i}{2\hbar}\hat{p}^{2}},\nonumber \\
\int d\rho\ket{\rho}\bra{\rho} & =\int d\rho\, e^{\frac{2\pi i\zeta}{\hbar}\hat{q}}e^{\frac{i}{2\hbar}\hat{q}^{2}}e^{\frac{i}{2\hbar}\hat{p}^{2}}\ket{\rho}\bra{\rho}e^{-\frac{i}{2\hbar}\hat{p}^{2}}e^{-\frac{i}{2\hbar}\hat{q}^{2}}e^{-\frac{2\pi i\zeta}{\hbar}\hat{q}},
\end{align}
and by using \eqref{eq:OpSim} and \eqref{eq:VecSim}, we obtain
\begin{align}
\zU[\tilde{N}][][-k][0]\left(\bm{m},-\zeta\right)= &\frac{1}{k^{\frac{k}{2}}} \frac{i^{-\frac{1}{2}k}e^{-i\theta_{\tilde{N}-N}-i\theta_{k}+i\pi\zeta^{2}}e^{-i\pi k\sum_{a=N+1}^{F}m_{a}^{2}-2\pi i\zeta\sum_{a=N+1}^{F}m_{a}}}{\prod_{a<a'}^{N}2\sinh\pi\left(m_{a}-m_{a'}\right)\prod_{a<a'}^{F-N}2\sinh\pi\left(m_{N+a}-m_{N+a'}\right)}\int d^{N}\kappa\int\frac{d^{\tilde{N}}\lambda}{\tilde{N}!}\nonumber \\
 &\times\int d^{F-N}\rho\prod_{a}^{N}\braket{\hbar m_{a}|e^{\frac{i}{2\hbar}\hat{p}^{2}}|\kappa_{a}}\frac{\prod_{a<a'}^{N}2\sinh\frac{\kappa_{a}-\kappa_{a'}}{2k}\prod_{j<j'}^{\tilde{N}}2\sinh\frac{\lambda_{j}-\lambda_{j'}}{2k}}{\prod_{a,j}^{N,\tilde{N}}2\cosh\frac{\kappa_{a}-\lambda_{j}}{2k}}\nonumber \\
 & \times\det\left(\begin{array}{cc}
\left[\braket{\lambda_{j}|\frac{1}{2\cosh\frac{\hat{q}+2\pi\zeta+i\pi k}{2}}|\rho_{a}}\right]_{j,a}^{\tilde{N}\times\left(\tilde{N}-k\right)} & \left[\braket{\lambda_{j}|-t_{k,\ell}-2\pi\zeta}\right]_{j,\ell}^{\tilde{N}\times k}\end{array}\right)\nonumber \\
 & \times\prod_{a}^{F-N}\braket{\rho_a|e^{-\frac{i}{2\hbar}\hat{p}^{2}}|\hbar m_{N+a}}.
\end{align}
After diagonalizing the matrix inside the determinant on the third line by using \eqref{eq:DiagForm}, we can perform the $d^{\tilde{N}} \lambda$ integral by using the delta functions. As a result, we obtain
\begin{align}
\zU[\tilde{N}][][-k][0]\left(\bm{m},-\zeta\right)= & \frac{1}{k^{\frac{k}{2}}}\frac{i^{-\frac{1}{2}k}e^{-i\theta_{\tilde{N}-N}-i\theta_{k}+i\pi\zeta^{2}}e^{-i\pi k\sum_{a=N+1}^{F}m_{a}^{2}-2\pi i\zeta\sum_{a=N+1}^{F}m_{a}}}{\prod_{a<a'}^{N}2\sinh\pi\left(m_{a}-m_{a'}\right)\prod_{a<a'}^{F-N}2\sinh\pi\left(m_{N+a}-m_{N+a'}\right)}\nonumber \\
 & \times\int d^{N}\kappa\int d^{F-N}\rho\prod_{a}^{N}\braket{\hbar m_{a}|e^{\frac{i}{2\hbar}\hat{p}^{2}}|\kappa_{a}}\frac{\prod_{a<a'}^{N}2\sinh\frac{\kappa_{a}-\kappa_{a'}}{2k}\prod_{j<j'}^{\tilde{N}}2\sinh\frac{\lambda_{j}-\lambda_{j'}}{2k}}{\prod_{a,j}^{N,\tilde{N}}2\cosh\frac{\kappa_{a}-\lambda_{j}}{2k}}\nonumber \\
 & \times\prod_{a}^{F-N}\frac{1}{2\cosh\frac{\rho_{a}+2\pi\zeta+i\pi k}{2}}\prod_{a}^{F-N}\braket{\rho_a|e^{-\frac{i}{2\hbar}\hat{p}^{2}}|\hbar m_{N+a}}.
\end{align}
where $\lambda_{j}$ are no longer independent variables but rather fixed parameters
\begin{equation}
\lambda_{j}=\begin{cases}
\rho_{j} & \left(1\leq j\leq F-N\right)\\
-t_{k,j-F}-2\pi\zeta. & \left(F-N+1\leq j\leq\tilde{N}\right)
\end{cases}.
\end{equation}
By inserting these expressions into the above matrix integral, we arrive at
\begin{align}
\zU[\tilde{N}][][-k][0]\left(\bm{m},-\zeta\right)= & \zpureU[k][-k]\frac{i^{-\frac{1}{2}\tilde{N}^{2}+\frac{1}{2}\left(F-N\right)^{2}}e^{-i\theta_{\tilde{N}-N}-i\theta_{k}+i\pi\zeta^{2}}e^{-i\pi k\sum_{a=N+1}^{F}m_{a}^{2}-2\pi i\zeta\sum_{a=N+1}^{F}m_{a}}}{\prod_{a<a'}^{N}2\sinh\pi\left(m_{a}-m_{a'}\right)\prod_{a<a'}^{F-N}2\sinh\pi\left(m_{N+a}-m_{N+a'}\right)}\nonumber \\
 & \times\int d^{N}\kappa\int d^{F-N}\rho\prod_{a}^{N}\braket{\hbar m_{a}|e^{\frac{i}{2\hbar}\hat{p}^{2}}|\kappa_{a}}\prod_{a}^{N}\frac{1}{\prod_{j}^{k}2\cosh\frac{\kappa_{a}+2\pi\zeta-t_{k,j}}{2k}}\Psi\left(\kappa,\rho\right)\nonumber \\
 & \times\prod_{a}^{F-N}i^{k}\frac{\prod_{j}^{k}2\sinh\frac{\rho_{a}-t_{k,j}}{2k}}{2\cosh\frac{\rho_{a}+2\pi\zeta+i\pi k}{2}}\prod_{a}^{F-N}\braket{\rho_a|e^{-\frac{i}{2\hbar}\hat{p}^{2}}|\hbar m_{N+a}}.
\end{align}
We can now compare this result to the electric one  \eqref{Final_Z_ele_NFk0}.
By using \eqref{eq:pureCS-Dual} and \eqref{eq:HW-1k}, we obtain
\begin{align}
\zU[\tilde{N}][][-k][0]\left(\bm{m},-\zeta\right)= & \frac{i^{-\frac{1}{2}\tilde{N}^{2}+\frac{1}{2}\left(F-N\right)^{2}}e^{-i\theta_{\tilde{N}-N}-i\theta_{k}+i\pi\zeta^{2}}e^{-i\pi k\sum_{a=N+1}^{F}m_{a}^{2}-2\pi i\zeta\sum_{a=N+1}^{F}m_{a}}}{\prod_{a<a'}^{N}2\sinh\pi\left(m_{a}-m_{a'}\right)\prod_{a<a'}^{F-N}2\sinh\pi\left(m_{N+a}-m_{N+a'}\right)}\int d^{N}\kappa\int d^{F-N}\rho\nonumber \\
 & \times\prod_{a}^{N}\braket{\hbar m_{a}|e^{\frac{i}{2\hbar}\hat{p}^{2}}|\kappa_{a}}\prod_{a}^{N}\frac{1}{2\cosh\frac{\kappa_{a}+2\pi\zeta}{2}}\Psi\left(\kappa,\rho\right)\prod_{a}^{F-N}\braket{\rho_a|e^{-\frac{i}{2\hbar}\hat{p}^{2}}|\hbar m_{N+a}}\nonumber \\
 =& i^{-\frac{1}{2}\tilde{N}^{2}+\frac{1}{2}\left(F-N\right)^{2}}e^{-i\theta_{\tilde{N}-N}-i\theta_{k}-i\theta_{2N-F}}e^{i\pi\zeta^{2}-i\pi k\sum_{a}^{F}m_{a}^{2}-2\pi i\zeta\sum_{a}^{F}m_{a}}\zU[][][][0]\left(\bm{m},\zeta\right).
\end{align}
Thus, we prove that the integral identity \eqref{eq:LRwoDiag} holds also for the $F > N$ case.

\subsection{The $U(N)_{0,0}$ ``ugly-good'' duality}\label{Z_CUG}
In this subsection, we study the $U(N)_{k,k+nN}$ Seiberg-like duality with $n=k=0$ and $F=2N-1$ by using the sphere partition function. The theory is called ``ugly'' since the monopole operator becomes a free field and decouples from the other sector \cite{Gaiotto:2008ak}. This is the duality studied in \cite{Kapustin:2010mh, Yaakov:2013fza, Assel:2017jgo} and known as the ``ugly-good'' duality. The duality claims that the 3d $\mathcal{N}=4$ $U(N)_{0,0}$ gauge theory with $2N-1$ fundamental flavors is infrared-dual to the 3d $\mathcal{N}=4$ $U(N-1)_{0,0}$ ``good'' gauge theory with $2N-1$ flavors and a free hypermultiplet. At the level of the partition functions, the duality leads to the following integral identity \cite{Kapustin:2010mh}:
\begin{equation}
\zU[][2N-1][0][0]\left(\bm{m},\zeta\right)=\frac{e^{2\pi i\zeta\sum_{a}^{2N-1}m_{a}}}{2\cosh\pi\zeta}\zU[N-1][2N-1][0][0]\left(\bm{m},-\zeta\right),\label{eq:GUwoDiag}
\end{equation}
where the exponential factor is a mixed CS term for the global symmetries. Since the good theory, the right-hand side of \eqref{eq:GUwoDiag}, is equipped with the free monopole hypermultiplet, the $\cosh$ function is located outside the integral.

In order to derive this equality, we elaborate on the matrix model $\zU[][F][0][0]\left(\bm{m},\zeta\right)$ in the region $F\geq 2N-1$.\footnote{
When $F\leq 2N-2$, the matrix integral \eqref{eq:Zele-Def} diverges since the integrand in this case does not approach to zero for $|\lambda_j|\rightarrow \infty$. In this way, the partition function also becomes ``bad''.
}
In what follows, we set $\hbar=1$. By inserting the identity $1=\int d^{F}\kappa\prod_{a}\delta\left(\kappa_{a}-m_{a}\right)$, the partition function of the ``ugly'' side becomes
\begin{align}
\zU[][][0][0]\left(\bm{m},\zeta\right)= & \frac{1}{\prod_{a<a'}^{F}2\sinh\pi\left(m_{a}-m_{a'}\right)}\int\frac{d^{N}\lambda}{N!}\int d^{F}\kappa\, e^{2\pi i\zeta\sum_{j}^{N}\lambda_{j}}\prod_{j<j'}^{N}
2\sinh\pi \left( \lambda_{j}-\lambda_{j'} \right)
\nonumber \\
 & \times\frac{\prod_{j<j'}^{N}2\sinh\pi\left(\lambda_{j}-\lambda_{j'}\right)\prod_{a<a'}^{F}2\sinh\pi\left(\kappa_{a}-\kappa_{a'}\right)}{\prod_{j,a}^{N,F}2\cosh\pi\left(\lambda_{j}-\kappa_{a}\right)}\prod_{a}^{F}\delta\left(\kappa_{a}-m_{a}\right).
\end{align}
By anti-symmetrizing the last factor $\prod_{a}^{F}\delta\left(\kappa_{a}-m_{a}\right)$ and using the determinant formula \eqref{eq:CauchyDet} with $\hbar=1$ and $k=\frac{1}{2\pi}$, we obtain
\begin{align}
\zU[][][0][0]\left(\bm{m},\zeta\right)= & \frac{1}{\prod_{a<a'}^{F}2\sinh\pi\left(m_{a}-m_{a'}\right)}\int\frac{d^{N}\lambda}{N!}\int\frac{d^{F}\kappa}{F!}e^{2\pi i\zeta\sum_{j}^{N}\lambda_{j}}
 \det\left(\left[\sqrt{2\pi}\bbraket{t_{N,\ell}|\lambda_{j}}\right]_{\ell,j}^{N\times N}\right)\nonumber \\
 & \times\det\left(\begin{array}{c}
\left[\braket{\lambda_{j}|\frac{1}{2\cosh\frac{\hat{p}-i\pi\left(F-N\right)}{2}}|\kappa_{a}}\right]_{j,a}^{N\times F}\\
\left[\sqrt{2\pi}\bbraket{t_{F-N,\ell}|\kappa_{a}}\right]_{\ell,a}^{\left(F-N\right)\times F}
\end{array}\right)\det\left(\left[\braket{\kappa_{a}|m_{b}}\right]_{a,b}^{F\times F}\right).
\end{align}
By bringing the exponential FI parameter into the inside of the first determinant, the integration variables $\lambda_j$ can be changed into the operator form. This results in the shift of the momentum vector as
\begin{equation}
\bbraket{t_{N,\ell}|e^{2\pi i\zeta\hat{q}}|\lambda_{j}}=\bbraket{t_{N,\ell}-2\pi\zeta|\lambda_{j}}.
\end{equation}
We can diagonalize the first and the second matrices by using \eqref{eq:DiagForm} and then evaluate the integral over $\kappa$ by using the resolution of the identity
\begin{equation}
\int d\kappa_{a}\ket{\kappa_{a}}\bra{\kappa_{a}}=1.
\end{equation}
On the other hand, we have to be careful of the $\lambda_{j}$ integral. After performing the similarity transformation
\begin{equation}
\int d\lambda_{j}\ket{\lambda_{j}}\bra{\lambda_{j}}=\int d\lambda_{j}\kket{\lambda_{j}}\bbra{\lambda_{j}},
\end{equation}
we obtain
\begin{align}
\zU[][][0][0]\left(\bm{m},\zeta\right)= & \frac{\left(2\pi\right)^{\frac{F}{2}}}{\prod_{a<a'}^{F}2\sinh\pi\left(m_{a}-m_{a'}\right)}\int\frac{d^{N}\lambda}{N!}\prod_{j}^{N}\delta\left(\lambda_{j}-t_{N,j}+2\pi\zeta\right)\nonumber \\
 & \times\prod_{j}^{N}\frac{1}{2\cosh\frac{\lambda_{j}-i\pi\left(F-N\right)}{2}}\det\left(\begin{array}{c}
\left[\bbraket{\lambda_{j}|m_{a}}\right]_{j,a}^{N\times F}\\
\left[\bbraket{t_{F-N,\ell}|m_{a}}\right]_{\ell,a}^{\left(F-N\right)\times F}
\end{array}\right).
\end{align}
To perform the $\lambda_{j}$ integral, we have to shift the integral contour from $\mathbb{R}$ to $\mathbb{R}+t_{N,j}$, so that we can evaluate the delta function. The ``not-bad'' condition $F\geq2N-1$ guarantees that the second line does not have any poles in the region around where we move the contour of integration, so that no pole hits the contour and then no residue appears. After evaluating the integral over $\lambda_{j}$ by using the delta functions, we obtain
\begin{align}
\zU[][][0][0]\left(\bm{m},\zeta\right)= & \frac{\left(2\pi\right)^{\frac{F}{2}}}{\prod_{a<a'}^{F}2\sinh\pi\left(m_{a}-m_{a'}\right)}\nonumber \\
 & \times\prod_{j}^{N}\frac{1}{2\cosh\frac{t_{N,j}-2\pi\zeta-i\pi\left(F-N\right)}{2}}\det\left(\begin{array}{c}
\left[\bbraket{t_{N,\ell}-2\pi\zeta|m_{a}}\right]_{\ell,a}^{N\times F}\\
\left[\bbraket{t_{F-N,\ell}|m_{a}}\right]_{\ell,a}^{\left(F-N\right)\times F}
\end{array}\right).
\end{align}
After some short calculations, we finally arrive at
\begin{align}
\zU[][][0][0]\left(\bm{m},\zeta\right)= & \frac{\left(-1\right)^{\frac{1}{2}N\left(N-1\right)}\left(2\pi\right)^{\frac{F}{2}}}{\prod_{a<a'}^{F}2\sinh\pi\left(m_{a}-m_{a'}\right)}\nonumber \\
 & \times\left(\frac{1}{2\cosh\left(\pi\zeta+\frac{i\pi}{2}\left(F-1\right)\right)}\right)^{N}\det\left(\begin{array}{c}
\left[\bbraket{t_{N,\ell}-2\pi\zeta|m_{a}}\right]_{\ell,a}^{N\times F}\\
\left[\bbraket{t_{F-N,\ell}|m_{a}}\right]_{\ell,a}^{\left(F-N\right)\times F}
\end{array}\right). \label{Z_final_form_ungly_bad}
\end{align}
Notice that the $\cosh \left( \pi \zeta +\frac{i\pi}{2}(F-1) \right)$ factor becomes zero for any even integer $F$ in the $\zeta \rightarrow 0$ limit since it is proportional to $\sinh \pi \zeta$. This is cured by the determinant factor which becomes zero in the $\zeta \rightarrow 0$ limit since its matrix has the same row. 

We find that the expression \eqref{Z_final_form_ungly_bad} is useful to prove the ``ugly-good'' duality and the corresponding integral identity \eqref{eq:GUwoDiag}. For $F=2N-1$, the electric side becomes
\begin{align}
\zU[][2N-1][0][0]\left(\bm{m},\zeta\right)= & \frac{\left(-1\right)^{\frac{1}{2}N\left(N-1\right)}\left(2\pi\right)^{\frac{2N-1}{2}}}{\prod_{a<a'}^{2N-1}2\sinh\pi\left(m_{a}-m_{a'}\right)}\nonumber \\
 & \times\left(\frac{1}{2\cosh\pi\zeta}\right)^{N}\det\left(\begin{array}{c}
\left[\bbraket{t_{N,\ell}-2\pi\zeta|m_{a}}\right]_{\ell,a}^{N\times\left(2N-1\right)}\\
\left[\bbraket{t_{N-1,\ell}|m_{a}}\right]_{\ell,a}^{\left(N-1\right)\times\left(2N-1\right)}
\end{array}\right),
\end{align}
while the magnetic side becomes
\begin{align}
\zU[N-1][2N-1][0][0]\left(\bm{m},-\zeta\right)= & \frac{\left(-1\right)^{\frac{1}{2}N\left(N-1\right)}\left(2\pi\right)^{\frac{2N-1}{2}}}{\prod_{a<a'}^{2N-1}2\sinh\pi\left(m_{a}-m_{a'}\right)}\nonumber \\
 & \times\left(\frac{1}{2\cosh\pi\zeta}\right)^{N-1}\det\left(\begin{array}{c}
\left[\bbraket{t_{N,\ell}+2\pi\zeta|m_{a}}\right]_{\ell,a}^{\left(N-1\right)\times\left(2N-1\right)}\\
\left[\bbraket{t_{N-1,\ell}|m_{a}}\right]_{\ell,a}^{N\times\left(2N-1\right)}
\end{array}\right).
\end{align}
Since these determinants satisfy
\begin{equation}
\det\left(\begin{array}{c}
\left[\bbraket{t_{N,\ell}-2\pi\zeta|m_{a}}\right]_{\ell,a}^{N\times\left(2N-1\right)}\\
\left[\bbraket{t_{N-1,\ell}|m_{a}}\right]_{\ell,a}^{\left(N-1\right)\times\left(2N-1\right)}
\end{array}\right)=e^{2\pi i\zeta\sum_{a}^{2N-1}m_{a}}\det\left(\begin{array}{c}
\left[\bbraket{t_{N,\ell}+2\pi\zeta|m_{a}}\right]_{\ell,a}^{\left(N-1\right)\times\left(2N-1\right)}\\
\left[\bbraket{t_{N-1,\ell}|m_{a}}\right]_{\ell,a}^{N\times\left(2N-1\right)}
\end{array}\right),
\end{equation}
we finally arrive at the integral identity \eqref{eq:GUwoDiag}.

\printbibliography

\end{document}